\documentclass[fleqn]{LMCS}

\def\doi{9(1:14)2013}
\lmcsheading%
{\doi}
{1--38}
{}
{}
{Jul.~\phantom02, 2010}
{Mar.~28, 2013}
{}

\usepackage{hyperref}
\usepackage{mymath,proof,latexsym}
\usepackage[usenames]{color}
\usepackage{amssymb,amsmath,proof}
\usepackage{txfonts}
\usepackage{enumerate} 

\begin{document}
\sloppy 
\hbadness=10000

\def\T#1{\leavevmode\hbox{\color{green}{$\clubsuit #1$}}}
\def\K#1{\leavevmode\hbox{\color{red}{$\spadesuit #1$}}}
\def\OT#1{\leavevmode\hbox{\color{yellow}{$\clubsuit #1$}}}
\def\OK#1{\leavevmode\hbox{\color{magenta}{$\spadesuit #1$}}}
\def\Skip#1{}

\def\cal{\mathcal} 
\def\red{\longrightarrow}
\def\Terms{\hbox{\bf Ter}}
\def\Sentences{\hbox{\bf Sen}}

\def\prooflineskip{\def\arraystretch{2.5}}
\def\<{\langle}
\def\>{\rangle}
\def\prove{\mathrel \vdash}
\def\lequiv{\mathrel \leftrightarrow}
\def\CBVDCmu{\hbox{{\tt CBV} $\DCmunu$}}
\def\CBNDCmu{\hbox{{\tt CBN} $\DCmunu$}}
\def\DCtwo{\hbox{${\tt DC}2$}}
\def\DCmu{\hbox{${\tt DC}\mu\nu$}}
\def\DC{{\hbox{${\tt DC}$}}}
\def\Itr{{\tt itr}}
\def\Coitr{{\tt coitr}}
\def\Pos{{\rm Pos}}
\def\Neg{{\rm Neg}}
\def\bottom{{\mathord \bot}}
\def\Zero{{\tt 0}}

\def\pared{\Rightarrow}
\def\wCBVred{\longrightarrow_{\tt wCBV}}
\def\wCBNred{\longrightarrow_{\tt wCBN}}
\def\CBVred{\longrightarrow_{\tt CBV}}
\def\CBNred{\longrightarrow_{\tt CBN}}
\def\nvalue{\mathcal{M}}
\def\nvalueb{\mathcal{N}}
\def\ncovalue{\mathcal{K}}
\def\snow{\circledast}

\def\Dcvdash{\prove}
\def\dcmid{\mathrel{\raisebox{-1.5pt}{$\rule{.5pt}{2ex}$}}}
\def\with{\&}
\def\cmid{\,\mid\,}
\def\implambda{\lambda^\to}

\def\Not{{\tt not}}
\def\Fst{{\tt fst}}
\def\Snd{{\tt snd}}
\def\Inl{{\tt inl}}
\def\Inr{{\tt inr}}
\def\Map{{\tt mono}}
\def\Inj{{\tt inj}}
\def\In{{\tt in}}
\def\Out{{\tt out}}
\def\Tyvars{{\tt TyVars}}
\def\Vars{{\tt Vars}}
\def\Covars{{\tt CoVars}}
\def\Nat{{\tt Nat}}
\def\Succ{{\tt succ}}
\def\List{{\tt List}}
\def\Nil{{\tt nil}}
\def\Stream{{\tt Stream}}
\def\Max{{\tt max}}
\def\Deg{{\tt deg}}
\def\endproof{$\Box$}
\def\SDCtoSSL{\dag}

\def\A{{\tt a}}
\def\E{{\tt e}}

\def\Duca{ {\tt DC} }
\def\SL2{ S\!\lambda 2 }
\def\DCmuV0{{\tt DC}\mu\nu_{cbv}}
\def\DCmuN0{{\tt DC}\mu\nu_{cbn}}
\def\CBVDCmu{{\tt CBV}\ {\tt DC}\mu\nu}
\def\CBNDCmu{{\tt CBN}\ {\tt DC}\mu\nu}

\def\SKIP{\hspace{5pt}}
\def\VSKIP{6pt}

\title[Call-by-Value and Call-by-Name Dual Calculi with Inductive and Coinductive Types]{Call-by-Value and Call-by-Name Dual Calculi with \\ Inductive and Coinductive Types\rsuper*}

\author[D.~Kimura]{Daisuke Kimura}
\address{National Institute of Informatics, 
2-1-2 Hitotsubashi, Tokyo 101-8430, Japan}
\email{\{kmr, tatsuta\}@nii.ac.jp}

\author[M.~Tatsuta]{Makoto Tatsuta}
\address{\vskip-6 pt}

\keywords{Curry-Howard isomorphism, Classical logic, Dual Calculus, Inductive definitions, Coinductive definitions}
\ACMCCS{[{\bf Theory of computation}]: Models of
  computation---Computability---Lambda calculus}
\subjclass{F.4.1}
\titlecomment{{\lsuper*}The conference version of this paper has appeared in 
\cite{Kimura09}}

\begin{abstract}
This paper extends the dual calculus with inductive types and
coinductive types.
The paper first introduces a non-deterministic dual calculus
with inductive and coinductive types.
Besides the same duality of the original dual calculus,
it has the duality of inductive and coinductive types,
that is, the duality of
terms and coterms for inductive and coinductive types,
and the duality of their reduction rules.
Its strong normalization is also proved, which
is shown by translating it into
a second-order dual calculus.
The strong normalization of the second-order dual calculus is
proved by translating it into the second-order symmetric lambda calculus.
This paper then introduces a call-by-value system and a call-by-name system
of the dual calculus with inductive and coinductive types, and
shows the duality of call-by-value and call-by-name, 
their Church-Rosser properties, and their strong normalization. 
Their strong normalization is proved by translating them into
the non-deterministic dual calculus with inductive and coinductive types.
\end{abstract}
\maketitle

\section{Introduction}
Dual Calculus $\DC$ given by Wadler~\cite{Wad03:01,Wad05:01} is a type system which 
corresponds to the classical sequent calculus LK  (see, for example, \cite{Girard87}). 
It represents computation induced by cut elimination in LK
by using its expressions and their reduction. 
The dual calculus has two nice properties: computation in
classical logic, and duality.

The computation of classical logic has been intensively studied, 
for example,
\cite{Bar96:01,Cur00:01,Griffin90,Herbelin05,Parigot97,Par00:01,Sel01:01,Wad03:01,Wad05:01}.
They all studied 
the Curry-Howard correspondence between 
classical logic and functional programming languages 
with sophisticated control structures like 
catch/throw and first-class continuations. 
This correspondence is an extension of 
the Curry-Howard correspondence between intuitionistic logic and 
the typed $\lambda$-calculus, which is well established. 

The classical sequent calculus LK has nice duality.
We have an involution that maps conjunction and disjunction to
each other, and maps the left and right rules of conjunction 
to the right and left rules of disjunction and vice versa.
This involution can be extended to the cut elimination procedure for LK.

The system $\DC$ inherits the duality of 
the classical sequent calculus LK.
Moreover, its proof terms called terms, coterms, and statements 
also have duality, since they correspond to proofs in LK.
This implies that its reduction relation can have duality since
the reduction relation is induced by the cut elimination procedure in LK. 
In this framework, Wadler gave the call-by-value and call-by-name strategies in $\DC$, 
and showed the duality of them~\cite{Wad03:01}.
He also showed that 
the equational correspondence between $\DC$ and Parigot's $\lambda\mu$-calculus~\cite{Par92:01}, 
and showed the duality between call-by-value and call-by-name of the $\lambda\mu$-calculus 
using the duality of the dual calculus~\cite{Wad05:01}. 
Since then, the dual calculus has been actively studied
\cite{Tzevelekos06,Kim07:01,Kimura09}. 

Inductive definitions are important in both mathematical logic and
computer science.
Inductive definitions strengthen expressiveness of logical systems
(for example, See \cite{Buc81:01}).
They are  central in programming and program verification
\cite{Paulin93,Nordstrom90,McDowell-Miller00} for 
handling recursive data structures such as lists and trees, and 
specification of recursive programs. 
Coinductive definitions are also important since 
they can represent streams, infinite trees, and 
bisimulation, for example, in \cite{Tat94:01}. 

This paper presents 
Dual Calculus $\DCmu$ with inductive types and coinductive types.
Our calculus extends the duality of $\DC$ to inductive types
and coinductive types.
The involution in $\DC$ is extended so that
it maps inductive types and coinductive types to each other. 
It also maps the left and right rules of inductive types to
the right and left rules of coinductive types and vice versa.
Because of the duality of the proof rules,
we will have cut elimination procedure that keeps
the duality of inductive types and coinductive types.
This induces the duality of
the reduction relations of proof terms for
inductive types and coinductive types.

Our main results are: 
(1) the duality between inductive types and coinductive types with reduction,
(2) strong normalization in $\DCmu$,
(3) strong normalization in the second-order Dual Calculus $\DCtwo$, 
(4) the duality between the call-by-value and call-by-name $\DCmu$, and 
(5) the Church-Rosser property and strong normalization of the call-by-value and call-by-name $\DCmu$. 

We will show strong normalization of $\DCmu$.
In order for proving the strong normalization,
we will first show the strong normalization 
of the second-order Dual Calculus $\DCtwo$ given by \cite{Tzevelekos06}
by
interpreting it in second-order symmetric lambda-calculus given in \cite{Par00:01}.
Then strong normalization of $\DCmu$
is proved by interpreting it in $\DCtwo$
by using second-order coding of inductive and coinductive
types.

We first introduce the system $\DCmu$ that does not have reduction strategies, since 
it is designed by the Curry-Howard correspondence for 
a standard cut elimination procedure in LK. 
The system can discuss non-deterministic aspects of computation in 
classical logic, since the execution of programs in $\DCmu$ is non-deterministic. 
It also works as a base framework for other variants of $\DCmu$ with 
specific reduction strategies such as call-by-value and call-by-name that will be given later. 

The duality between call-by-value and call-by-name is first suggested by Filinski~\cite{Filinski89}. 
The dual calculus gives a clear explanation for this duality by using the logical duality of classical logic. 
We will show the duality of call-by-value and call-by-name
in the dual calculus extended with inductive types and coinductive types. 
We extend 
the call-by-value $\DC$ and the call-by-name $\DC$ given in \cite{Wad03:01} with inductive types and 
coinductive types, and
introduce the systems $\CBVDCmu$ and $\CBNDCmu$.
They are obtained from $\DCmu$ by restricting its non-deterministic reduction 
to the call-by-value or call-by-name strategies, 
and also by adding some strategy-specific reduction rules. 
In the same way as \cite{Wad03:01}, we show
the duality of call-by-value and call-by-name in the dual calculus
with inductive types and coinductive types.
We will show the Church-Rosser property as well as strong normalization
for $\CBVDCmu$ and $\CBNDCmu$.
The strong normalization will be shown by translating
$\CBVDCmu$ and $\CBNDCmu$ into $\DCmu$.


 In \cite{Baelde12}, the duality between inductive types and
coinductive types in linear logic is studied.
Our system $\DCmu$ shows the duality in ordinary sequent calculus LK.

Momigliano and Tiu~\cite{Mom03:01,Tiu10} discussed 
an intuitionistic sequent calculus with inductive definitions and
coinductive definitions and showed its cut elimination theorem.  
Our system $\DCmu$ is a classical system and
our strong normalization shows the cut elimination theorem of
the classical sequent calculus.
Our cut elimination procedure is not closed in an intuitionistic fragment 
because it keeps the duality and 
we have the corresponding proof rule that manipulates a succedent if 
we have some proof rule that manipulates an antecedent. 
So we cannot directly compare our method and their method.

In category theory, inductive definitions are represented by
initial algebras and coinductive definitions are represented by
final coalgebras~\cite{Geu92:01}, and their duality in category
theory is known.
Our system $\DCmu$ enables us to show the duality in a clear syntactic way
by using a type system.

Several papers for dual calculus investigated the duality of computation. 
Wadler showed 
the duality between values and continuations, 
and the duality between call-by-value computation and 
call-by-name computation  by using the explicit duality of $\DC$~\cite{Wad03:01,Wad05:01}. 
The first author of this paper showed 
the duality between 
the call-by-name fixed point operator and the call-by-value loop operator 
by extending $\DC$~\cite{Kim07:02}. 
The first author also showed the duality of reduction
between call-by-value computation and call-by-name computation
in $\lambda\mu$-calculus by using $\DC$ \cite{Kim07:01} 
to answer the open question presented in Wadler's 
invited talk at RTA2005~\cite{Wad05:01}, which asked 
whether the duality between call-by-value and call-by-name in his equation systems 
would be refined in reduction systems. 
Tzevelekos~\cite{Tzevelekos06} investigated the dual calculus given in~\cite{Wad03:01}. 
He assumed some additional conditions on reductions, 
and showed both Church-Rosser property and strong normalization hold under his conditions. 
He also investigated 
the relationship between $\DC$ and 
the symmetric $\lambda$-calculus by Barbanera and Berardi~\cite{Bar96:01}. 
A second-order extension of $\DC$ is also considered in \cite{Tzevelekos06}. 

The system $\bar \mu \tilde \mu$ in \cite{Cur00:01} is a system with implication and subtraction, 
and also has duality. 
Their calculus with negation, conjunction, and disjunction is
called $\mu \tilde\mu ^{\land_a\lor_a\neg}$ and
the correspondence between it and the dual calculus is discussed 
in \cite{Herbelin05}. 

A semantical approach to the duality between call-by-value and call-by-name was studied 
by Selinger~\cite{Sel01:01}. 
He gave a categorical semantics of the $\lambda\mu$-calculus, 
and explained the duality by using the categorical duality. 
This approach is extended to the duality between the fixed point operator and the loop operator 
by Kakutani~\cite{Kak02:01}. 

Section 2 gives a definition of $\DC$ and states its duality. 
Section 3 introduces $\DCmu$ and shows its duality.
Section 4 gives examples.
In section 5, we 
give $\DCtwo$ and show its strong normalization.
Section 6 proves strong normalization for $\DCmu$.
Section 7 introduces $\CBVDCmu$ and $\CBNDCmu$ and shows
their Church-Rosser properties and strong normalization.

\section{The Dual Calculus \texorpdfstring{$\DC$}{}}

This section defines Dual Calculus $\DC$ and states its duality.
This system is obtained from the original Dual Calculus given in \cite{Wad03:01} 
by removing reduction strategies in reduction rules. 
This system gives us a base framework for several variants
of dual calculi. 

\begin{defi}[Types and Expressions of $\DC$]
Let $X, Y, Z, \ldots$ range over type variables, $A, B, \ldots$ range over types, 
The symbols $x, y, z, \ldots$ range over variables, and $\alpha, \beta, \gamma, \ldots$ range over covariables. 
We assume an involution $(-)'$ between variables and covariables, 
which satisfies $x'' = x$ and $\alpha'' = \alpha$. 
An expression (denoted by $D, E, \ldots$)
is either a term (denoted by $M, N, \ldots$), a coterm (denoted by $K, L, \ldots$), 
or a statement (denoted by $S, T, \ldots$). 
We define them as follows: 
\begin{center}
\begin{tabular}{ll}
Types 
& 
$A\ \Coloneqq\ X \mid A\land A \mid A \vee A \mid \neg A$,\\
Expressions
&
$D\ \Coloneqq  M \mid K \mid S$,\\
Terms
&
$M\ \Coloneqq\ x \mid \langle M, M\rangle \mid \langle M\rangle\Inl \mid \langle M\rangle\Inr 
\mid [K]\Not \mid (S).\alpha$,\\
Coterms
&
$K\ \Coloneqq\ \alpha \mid [K, K] \mid \Fst[K] \mid \Snd[K] 
\mid \Not\langle M\rangle \mid x.(S)$,\\
Statements 
&
$S\ \Coloneqq\ M\bullet K$.
\end{tabular}
\end{center}
The term $(S).\alpha$ binds the covariable $\alpha$ in $S$. 
The coterm $x.(S)$ binds the variable $x$ in $S$. 
We write ${\it FV}(D)$ for the set of free variables in $D$. 
We also write ${\it FCV}(D)$ for the set of free covariables in $D$. 
We will use $\_[\_/\_]$ for substitution. For example,
the substitution $S[M/x]$ denotes the statement obtained from $S$
by replacing $x$ by $M$. 
\end{defi}

The type $A \land B$ denotes a conjunction,
$A \lor B$ denotes a disjunction, and
$\neg A$ denotes a negation.
A variable means an ordinary variable.
A covariable means an output port and gets some value after
computation.
A term represents an ordinary computation which becomes 
a value or puts values at output ports after computation.
The term $\langle M,N\rangle$ means a pair.
The terms $\langle M\rangle\Inl$ and
$\langle M\rangle\Inr$ 
mean the left injection and  the right injection to a disjoint sum,
respectively.
When $[K]\Not$ gets its input, it gives the input to $K$ and computes $K$.
The term $(S).\alpha$ is an abstraction of $S$ by $\alpha$. 
It computes $S$ and its value is the value at the output port $\alpha$.
A coterm represents continuation which puts values at
output ports after computation when it gets its input.
The coterm $[K,L]$ gets an input of a disjoint sum. If the input is $\< M \>\Inl$, 
it gives $M$ to $K$ and computes $K$. 
If the input is $\< M \>\Inr$, it gives $M$ to $L$ and computes $L$.
The coterm $\Fst[K]$ gets an input of a pair. 
If the input is $\langle M,N\rangle$, then it gives $M$ to $K$ and computes
$K$.
The coterm $\Snd[K]$ also gets an input of a pair. 
If the input is $\langle M,N\rangle$, then it gives $N$ to $K$ and computes
$K$.
The coterm $\Not\langle M\rangle$ gets a continuation as its input.
It gives $M$ to the continuation and computes the continuation.
The coterm $x.(S)$ is an abstraction of $S$ by $x$. If it gets an input, 
it puts the input in $x$ and computes $S$.
The statement $M\bullet K$ means the computation of $K$ with the input $M$ that
may put values at output ports. 

A typing judgment (denoted by $J$) of $\DC$ takes either the form 
$\Gamma \prove \Delta \dcmid M:A$, 
the form $K:A \dcmid \Gamma \prove \Delta$, or 
the form $\Gamma \dcmid S \prove \Delta$, 
where $\Gamma$ denotes a context 
$x_1:A_1, \ldots, x_n:A_n$ 
that is a set of variable declarations,
and $\Delta$ denotes a cocontext 
$\alpha_1:B_1,\ldots, \alpha_m:B_m$
that is a set of covariable declarations. 
We will call $M$, $K$, and $S$ a principal expression in those judgments. 
The domain of $\Gamma$ (denoted by ${\it dom}(\Gamma)$) is the set of variables $\{x_1, \ldots, x_n\}$ 
if $\Gamma$ is $x_1:A_1, \ldots, x_n:A_n$. 
The domain of $\Delta$ (denoted by ${\it dom}(\Delta)$) is the set of covariables $\{\alpha_1, \ldots, \alpha_m\}$ 
if $\Delta$ is $\alpha_1:B_1,\ldots, \alpha_m:B_m$. 

We intuitively explain the typing judgments.  There can be other ways 
of intuitive explanation, for example, \cite{Tzevelekos06}.  
In order to give an intuitive idea in general, 
we assume an evaluation strategy for expressions, and a notion of values for the strategy. 
For example, when we take call-by-name,
the values will be canonical form, and the computation will be lazy
evaluation. The focus $|$ is used only for denoting which part contains
a term, a coterm, or a statement in a judgment, and 
when we think the corresponding sequent in ordinary sequent calculus,
we will erase it. 
The typing judgment $x_1:A_1,\ldots,x_n:A_n \prove \alpha_1:B_1,\ldots,\alpha_m:B_m \dcmid M:A$ means
that when each $x_i$ has a value of type $A_i$, 
and $M$ is computed, then $M$ returns a value of type $A$ or some $\alpha_i$ 
gets a value of type $B_i$.
The judgment
$K:A \dcmid x_1:A_1,\ldots,x_n:A_n \prove \alpha_1:B_1,\ldots,\alpha_m:B_m$ means
that when each $x_i$ has a value of type $A_i$,
an input of type $A$ is given to $K$, and 
$K$ is computed, then 
some $\alpha_i$ gets a value of type $B_i$.
The judgment
$x_1:A_1,\ldots,x_n:A_n \dcmid S \prove \alpha_1:B_1,\ldots,\alpha_m:B_m$ means
that when each $x_i$ has a value of type $A_i$ and
$S$ is computed, then some $\alpha_i$ gets a value of type $B_i$. 
We sometimes use the symbol $\prove_{\Duca}$ instead of the symbol $\prove$ 
that appears in a judgment 
in order to explicitly show it is a judgment of $\Duca$. 
That is, 
we write $\Gamma \prove_{\Duca} \Delta \dcmid M\colon A$ for 
the judgment $\Gamma \prove \Delta \dcmid M\colon A$. 
Similarly, we write 
$K\colon A \dcmid \Gamma \prove_{\Duca} \Delta$ and 
$\Gamma \dcmid S \prove_{\Duca} \Delta$. 

The typing rules are given in Figure~\ref{Typing_DC}. 
If we erase terms, coterms, statements, and the symbol $|$, 
the system becomes logically equivalent 
to a fragment of classical sequent calculus LK,
whose definition is given in, for example, \cite{Girard87}. 

\begin{figure}[t]
\begin{center}
\begin{tabular}{c@{\SKIP\SKIP}c}
$\infer[({\it AxR})]{
        \Gamma, x:A \prove \Delta \dcmid x:A
        }{}$
&
$\infer[({\it AxL})]{
        \alpha:A \dcmid \Gamma \prove \Delta, \alpha:A
        }{}$\\[\VSKIP]
$\infer[(\land R)]{
        \Gamma \prove \Delta \dcmid \langle M,N\rangle:A\land B
        }{
        \Gamma \prove \Delta \dcmid M:A
        &
        \Gamma \prove \Delta \dcmid N:B
        }$
&
$\infer[(\vee L)]{
        [K, L]:A\vee B \dcmid \Gamma \prove \Delta
        }{
        K:A \dcmid \Gamma \prove \Delta
        &
        L:B \dcmid \Gamma \prove \Delta
        }
        $\\[\VSKIP]
$\infer[(\vee R1)]{
        \Gamma \prove \Delta \dcmid \langle M\rangle\Inl:A\vee B
        }{
        \Gamma \prove \Delta \dcmid M:A
        }$
&
$\infer[(\land L1)]{
        \Fst[K]:A\land B \dcmid \Gamma \prove \Delta
        }{
        K:A \dcmid \Gamma \prove \Delta
        }$\\[\VSKIP]
$\infer[(\vee R2)]{
        \Gamma \prove \Delta \dcmid \langle M\rangle\Inr:A\vee B
        }{
        \Gamma \prove \Delta \dcmid M:B
        }$
&
$\infer[(\land L2)]{
        \Snd[K]:A\land B \dcmid \Gamma \prove \Delta
        }{
        K:B \dcmid \Gamma \prove \Delta
        }$\\[\VSKIP]
$\infer[(\neg R)]{
        \Gamma \prove \Delta \dcmid [K]\Not:\neg A
        }{
        K:A \dcmid \Gamma \prove \Delta
        }$
&
$\infer[(\neg L)]{
        \Not\langle M\rangle:\neg A\dcmid \Gamma \prove \Delta
        }{
        \Gamma \prove \Delta \dcmid M:A
        }$\\[\VSKIP]
$\infer[({\it IR})]{
        \Gamma \prove \Delta \dcmid (S).\alpha:A
        }{
        \Gamma \dcmid S \prove \Delta, \alpha:A
        }$
&
$\infer[({\it IL})]{
        x.(S):A \dcmid \Gamma \prove \Delta
        }{
        \Gamma, x:A \dcmid S \prove \Delta
        }$\\[\VSKIP]
\multicolumn{2}{c}{
$\infer[({\it Cut})]{
        \Gamma \dcmid M\bullet K \prove \Delta
        }{
        \Gamma \prove \Delta \dcmid M:A
        &
        K:A \dcmid \Gamma \prove \Delta
        }$
}
\end{tabular}
\end{center}
\hrule
\caption{Typing rules of \Duca}
\label{Typing_DC}
\end{figure}

\begin{defi}[Reduction]
The reduction relation $\longrightarrow_\Duca$ is defined as the compatible closure of the following reduction rules:
\[
\begin{array}{ll}
(\beta\land_1)
&
\langle M,N\rangle\bullet \Fst[K]  \longrightarrow_\Duca M\bullet K, 
\\
(\beta\land_2)
&
\langle M,N\rangle\bullet \Snd[K] \longrightarrow_\Duca N\bullet K,
\\
(\beta\vee_1)
&
\langle M\rangle\Inl \bullet [K,L] \longrightarrow_\Duca M\bullet K,
\\
(\beta\vee_2)
&
\langle M\rangle\Inr \bullet [K,L] \longrightarrow_\Duca M\bullet L,
\\
(\beta\neg)
&
[K]\Not \bullet \Not\langle M\rangle \longrightarrow_\Duca M\bullet K,
\\
(\beta R)
&
(S).\alpha \bullet K \longrightarrow_\Duca S[K/\alpha],
\\
(\beta L)
&
M\bullet x.(S) \longrightarrow_\Duca S[M/x],
\\
(\eta R)
&
(M\bullet \alpha).\alpha \longrightarrow_\Duca M,
\\
(\eta L)
&
x.(x\bullet K) \longrightarrow_\Duca K,
\end{array}
\]
where $x$ and $\alpha$ are fresh in $(\eta L)$ and $(\eta R)$, respectively. 
\end{defi}
The rules $(\eta R)$ and $(\eta L)$ are necessary to get
the results of computation of
terms and coterms from computation of statements inside them. 
We do not include the $\eta$-rules for logical connectives 
that are given in \cite{Wad05:01}, since
these break the confluence property for call-by-value and call-by-name
systems, which we will study in Section 7.
In order to study a base framework, 
we first consider a
non-deterministic rewriting system that does not commit to either the
call-by-name or call-by-value theory.

The system $\Duca$ we consider first is obtained from 
the original dual calculus given in~\cite{Wad03:01} by 
omitting evaluation strategies, 
dropping $(\varsigma)$-rules that provide strong evaluation under call-by-value and call-by-name strategies, 
and replacing $(\eta L)$ and $(\eta R)$-expansion rules by $(\eta L)$ and $(\eta R)$-reduction rules. 

The role of $(\eta L)$ and $(\eta R)$-reduction rules are to simplify logical proofs 
without changing any proof structure. 
In the last section, we also give the call-by-value and call-by-name variants of $\DCmu$. 
The role of these rules become clearer in that section 
since they are necessary to obtain a value as the result of a computation under some strategy. 

The type of an expression is preserved by reduction. 
\begin{prop}[Subject reduction of $\Duca$]\label{SR_DC}
The following claims hold. 
\begin{enumerate}[\rm(1)]
\item[\rm(1)] If $\Gamma \prove_{\Duca} \Delta \dcmid M\colon A$ and $M \longrightarrow_{\Duca} N$, then 
$\Gamma \prove_{\Duca} \Delta \dcmid N\colon A$ holds. 

\item[\rm(2)] If $K\colon A \dcmid \Gamma \prove_{\Duca} \Delta$ and $K \longrightarrow_{\Duca} L$, then 
$L\colon A \dcmid \Gamma \prove_{\Duca} \Delta$ holds. 

\item[\rm(3)] If $\Gamma \dcmid S \prove_{\Duca} \Delta$ and $S \longrightarrow_{\Duca} T$, then 
$\Gamma \dcmid T \prove_{\Duca} \Delta$ holds. 
\end{enumerate}
\end{prop}

This proposition is shown by induction on reduction using the following substitution lemma. 

\begin{lem}[Substitution lemma]\label{lem:Substitution_DC}
The following claims hold. 
\begin{enumerate}[\rm(1)]
\item
Suppose $\Gamma \prove_{\Duca} \Delta \dcmid N\colon A$ is derivable. 
Then we have the following. 
\begin{enumerate}[\rm(1a)]
\item[\rm(1a)]
If $\Gamma, x\colon A \prove_{\Duca} \Delta \dcmid M\colon B$, then 
$\Gamma \prove_{\Duca} \Delta \dcmid M[N/x]\colon B$,
\item[\rm(1b)]
if $K\colon B \dcmid \Gamma, x\colon A \prove_{\Duca} \Delta$, then 
$K[N/x] \colon B \dcmid \Gamma \prove_{\Duca} \Delta$, and 
\item[\rm(1c)]
if $\Gamma, x\colon A \dcmid S \prove_{\Duca} \Delta$, then 
$\Gamma \dcmid S[N/x] \prove_{\Duca} \Delta$. 
\end{enumerate}

\item[\rm(2)]
Suppose $L\colon A \dcmid \Gamma \prove_{\Duca} \Delta$ is derivable. 
Then we have the following. 
\begin{enumerate}[(1a)]
\item[\rm(2a)]
If $\Gamma \prove_{\Duca} \Delta, \alpha\colon A \dcmid M\colon B$, then 
$\Gamma \prove_{\Duca} \Delta \dcmid M[L/\alpha]\colon B$,
\item[\rm(2b)]
if $K\colon B \dcmid \Gamma \prove_{\Duca} \Delta, \alpha\colon A$, then 
$K[L/\alpha] \colon B \dcmid \Gamma \prove_{\Duca} \Delta$, and
\item[\rm(2c)]
if $\Gamma \dcmid S \prove_{\Duca} \Delta, \alpha\colon A$, then 
$\Gamma \dcmid S[L/\alpha] \prove_{\Duca} \Delta$. 
\end{enumerate}
\end{enumerate}
\end{lem}
\proof
The claims (1a),(1b), and (1c) are shown simultaneously by induction on $M$, $K$, and $S$. 
The claims (2a),(2b), and (2c) are also shown simultaneously by induction on $M$, $K$, and $S$. 
\qed

The following duality transformation extends the duality in the sequent
calculus LK to terms, coterms, and statements.
\begin{defi}[Duality Transformation]
The duality transformation $(-)^\circ$ from $\Duca$ into itself is defined
for types and expressions as follows:
\[\begin{array}{ll}
(X)^\circ=X,
\quad
(\neg A)^\circ=\neg (A)^\circ, 
\qquad
(A\land B)^\circ=(A)^\circ \vee (B)^\circ,
\quad
(A\vee B)^\circ=(A)^\circ \land (B)^\circ, \span\\
(x)^\circ=x', 
&
(\alpha)^\circ=\alpha', 
\\
(\langle M,N\rangle)^\circ=[(M)^\circ, (N)^\circ], 
&
([K,L])^\circ=\langle (K)^\circ, (L)^\circ\rangle,
\\
(\langle M\rangle\Inl)^\circ=\Fst[(M)^\circ], 
&
(\Fst[K])^\circ=\langle (K)^\circ\rangle\Inl,
\\
(\langle M\rangle\Inr)^\circ=\Snd[(M)^\circ], 
&
(\Snd[K])^\circ=\langle (K)^\circ\rangle\Inr,
\\
([K]\Not)^\circ=\Not\langle (K)^\circ\rangle, 
&
(\Not\langle M\rangle)^\circ=[(M)^\circ]\Not,
\\
((S).\alpha)^\circ=\alpha'.((S)^\circ), 
&
(x.(S))^\circ=((S)^\circ).x', 
\\
(M\bullet K)^\circ = (K)^\circ \bullet (M)^\circ.
\span
\end{array}\]

\noindent Note that a type and a statement are mapped to themselves.
A term and a coterm are mapped to each other.

We also define transformation for judgments.
If $\Gamma$ is $x_1\colon A_1, \ldots, x_n\colon A_n$,
then $(\Gamma)^\circ$ is defined as 
$(x_1)^\circ\colon (A_1)^\circ,\ldots,(x_n)^\circ\colon (A_n)^\circ$.
If $\Delta$ is $\alpha_1\colon B_1,\ldots, \alpha_m\colon B_m$,
then $(\Delta)^\circ$ is defined as
$(\alpha_1)^\circ\colon(B_1)^\circ,\ldots, (\alpha_m)^\circ\colon (B_m)^\circ$.
The judgment
$(\Gamma \prove \Delta \dcmid M\colon A)^\circ$ is defined as
$(M)^\circ\colon (A)^\circ \dcmid (\Delta)^\circ \prove (\Gamma)^\circ$.
The judgment
$(K\colon A \dcmid \Gamma \prove \Delta)^\circ$ is defined as
$(\Delta)^\circ  \prove (\Gamma)^\circ \dcmid (K)^\circ: (A)^\circ$.
The judgment
$(\Gamma \dcmid S \prove \Delta)^\circ$ is defined as
$(\Delta)^\circ \dcmid (S)^\circ \prove (\Gamma)^\circ$.

We also define transformation for inference rule names as follows:
$(AxR)^\circ = (AxL)$,
$(AxL)^\circ = (AxR)$,
$(\lor R1)^\circ = (\land L1)$,
$(\land L1)^\circ = (\lor R1)$,
$(\land R)^\circ = (\lor L)$,
$(\lor L)^\circ = (\land R)$,
$(\lor L2)^\circ = (\land R2)$,
$(\lor R2)^\circ = (\land L2)$,
$(\neg L)^\circ = (\neg R)$,
$(\neg R)^\circ = (\neg L)$,
$(IR)^\circ = (IL)$,
$(IL)^\circ = (IR)$,
and
$(Cut)^\circ = (Cut)$.
\end{defi}

This duality transformation preserves substitution of terms and
coterms. \newpage

\begin{lem}\label{Lemma:Duality_Substitution_DC}
The following claims hold. 
\begin{enumerate}[\rm(1)] 
\item[\rm(1)] 
$(D[M/x])^\circ = (D)^\circ[(M)^\circ/x']$. 
\item[\rm(2)]
$(D[K/\alpha])^\circ = (D)^\circ[(K)^\circ/\alpha']$. 
\end{enumerate}
\end{lem}
\proof
The claims (1) and (2) are shown by induction on $D$.
We treat the first case of (1): the case of $D$ is $x$. 
$(x[M/x])^\circ = (M)^\circ = x'[(M)^\circ/x'] = (x)^\circ[(M)^\circ/x']$. 
The other cases are straightforwardly proved by the induction hypothesis. 
\qed

This duality transformation is shown to preserve typing and reduction, 
and to be an
involution. This transformation is a homomorphism for
this system in the sense that it preserves typing and reduction.
An important feature of $\Duca$ is its duality by this transformation.
A term is dual to a coterm by this homomorphism.
\begin{prop}[Duality of ${\tt DC}$]\label{prop:dual}
The followings hold. 
\begin{enumerate}[\rm(1)]
\item[\rm(1)]
If $J$ is derived from $J_1,\ldots,J_n$ ($n = 1$ or $2$) by an inference rule $R$, 
then 
$(J)^\circ$ is derived from $(J_n)^\circ,\ldots,(J_1)^\circ$ by the inference rule $(R)^\circ$. 
\item[\rm(2)]
$D \longrightarrow_\Duca E$ implies  $(D)^\circ \longrightarrow_\Duca (E)^\circ$. 
\item[\rm(3)]
$((A)^\circ)^\circ = A$, 
\SKIP 
$((D)^\circ)^\circ = D$, 
and 
$((J)^\circ)^\circ = J$ hold. 
\end{enumerate}
\end{prop}
\proof
The claim (1) is proved by case analysis of the inference rules.
The claim (2) is proved by induction on the generation of $\longrightarrow_{\Duca}$ using Lemma~\ref{Lemma:Duality_Substitution_DC}. 
The claim (3) is proved by induction on types and expressions.
\qed

\begin{rem}
The $(-)^\circ$ transformation maps dual reduction rules to each other. That is,
if $D \longrightarrow_\Duca E$ is the reduction rules 
$(\beta\land_1)$, 
$(\beta\land_2)$, 
$(\beta\vee_1)$, 
$(\beta\vee_2)$, 
$(\beta\neg)$,
$(\beta R)$, 
$(\beta L)$, 
$(\eta R)$, and 
$(\eta L)$, 
then 
$(D)^\circ \longrightarrow_\Duca (E)^\circ$ is the reduction rules 
$(\beta\vee_1)$, 
$(\beta\vee_2)$, 
$(\beta\land_1)$, 
$(\beta\land_2)$, 
$(\beta\neg)$,
$(\beta L)$, 
$(\beta R)$, 
$(\eta L)$, and 
$(\eta R)$, 
respectively.
\end{rem}

Implication $\supset$ can be defined by $\neg$ and $\lor$
in the same way as \cite{Wad03:01}. 
\begin{defi}\label{def:imp}
We write $A\supset B$ for $\neg A \vee B$.
We also write $\lambda x.M$ for 
$( \langle [x.(\langle M\rangle\Inr\bullet \gamma)]\Not\rangle\Inl\bullet \gamma ).\gamma$.
We also write $N@K$ for $[\Not\langle N\rangle, K]$.
\end{defi}

The constructor $@$ simulates the application in $\lambda$-calculus together 
with $\bullet$.
The following holds from the definition.
\begin{prop}
The following typing inference rules and reduction rule are derivable.
{\def\arraystretch{1.5}
\[\begin{array}{l}
\infer[(\supset R)]{
        \Gamma \prove \Delta \dcmid \lambda x.M:A\supset B
        }{
        \Gamma, x:A \prove \Delta \dcmid M:B
        }
\qquad
\infer[(\supset L)]{
        M@K:A\supset B\dcmid \Gamma \prove \Delta
        }{
        \Gamma \prove \Delta \dcmid M:A
        &
        K:B \dcmid \Gamma \prove \Delta
        }
\end{array}\]
\[\begin{array}{l}
(\beta\supset)\quad
\lambda x.M\bullet (N@K) \longrightarrow_{\Duca} M[N/x]\bullet K
\end{array}\]
}
\end{prop}

\section{
The Dual Calculus \texorpdfstring{$\DCmu$}{} with Inductive and Coinductive Types
}\label{Sect:(co)ind_dual}

In this section, we present $\DCmu$, which is an extension of $\Duca$ 
with inductive types and coinductive types. 
We first extend the definition of types of $\Duca$ to
inductive types $\mu X.A$ and coinductive types $\nu X.A$, 
and then extend expressions and reduction. 

In Section~\ref{Sect:DCtwo}, we will introduce the second-order system $\DC2$.
The system $\DCmu$ is worth to be studied as well as $\DC2$,
since $\DCmu$ is within a first-order logic.

We first define types, their positive type variables, and
their negative type variables.
A positive type variable in a type does not occur negatively in the
type in the usual sense.
A negative type variable in a type does not occur positively in the type. 
\begin{defi}
The set of type variables is written by $\Tyvars$.
We define the types of $\DCmu$ (denoted by $A, B, \ldots$) and 
the set ${\rm Pos}(A)$ of {\it positive type variables}  in the type $A$ and the set ${\rm Neg}(A)$ of
{\it negative type variables}  in the type $A$
as follows:
\[
A\ \Coloneqq\ X \mid A\land A \mid A \vee A \mid \neg A \mid \mu X.A \mid \nu X.A
\]
where $\mu X.A$ and $\nu X.A$ are defined when 
the type variable $X$ is in ${\rm Pos}(A)$.
\[\begin{array}{l}
{\rm Pos}(X)=\Tyvars, \\
{\rm Neg}(X)=\Tyvars \setminus \{X\}, \\
{\rm Pos}(A_1\land A_2)={\rm Pos}(A_1\vee A_2)={\rm Pos}(A_1) \cap {\rm Pos}(A_2), \\
{\rm Neg}(A_1\land A_2)={\rm Neg}(A_1\vee A_2)={\rm Neg}(A_1) \cap {\rm Neg}(A_2), \\
{\rm Pos}(\neg B)={\rm Neg}(B), \\
{\rm Neg}(\neg B)={\rm Pos}(B), \\
{\rm Pos}(\mu X.B)={\rm Pos}(\nu X.B)={\rm Pos}(B) \cup \{X\}, \\
{\rm Neg}(\mu X.B)={\rm Neg}(\nu X.B)={\rm Neg}(B) \cup \{X\}.\\
\end{array}\]
The types $\mu X.A$ and $\nu X.A$ bind $X$ in $A$.
\end{defi}

When we think standard semantics of the propositional logic
with inductive and coinductive definitions,
$\mu X.A$ and $\nu X.A$ are interpreted by 
the least fixed point 
and the greatest fixed point of the monotone function $\mathcal{P}$ 
respectively,
where 
$\mathcal{P}$ is the function which maps
a set $U$ to the set $A[U/X]$.
Let $\mu$ be $\mu X.A$ and $\nu$ be $\nu X.A$.
They will have the following properties:
(a) $A[\mu/X] \subseteq \mu$, 
(b) $A[B/X] \subseteq B$ implies $\mu \subseteq B$, 
(c) $\nu \subseteq A[\nu/X]$, and 
(d) $B \subseteq A[B/X]$ implies $B \subseteq \nu$. 
Based on this meaning,
we will introduce terms, coterms, and their reduction for inductive and
coinductive types  in the same way as \cite{Mendler91}.
\begin{defi}
The terms, coterms, and statements of $\DCmu$ are defined as follows: 
\[\begin{array}{l}
M\ \Coloneqq\ x \mid \langle M, M\rangle \mid \langle M\rangle\Inl \mid \langle M\rangle\Inr 
\mid [K]\Not \mid (S).\alpha \mid \In^{\mu X.A}\langle M\rangle\mid \Coitr^A_x\langle M, M\rangle,
\\
K\ \Coloneqq\ \alpha \mid [K, K] \mid \Fst[K] \mid \Snd[K] 
\mid \Not\langle M\rangle \mid x.(S) \mid \Out^{\nu X.A}[K] \mid \Itr^A_\alpha[K,K], 
\\
S\ \Coloneqq\ M\bullet K.
\end{array}\]
The term $\Itr^A_\alpha[K,L]$ binds $\alpha$ in $K$. 
The coterm $\Coitr^A_x\langle M,N\rangle$ binds $x$ in $M$. 
\end{defi}

The expressions $\In^{\mu X.A}\langle M\rangle$ and $\Itr^A_\alpha[K,L]$ are the expressions for inductive types. 
The constructor $\In^{\mu X.A}$
maps a term of type $A[\mu X.A/X]$ to that of $\mu X.A$.
The coterm $\Itr^B_\alpha[K,L]$ is an iterator having an input of type $\mu X.A$ where
$L$ is a postprocessor after iteration.
When it gets the input of type $\mu X.A$,
first a value of type $A[\mu X.A/X]$ is computed according to the input, 
next a value of type $A[B/X]$ is computed by recursive invocation of the iterator, 
then it is given to $K$ and $K$ is computed to get a value of type $B$, 
and finally the value is given to $L$ and $L$ is computed. 
Dually, $\Out^{\nu X.A}[K]$ and $\Coitr^A_x\langle M, N\rangle$ are defined for coinductive types. 
The constructor $\Out^{\nu X.A}$
maps a coterm of type $\nu X.A$ to that of $A[\nu X.A/X]$.
When the coterm $\Out^{\nu X.A}[K]$ gets the input of type $\nu X.A$, first the input is transformed into
a value of type $A[\nu X.A/X]$, then the value is given to $K$,
and finally $K$ is computed.
The term $\Coitr^B_x\langle M,N\rangle$ 
is a coiterator of type $\nu X.A$.
It transforms $N$ of type $B$ into a value of $\nu X.A$ according to $M$.
Type annotations will be necessary for defining reduction rules.

\begin{defi}
The typing rules of $\DCmu$ are defined by those of $\Duca$ and the following 
rules:\\

{\prooflineskip
$\infer[(\mu R)]{
        \Gamma \prove \Delta \dcmid \In^{\mu X.A}\langle M\rangle:\mu X.A
        }{
        \Gamma \prove \Delta \dcmid M:A[\mu X.A/X]
        }$
\qquad
$\infer[(\mu L)]{
        \Itr^B_\alpha[K,L]:\mu X.A \dcmid \Gamma \prove \Delta
        }{
        K:A[B/X] \dcmid \Gamma \prove \Delta, \alpha:B
        &
        L:B \dcmid \Gamma \prove \Delta
        }$\\[5pt]

$\infer[(\nu L)]{
        \Out^{\nu X.A}[K]:\nu X.A \dcmid \Gamma \prove \Delta
        }{
        K:A[\nu X.A/X] \dcmid \Gamma \prove \Delta
        }$
\qquad
$\infer[(\nu R)]{
        \Gamma \prove \Delta \dcmid \Coitr^B_x\langle M, N \rangle:\nu X.A
        }{
        \Gamma, x:B \prove \Delta \dcmid M:A[B/X]
        &
        \Gamma \prove \Delta \dcmid N:B
        }$
}\smallskip
\end{defi}

\noindent We sometimes use the symbol $\prove_{\Duca\mu\nu}$ 
instead of the symbol $\prove$ in a judgment 
in order to explicitly show it is a judgment of $\DCmu$. 
That is, 
we write $\Gamma \prove_{\Duca\mu\nu} \Delta \dcmid M\colon A$ for 
the judgment $\Gamma \prove \Delta \dcmid M\colon A$. 
Similarly, we write 
$K\colon A \dcmid \Gamma \prove_{\Duca\mu\nu} \Delta$ and 
$\Gamma \dcmid S \prove_{\Duca\mu\nu} \Delta$. 

The system $\Duca\mu\nu$ satisfies the following basic lemmas. 

\begin{lem}[Weakening lemma of $\Duca\mu\nu$]\label{lem:Weakening_DCmu}\rm
Let $\Gamma \subseteq \Gamma'$ and $\Delta \subseteq \Delta'$. Then 
\begin{enumerate}[\rm(1)]
\item[\rm(1)]
if $\Gamma \prove_{\Duca\mu\nu} \Delta \dcmid M\colon A$ is provable, then 
$\Gamma' \prove_{\Duca\mu\nu} \Delta' \dcmid M\colon A$ holds, 
\item[\rm(2)]
if $K\colon A \dcmid \Gamma \prove_{\Duca\mu\nu} \Delta$ is provable, then 
$K\colon A \dcmid \Gamma' \prove_{\Duca\mu\nu} \Delta'$ holds, and 
\item[\rm(3)]
if $\Gamma \dcmid S \prove_{\Duca\mu\nu} \Delta$ is provable, then 
$\Gamma' \dcmid S \prove_{\Duca\mu\nu} \Delta'$ holds. 
\end{enumerate}
\end{lem}
\proof
They are shown simultaneously by induction on $M$, $K$, and $S$. 
\qed

\begin{lem}\label{lem:Thinning_DCmu}
Let $\Gamma' \subseteq \Gamma$ and $\Delta' \subseteq \Delta$. Then the following claims hold 
in $\DCmu$. 
\begin{enumerate}[\rm(1)]
\item[\rm(1)]
If $FV(M) \subseteq dom(\Gamma')$ and $FCV(M) \subseteq dom(\Delta')$, 
then $\Gamma \prove \Delta \dcmid M\colon A$ implies 
$\Gamma' \prove \Delta' \dcmid M\colon A$. 
\item[\rm(2)]
If $FV(K) \subseteq dom(\Gamma')$ and $FCV(K) \subseteq dom(\Delta')$, 
then $K\colon A \dcmid \Gamma \prove \Delta$ implies 
$K\colon A \dcmid \Gamma' \prove \Delta'$. 
\item[\rm(3)]
If $FV(S) \subseteq dom(\Gamma')$ and $FCV(S) \subseteq dom(\Delta')$, 
then $\Gamma \mid S \prove \Delta$ implies 
$\Gamma' \mid S \prove \Delta'$. 
\end{enumerate}
\end{lem}
\proof
They are shown simultaneously by induction on $M$, $K$, and $S$. 
\qed

\begin{lem}[Substitution lemma of $\Duca\mu\nu$]\label{lem:Substitution_DCmu}
The following claims hold.
\begin{enumerate}[\rm(1)]
\item[\rm(1)]
Suppose $\Gamma \prove_{\Duca\mu\nu} \Delta \dcmid N\colon A$ is derivable. 
Then the following hold. 
\begin{enumerate}[\rm(1a)]
\item[\rm(1a)]
If $\Gamma, x\colon A \prove_{\Duca\mu\nu} \Delta \dcmid M\colon B$, then 
$\Gamma \prove_{\Duca\mu\nu} \Delta \dcmid M[N/x]\colon B$,
\item[\rm(1b)]
if $K\colon B \dcmid \Gamma, x\colon A \prove_{\Duca\mu\nu} \Delta$, then 
$K[N/x] \colon B \dcmid \Gamma \prove_{\Duca\mu\nu} \Delta$, and 
\item[\rm(1c)]
if $\Gamma, x\colon A \dcmid S \prove_{\Duca\mu\nu} \Delta$, then 
$\Gamma \dcmid S[N/x] \prove_{\Duca\mu\nu} \Delta$. 
\end{enumerate}

\item[\rm(2)]
Suppose $L\colon A \dcmid \Gamma \prove_{\Duca\mu\nu} \Delta$ is derivable. 
Then the following hold. 
\begin{enumerate}[\rm(1a)]
\item[\rm(2a)]
If $\Gamma \prove_{\Duca\mu\nu} \Delta, \alpha\colon A \dcmid M\colon B$, then 
$\Gamma \prove_{\Duca\mu\nu} \Delta \dcmid M[L/\alpha]\colon B$, 
\item[\rm(2b)]
if $K\colon B \dcmid \Gamma \prove_{\Duca\mu\nu} \Delta, \alpha\colon A$, then 
$K[L/\alpha] \colon B \dcmid \Gamma \prove_{\Duca\mu\nu} \Delta$, and 
\item[\rm(2c)]
if $\Gamma \dcmid S \prove_{\Duca\mu\nu} \Delta, \alpha\colon A$, then 
$\Gamma \dcmid S[L/\alpha] \prove_{\Duca\mu\nu} \Delta$. 
\end{enumerate}
\end{enumerate}
\end{lem}
\proof
The claims (1a), (1b), and (1c) are shown simultaneously by induction on $M$, $K$, and $S$. 
The claims (2a), (2b), and (2c) are also shown simultaneously by induction on $M$, $K$, and $S$. 
\qed

The duality transformation can be extended from $\DC$ to $\DCmu$. 
\begin{defi}[Duality Transformation]\label{def:trans}
The duality transformation 
for types, terms, coterms, statements, and inference rule names of $\DCmu$ 
is defined by those of $\Duca$ and 
the following equations:
\[\begin{array}{l}
(\mu X.A)^\circ =\nu X.(A)^\circ,
\qquad 
(\nu X.A)^\circ=\mu X.(A)^\circ. 
\end{array}\]
\[\begin{array}{l}
(\In^{\mu X.A}\langle M\rangle)^\circ=\Out^{\nu X.(A)^\circ}[(M)^\circ], \\
(\Out^{\nu X.A}[K])^\circ=\In^{\mu X.(A)^\circ}\langle (K)^\circ\rangle,\\
(\Itr^A_\alpha[K,L])^\circ=\Coitr^{(A)^\circ}_{\alpha'}\langle (K)^\circ, (L)^\circ\rangle, \\ 
(\Coitr^A_x\langle M, N\rangle)^\circ=\Itr^{(A)^\circ}_{x'}[(M)^\circ, (N)^\circ].
\end{array}\]
\[\begin{array}{l}
(\mu R)^\circ = (\nu L), 
\qquad
(\nu L)^\circ = (\mu R), 
\qquad
(\mu L)^\circ = (\nu R), 
\qquad
(\nu R)^\circ = (\mu L). \vspace{6 pt}
\end{array}\]
\end{defi}

\noindent The above duality transformation is well-defined. 
\begin{lem}
The type $(A)^\circ$ is defined, and
${\rm Pos}(A) = {\rm Pos}((A)^\circ)$ and 
${\rm Neg}(A) = {\rm Neg}((A)^\circ)$ hold. 
\end{lem}
\proof
These claims are shown by induction on $A$. 
We consider the cases of $\mu X.B$ and $\nu X.B$. 
The other cases are straightforwardly proved by the induction hypothesis. 

The case of $\mu X.B$: 
Suppose that $\mu X.B$ is defined. Then we have $X$ is in ${\rm Pos}(B)$. 
By the induction hypothesis, $(B)^\circ$ is defined and $X$ occurs positively in $(B)^\circ$. 
Therefore $\nu X.(B)^\circ$ is defined, and 
we have ${\rm Pos}(\mu X.B) = {\rm Pos}(\nu X.(B)^\circ)$ and 
${\rm Neg}(\mu X.B) = {\rm Neg}(\nu X.(B)^\circ)$ by the induction hypothesis. 

The case of $\nu X.B$ can be shown in the similar way to the case of $\mu X.B$. 
\qed

This duality transformation alternates free variables and free covariables that occur in terms and coterms. 
Let $\mathcal{V}$ be a set of variables, and $\mathcal{C}$ be a set of covariables. 
Then a set of covariables $(\mathcal{V})^\circ$ is defined by $\{ x' \mid x \in \mathcal{V}\}$. 
A set of variables $(\mathcal{C})^\circ$ is also defined by $\{ \alpha' \mid \alpha \in \mathcal{C}\}$. 

\begin{lem}\label{lem:Duality_free_vars_covars_DCmu}
Let $D$ be an expression of $\DCmu$. 
Then $FV((D)^\circ) = (FCV(D))^\circ$ and $FCV((D)^\circ) = (FV(D))^\circ$ hold. 
\end{lem}
\proof
The claims are shown by induction on $D$. 
\qed

This duality transformation preserves substitution of types, terms, and coterms. 

\begin{lem}\label{lem:Duality_Substitution_DCmu}
Let $A$ and $B$ be types, $D$ be an expression, $M$ be a term, and $K$ be a coterm of $\DCmu$. 
Then the following hold. 
\begin{enumerate}[\rm(1)]
\item[\rm(1)]
 $(A[B/X])^\circ = (A)^\circ[(B)^\circ/X]$. 
\item[\rm(2)]
$(D[M/x])^\circ = (D)^\circ[(M)^\circ/x']$. 
\item[\rm(3)]
$(D[K/\alpha])^\circ = (D)^\circ[(K)^\circ/\alpha']$. 
\end{enumerate}
\end{lem}
\proof
The claim (1) is shown by induction on $A$. 
The claims (2) and (3) are shown by induction on $D$. 
\qed

The extended duality transformation preserves typing, and is an involution in $\DCmu$. 
\begin{prop}\label{prop:duality_of_typing_DCmu}
The following claims hold. 
\begin{enumerate}[\rm(1)]
\item[\rm(1)]
If $J$ is derived from $J_1,\ldots,J_n$ ($n = 1$ or $2$) by an inference rule $R$, 
then 
$(J)^\circ$ is derived from $(J_n)^\circ,\ldots,(J_1)^\circ$ by the inference rule $(R)^\circ$.
\item[\rm(2)]
        $((A)^\circ)^\circ = A$, 
        $((D)^\circ)^\circ = D$, and $((J)^\circ)^\circ = J$ hold
for any type $A$, expression $D$, and judgment $J$ of $\DCmu$. 
\end{enumerate}
\end{prop}
\proof
The claim (1) is shown by case analysis of the inference rules of $\DCmu$ using Lemma~\ref{lem:Duality_Substitution_DCmu} (1). 
The claim (2) is shown by induction on types and expressions. 
\qed

Our reduction rules for inductive and coinductive types will be defined
so that they correspond to cut elimination procedures in the classical sequent calculus LK 
extended with inductive definitions and coinductive definitions.
In the following proof figures,
we will write $\mu$, $\nu$, and $A[B]$ for $\mu X.A$, $\nu X.A$, and 
$A[B/X]$ respectively.  
In the logical system,
when the cut formula is an inductive type,
the cut elimination procedure reduces the proof 

{\small
$\infer[(Cut)]{
        \Gamma \vdash \Delta
        }{
        \infer[(\mu R)]{
                \Gamma \vdash \Delta, \mu
                }{
                \infer*{
                        \Gamma \vdash \Delta, A[\mu]
                        }{}
                }
        &
        \infer[(\mu L)]{
                \mu, \Gamma \vdash \Delta
                }{
                \infer*{
                        A[B], \Gamma \vdash \Delta, B
                        }{}
                &
                \infer*{
                        B, \Gamma \vdash \Delta
                        }{}
                }
        }$
}

to the proof 

{\small
$\infer[(Cut)]{
  \Gamma \vdash \Delta
}{
  \infer[(Cut)]{
    \Gamma \vdash \Delta, B
  }{
        \infer[(Wk)]{
                \Gamma \vdash \Delta, B, A[\mu]
                }{
                \infer*{
                      \Gamma \vdash \Delta, A[\mu]
                    }{}
                }
            &
            \infer=[(mono)]{
              A[\mu], \Gamma \vdash \Delta, B
            }{
              \infer[(\mu L)]{
                \mu, \Gamma \vdash \Delta, B
              }{
                \infer*{
                  A[B], \Gamma \vdash \Delta, B
                }{}
                &
                \infer{
                  B, \Gamma \vdash \Delta, B
                }{}
                }
              &
              \infer*{
        	A[B], \Gamma \vdash \Delta, B
              }{}
            }
        }
          &
          \infer*{
            B, \Gamma \vdash \Delta
          }{}
        }$
}

\noindent We can intuitively understand the rule $(mono)$ as follows:
$\mu \prove B$ implies $A[\mu] \prove A[B]$,
so we have $A[\mu] \prove B$ by combining it with $A[B] \prove B$.
This rule will be formally shown in Lemma \ref{lem:Map} (2a). 
This reduction changes the cut formula from $\mu$ to $A[\mu]$.
We do not have to count the cut formula $B$, 
since that cut is auxiliary.
When the cut formula is a coinductive type, 
the cut elimination procedure reduces a proof in a dual way to the above reduction.

When we have a function $\lambda x.M$ from $A$ to $B$ and
the variable $X$ is in $\Pos(C)$,
we can define the function from $C[A/X]$ to $C[B/X]$ by
extending $\lambda x.M$.
We will use $\Map^{X.C}_{A,B,x.M}\{N\}$ so that
this function maps $z$ to $\Map^{X.C}_{A,B,x.M}\{z\}$.
We will define $\Map^{X.C}_{A,B,x.M}\{N\}$ by induction on
the measure $|| C ||_X$ for a type $C$ and a type variable $X$,
which is defined by induction on $C$ as follows:
If $X$ is not free in $A$, then $|| A ||_X = 0$.
In the other cases, we assume that some $X$ occurs in $A$ and we define
\[
|| X ||_X = 1, \\
|| A\land B ||_X = || A \vee B ||_X = || A ||_X + || B ||_X + 1, \\
|| \neg A ||_X = || A ||_X + 1, \\
|| \mu Y.A ||_X = || \nu Y.A ||_X = || A ||_X + || A||_Y + 1.
\]
Note that if $X$ is not free in $B$ 
and we have $X \not= Y$, then $\|\,A\,\|_X = \|\,A[B/Y]\,\|_X$.

The number $||A||_X$ will also be used for
evaluating the size of $\Map^{X.C}_{A,B,x.M}\{N\}$ by using $M$, $N$, and $C$ 
(see Lemma~\ref{lem:evaluation_of_map}).
If we replaced $|| A ||_X + || A||_Y + 1$ by
$|| A ||_X + 1$ in the definition 
of $|| \mu Y.A ||_X$ and $|| \nu Y.A ||_X$,
it would not work for this purpose.

\begin{defi}
Assume a type variable $X$ and types $A,B,C$ are given and
$X$ is not free in $A$ and $B$.
For a variable $x$ and terms $M$ and $N$,
we define the term $\Map^{X.C}_{A,B,x.M}\{N\}$ by induction on $\|\,C\,\|_X$ as follows:
\begin{align*}
        \Map^{X.X}_{A,B,x.M}\{N\}&= (N \bullet x.(M \bullet \alpha)).\alpha,\\
        \Map^{X.C}_{A,B,x.M}\{N\}&= N 
\qquad \hbox{($X$ does not occur in $C$)},\\
        \Map^{X.C \land D}_{A,B,x.M}\{N\}&
                = 
                        \langle\,
                                \Map^{X.C}_{A,B,x.M}\{(N\bullet \Fst[\alpha]).\alpha\}, \,
                                \Map^{X.D}_{A,B,x.M}\{(N\bullet \Snd[\beta]).\beta\}\,
                        \rangle,\\
        \Map^{X.C \lor D}_{A,B,x.M}\{N\}&
                =
                (
                                N\bullet
                                [\,
                                        y.(
                                                \langle
                                                        \Map^{X.C}_{A,B,x.M}\{ y\}\,
                                                \rangle\Inl \bullet \gamma
                                        ), 
                                        z.(
                                                \langle
                                                        \Map^{X.D}_{A,B,x.M}\{ z\}\,
                                                \rangle\Inr \bullet \gamma
                                        )\,
                                ]
                ).\gamma,\\
        \Map^{X.\neg C}_{A,B,x.M}\{N\}&
                =
                [\,
                        z.(
                                N\bullet \Not\langle\, \Map^{X.C}_{B,A,x.M}\{z\}\,\rangle\,
                        )\,
                ]\Not,\\
        \Map^{X.\mu Y.C}_{A,B,x.M}\{N\}&
                =
                (\,
                        N\bullet \Itr^{\mu Y.C[B/X]}_\alpha[\,
                                z.(\,
                                        \In^{\mu Y.C[B/X]}\langle\,
                                                \Map^{X.C[\mu Y.C[B/X]/Y]}_{A,B,x.M}\{\,z\,\}\,
                                        \rangle\bullet \alpha\,
                                \,),\,\beta
                        \,]
                \,).\beta,\\
        \Map^{X.\nu Y.C}_{A,B,x.M}\{N\}&
                =
                {\Coitr^{\nu Y.C[A/X]}_z}\langle\,
                \Map^{X.C[\nu Y.C[A/X]/Y]}_{A,B,x.M}\{\,
                                                                                        (z\bullet \Out^{\nu Y.C[A/X]}[\alpha]).\alpha\,
                                                                                \}, \,N\,
                \rangle.
\end{align*}
For a covariable $\alpha$ and coterms $K$ and $L$,
we also define
\[
\Map^{X.C}_{A,B,\alpha.K}\{L\} 
        = (\Map^{X.(C)^\circ}_{(B)^\circ,(A)^\circ,\alpha'.(K)^\circ}\{(L)^\circ\})^\circ.
\]\vspace{1 pt}
\end{defi}

\noindent Note that $||\mu Y.C||_X > ||C[\mu Y.C[B/X]/Y]||_X$ and
$||\nu Y.C||_X > ||C[\nu Y.C[A/X]/Y]||_X$ hold since $X$ is not free in
$\mu Y.C[B/X]$ and $\nu Y.C[A/X]$.
We cannot replace $C[\mu Y.C[B/X]/Y]$ by $C$ in the definition of
$\Map^{X.\mu Y.C}_{A,B,x.M}\{N\}$ because of the type annotation for $\In$.
For readability, we sometimes write \ 
$\Map^{X.C}_{A,B,x}\{M, N\}$ \ and \ $\Map^{X.C}_{A,B,\alpha}\{K, L\}$ \ 
for
$\Map^{X.C}_{A,B,x.M}\{N\}$ and $\Map^{X.C}_{A,B,\alpha.K}\{L\}$, respectively.

The paper \cite{Tiu10} 
studied an intuitionistic logical system with strictly-positive
inductive definitions, and on the other hand
we study a classical logical system with positive inductive definitions.
Our cut elimination for inductive types is the same as theirs, and
on the other hand our cut elimination for coinductive types is
different from theirs.
They can avoid the use of $\Map$.
However, we cannot straightforwardly compare our method and their method,
since our system is strictly larger than their system.

Our method works only for classical logic and 
does not work for an intuitionistic logic. 
This is 
because our cut elimination procedure keeps the duality and
we have the corresponding proof rule that manipulates a succedent if
we have some proof rule that manipulates an antecedent.
In particular, we define the operator $\Map$ for coterms as the dual of 
the operator $\Map$ for terms.
Roughly speaking,
in the proof of the next lemma,
when we show the properties of $\Map$ for negation by using the derivation
\[
\infer{\neg B \prove \neg A}{
\infer{A,\neg B \prove}{
A \prove B
}},
\]
we need the following derivation in order 
to show the properties of its dual:
\[
\infer{\neg A \prove \neg B}{
\infer{\prove A,\neg B }{
B \prove A
}}
\]
which uses a non-intuitionistic sequent.

\begin{lem}\label{lem:freevar_of_map}
The following claims hold. 
\begin{enumerate}[\rm(1a)]
\item[\rm(1a)]
$FV(\Map^{X.C}_{A,B,x}\{M,N\}) \subseteq (FV(M)\setminus\{x\})\cup FV(N)$.
\item[\rm(1b)]
$FCV(\Map^{X.C}_{A,B,x}\{M,N\}) \subseteq FCV(M)\cup FCV(N)$.
\item[\rm(2a)]
$FV(\Map^{X.C}_{A,B,\alpha}\{K,L\}) \subseteq FV(K) \cup FV(L)$.
\item[\rm(2b)]
$FCV(\Map^{X.C}_{A,B,\alpha}\{K,L\}) \subseteq (FCV(K)\setminus\{\alpha\}) \cup FCV(L)$.
\end{enumerate}
\end{lem}
\proof
The claims (1a) and (1b) are shown by induction on $\|\,C\,\|_X$. 
The claims (2a) and (2b) are shown by using (1a), (2b), and Lemma~\ref{lem:Duality_free_vars_covars_DCmu}. 
\qed

\begin{lem}\label{lem:Map}
Assume $X$ is in ${\rm Pos}(C)$ and ${\rm Neg}(D)$. 
Then the following hold: 
\begin{enumerate}[\rm(1a)]
\item[\rm(1a)]
$\Gamma, x\colon A \prove \Delta \mid M\colon B$ and 
$\Gamma \prove \Delta \mid N\colon C[A]$ implies 
$\Gamma \prove \Delta \mid \Map^{X.C}_{A,B, x.M}\{N\}\colon C[B]$, 

\item[\rm(1b)]
$\Gamma, x\colon B \prove \Delta \dcmid M\colon A$ and 
$\Gamma \prove \Delta \dcmid N\colon D[A]$ implies 
$\Gamma \prove \Delta \dcmid \Map^{X.D}_{A,B, x.M}\{N\}\colon D[B]$, 

\item[\rm(2a)]
$K\colon A \dcmid \Gamma \prove \Delta, \alpha\colon B$ and 
$L\colon C[B] \dcmid \Gamma \prove \Delta$ implies 
$\Map^{X.C}_{A,B, \alpha.K}\{L\}\colon C[A] \dcmid \Gamma \prove \Delta$, 

\item[\rm(2b)]
$K\colon B \dcmid \Gamma \prove \Delta, \alpha\colon A$ and 
$L\colon D[B] \dcmid \Gamma \prove \Delta$, implies  
$\Map^{X.D}_{A,B, \alpha.K}\{L\}\colon C[A] \dcmid \Gamma \prove \Delta$, \\
\end{enumerate}
where $C[A]$ and $D[A]$ are abbreviations of $C[A/X]$ and $D[A/X]$, respectively. 
\end{lem}
\proof
The claims (1a) and (1b) are shown simultaneously by induction on $|| C ||_X$ and $|| D ||_X$. 
The claims (2a) and (2b) are shown by using (1a), (1b), and Proposition~\ref{prop:duality_of_typing_DCmu}. 
\qed

The following proposition is obtained as a special case of the above lemma. 

\begin{prop}
Assume $X$ is in ${\rm Pos}(C)$.
The following are derivable: 
\[\begin{array}{l}
\infer{
        \Gamma, z:C[A] \Dcvdash \Delta \mid \Map^{X.C}_{A,B,x.M}\{z\}:C[B]
        }{
        \Gamma, x:A \Dcvdash \Delta \mid M:B
        }
\hspace{1cm}
\infer{
        \Map^{X.C}_{A,B,\alpha.K}\{\beta\}\colon C[A] \dcmid \Gamma \Dcvdash \Delta, \beta\colon C[B]
        }{
        K\colon A \dcmid \Gamma \Dcvdash \Delta, \alpha\colon B
        }
\end{array}\]
\end{prop}

\begin{defi}
We define the one-step reduction relation $\longrightarrow_{\Duca\mu\nu}$ of $\DCmu$ 
as the compatible closure of the reduction rules of $\DC$
and the following reduction rules:\\[5pt]
$\begin{array}{ll}
(\beta\mu)
&
\In^{\mu X.C}\langle M\rangle \bullet \Itr^A_\alpha[K,L]
\longrightarrow_{\Duca\mu\nu}
(M\bullet \Map^{X.C}_{\mu X.C, A,\beta}\{\, \Itr^A_\alpha[K,\beta],\, K\,\}).\alpha \bullet L,\\
(\beta\nu)
&
\Coitr^A_x\langle M,N\rangle \bullet \Out^{\nu X.C}[K]
\longrightarrow_{\Duca\mu\nu}
N \bullet x.(\Map^{X.C}_{A,\nu X.C,z}\{\,\Coitr^A_x\langle M, z\rangle,\, M\,\}\bullet K). 
\end{array}$
\end{defi}

This system has subject reduction. 

\begin{prop}[Subject reduction of $\DCmu$]\label{prop:SubjectReduction_DCmu}
The following claims hold. 
\begin{enumerate}[\rm(1)]
\item[\rm(1)] If $\Gamma \prove_{\Duca\mu\nu} \Delta \dcmid M\colon A$ and 
$M \longrightarrow_{\Duca\mu\nu} N$, then 
$\Gamma \prove_{\Duca\mu\nu} \Delta \dcmid N\colon A$ holds. 

\item[\rm(2)] If $K\colon A \dcmid \Gamma \prove_{\Duca\mu\nu} \Delta$ and 
$K \longrightarrow_{\Duca\mu\nu} L$, then 
$L\colon A \dcmid \Gamma \prove_{\Duca\mu\nu} \Delta$ holds. 

\item[\rm(3)] If $\Gamma \dcmid S \prove_{\Duca\mu\nu} \Delta$ and 
$S \longrightarrow_{\Duca\mu\nu} T$, then 
$\Gamma \dcmid T \prove_{\Duca\mu\nu} \Delta$ holds. 
\end{enumerate}
\end{prop}
\proof
They are shown simultaneously by induction on the generation of $\longrightarrow_{\Duca\mu\nu}$ 
using 
Lemma~\ref{lem:Weakening_DCmu}, 
\ref{lem:Thinning_DCmu}, 
\ref{lem:Substitution_DCmu}, and 
\ref{lem:Map}. 
We consider the cases of $(\beta\mu)$ and $(\beta\nu)$. 

Case of $(\beta\mu)$. 
Assume $\Gamma \dcmid \In^{\mu X.C}\langle M\rangle \bullet \Itr^A_\alpha[K,L] \prove \Delta$ 
is derivable in $\Duca\mu\nu$. 
We use $\mu$ and $C[A]$ as abbreviations of $\mu X.C$ and $C[A/X]$, respectively. 
The last rule of the derivation must be $(Cut)$ rule. 
Then $\Gamma \prove \Delta \dcmid \In^{\mu X.C}\langle M\rangle\colon D$ and 
$\Itr^A_\alpha[K,L]\colon D \dcmid \Gamma \prove \Delta$ are derivable for some type $D$. 
Since the last rules of these derivations must be $(\mu R)$ and $(\mu L)$, 
we obtain $D$ is $\mu X.C$, and the derivations of 
$\Gamma \prove \Delta \dcmid M \colon C[\mu]$, the judgment
$K\colon C[A] \dcmid \Gamma \prove \Delta, \alpha\colon B$, and 
$L\colon A \dcmid \Gamma \prove \Delta, \alpha\colon B$. 
Hence we have 
$\Itr^A_\alpha[K, \beta]\colon \mu \dcmid \Gamma \prove \Delta, \beta\colon A$ by $(AxL)$ and $(\mu L)$ rules, 
and then 
$\Map^{X.C}_{\mu, A,\beta}\{\, \Itr^A_\alpha[K,\beta],\, K\,\} \colon C[\mu] \dcmid \Gamma \prove \Delta, \alpha \colon A$ is derivable by Lemma~\ref{lem:Map}. 
Therefore we have 
$\Gamma \dcmid  (M \bullet \Map^{X.C}_{\mu, A,\beta}\{\, \Itr^A_\alpha[K,\beta],\, K\,\}).\alpha \bullet L\prove \Delta$ by using $(IR)$, $(Cut)$ rules. 

The case of $(\beta\nu)$ is shown similarly to the case of $(\beta\mu)$. 

The other cases are straightforwardly proved by the induction
hypothesis.\vspace{1 pt}\qed

\noindent The duality transformation $(-)^\circ$ preserves reduction. 

\begin{thm}[Duality of $\DCmu$]\label{thm:duality_of_DCmu}
$D \longrightarrow_{\Duca\mu\nu} E$ implies $(D)^\circ \longrightarrow_{\Duca\mu\nu} (E)^\circ$ 
for any expressions $D$ and $E$.
\end{thm}
\proof
This is proved by induction on the generation of $\longrightarrow_{\Duca\mu\nu}$. 
\qed

\begin{prop}\label{prop:dual-reduction}
If $D \longrightarrow_{\Duca\mu\nu} E$ is the rules $(\beta\mu)$ and $(\beta\nu)$, then 
$(D)^\circ \longrightarrow_{\Duca\mu\nu} (E)^\circ$ is $(\beta\nu)$ and $(\beta\mu)$ respectively. 
\end{prop}

We have shown the duality of inductive types and coinductive types. 
Proposition~\ref{prop:duality_of_typing_DCmu} and 
Theorem~\ref{thm:duality_of_DCmu} show that the duality transformation is a 
homomorphic involution.
The description of a type can be defined as the set of
the type itself, its terms, its coterms, and their reduction.
The duality transformation maps the description of an inductive type and
that of a coinductive type to each other.
That is, we have the following. (1) Definition \ref{def:trans} shows that
the inductive type $\mu X.A$ is mapped to the coinductive type $\nu X.(A)^\circ$,
the term constructed by $\In$ for the inductive type is mapped to 
the coterm constructed by $\Out$ for the coinductive type,
and
the coterm constructed by $\Itr$ for the inductive type is mapped to
the term constructed by $\Coitr$ for the coinductive type.
(2) Proposition \ref{prop:dual-reduction} shows that
the cut elimination procedure of the inductive type is mapped to
the cut elimination procedure of the coinductive type. 
(3) the coinductive type is mapped to the inductive type in a similar way
to (1) and (2). 

\begin{rem}
We cannot define our typing system by using
\[
\infer[(\mu L')]{
        \Itr^A_\alpha[K,\beta]:\mu X.C \dcmid \Gamma \Dcvdash \Delta, \beta:A
        }{
        K:C[A/X] \dcmid \Gamma \Dcvdash \Delta, \alpha:A
        }
\]
instead of the typing rule $(\mu L)$.
If we used $(\mu L')$, the set of terms would not be
closed under substitution, because $\Itr^A_\alpha[K,L]$ 
would not have typing rules for it and hence it
would not be a term,
though it is obtained from $\Itr^A_\alpha[K,\beta]$ by substituting $L$ for
$\beta$.
\end{rem}

\section{Examples}\label{Sect:applications}

In this section we show some examples of inductive and coinductive
types in $\DCmu$. 
Let $X_0$ be a distinguished type variable.
We use the following abbreviations: 
\[\begin{array}{l}
        \top = \neg X_0 \vee X_0, \qquad
        \bot = \neg X_0 \land X_0, \quad
	\mbox{ and } \quad
        * = \lambda x.x.
\end{array}\]
The type $\Nat$ of natural numbers can be represented by:
\begin{align*}
\Nat &= \mu X.(\top \vee X), \\
{\tt 0} &= \In^{\Nat}\langle\,\langle *\rangle\Inl\,\rangle,\\
{\tt succ}\langle M\rangle &= 
	\In^{\Nat}\langle\,\langle M \rangle\Inr\,\rangle, 
\end{align*}
where ${\tt 0}$ is the zero and ${\tt succ}$ is the successor. 
We can prove $\Gamma \Dcvdash \Delta \dcmid {\tt 0}\colon \Nat$. 
We can also prove $\Gamma \Dcvdash \Delta \dcmid {\tt succ}\langle M\rangle\colon \Nat$ 
from $\Gamma \Dcvdash \Delta \dcmid M\colon \Nat$. 
The $n$-th natural number $\tilde n$ 
is represented by $\Succ \< \Succ \<\ldots \Succ \< \Zero \> \ldots \> \>$
\quad ($n$ times of $\Succ$).
We will write $M[\_/x]^n(N)$ for
$M[M[ \ldots [M[N/x]/x] \ldots /x]/x]$ \quad ($n$ times of $M$).
We define a coterm ${\tt Itr}^B[F, N, K]$ of type $\Nat$ by 
$\Itr^B_\alpha\bigl[\, [y.(N\bullet \alpha), x.(F\bullet (x@\alpha))], K\,\bigr]$, 
where $y$ is not free in $N$, the term $F$ has type $B \supset B$, and 
$N$ and $K$ are of type $B$.
When the coterm ${\tt Itr}^B[F, N, K]$ gets ${\tilde n}$ as its input, 
it computes $n$-time iterations of applying the function $F$ to $N$,
and passes the output to $K$.
This reduces ${\tilde n} \bullet {\tt Itr}^B[\lambda x.M, N, K]$ to 
$M[\_/x]^n(N)\bullet K$. 

The type $\List(A)$ of lists of elements of type $A$ is represented by: 
\begin{align*}
\List(A) &= \mu X.(\top \vee (A\land X)), \\
\Nil &= \In^{\List(A)}\langle\,\langle *\rangle\Inl\,\rangle,\\
M::Nl &= 
\In^{\List(A)}\langle\,\langle\,\langle M,Nl\rangle\,\rangle\Inr\,\rangle. 
\end{align*}

The term $\Nil$ is the empty list and $(::)$ is the list constructor. 
In $\DCmu$, the judgment $\Gamma \Dcvdash \Delta \dcmid \Nil\colon \List(A)$ is provable. 
The judgment $\Gamma \Dcvdash \Delta \dcmid M::Nl \colon \List(A)$ is also provable 
from $\Gamma \Dcvdash \Delta \dcmid M\colon A$ and $\Gamma \Dcvdash \Delta \dcmid Nl\colon \List(A)$. 

We note that the above examples can be considered under the call-by-value setting (section~\ref{CbvCbnDCmu}) 
if we restrict terms in the above examples to values. 

We can also define the type $\Stream(A)$ of streams of elements of type $A$ by: 
\begin{align*}
\Stream(A) &= \nu X.(A \land X), \\
{\tt cons}\langle M, {\it Ns}\rangle &= 
\Coitr^{A \land \Stream(A)}_x\langle\,\langle \pi_1(x), (\pi_2(x)\bullet \Out^{\Stream(A)}[\alpha]).\alpha\rangle, \langle M, {\it Ns}\rangle \,\rangle, \\
{\tt hd}[K] &= \Out^{\texttt{Stream}(A)}[\,\Fst[K]\,], \\
{\tt tl}[L] &= \Out^{\texttt{Stream}(A)}[\,\Snd[L]\,], 
\end{align*}
where $\pi_1(M)$ is the first projection of $M$ defined by $(M\bullet \Fst[\alpha]).\alpha$, and 
$\pi_2(M)$ is the second projection of $M$ defined by $(M\bullet \Snd[\alpha]).\alpha$. 
The term ${\tt cons}\langle M, {\it Ns}\rangle$ constructs a new stream from a given element $M$ and 
a given stream ${\it Ns}$. 
The coterm ${\tt hd}[K]$ receives the first element from a given stream 
and gives it to $K$. 
The coterm ${\tt tl}[L]$ removes the first element from a given stream
and gives the resulting stream to $L$. 
We can prove $\Gamma \Dcvdash \Delta \dcmid {\tt cons}\langle M, {\it Ns} \rangle\colon \Stream(A)$ 
from $\Gamma \Dcvdash \Delta \dcmid M\colon A$ and $\Gamma \Dcvdash \Delta \dcmid {\it Ns}\colon \Stream(A)$. 
We can also prove ${\tt hd}[K]\colon \Stream(A) \dcmid \Gamma \Dcvdash \Delta$ 
from $K\colon A \dcmid \Gamma \Dcvdash \Delta$. 
We can also prove ${\tt tl}[L]\colon \Stream(A) \dcmid \Gamma \Dcvdash \Delta$ 
from $L\colon \Stream(A) \dcmid \Gamma \Dcvdash \Delta$. 
This reduces
${\tt cons}\langle M, {\it Ns}\rangle\bullet {\tt hd}[K]$ to $M\bullet K$. 
We also reduce
${\tt cons}\langle M, {\it Ns}\rangle \bullet {\tt tl}^{n+1}[{\tt hd}[K]]$ to 
${\it Ns} \bullet {\tt tl}^n[{\tt hd}[K]]$,
where ${\tt tl}^n[{\tt hd}[K]]$ is 
defined by ${\tt tl}[{\tt tl}[\ldots{\tt tl}[\,{\tt hd}[K]\,]\ldots]]$\quad ($n$ times of ${\tt tl}$). 
Hence the coterm ${\tt tl}^n[{\tt hd}[K]]$ receives the $n$-th element of a given stream and gives it to $K$. 
Let $M$ be a term of type $A$. 
The stream of infinite number of $M$ is represented by $\Coitr^\top_x\< \<M, x\>, *\>$, 
where $x$ is a fresh variable. 
We will write ${\tt stream}(M)$ for $\Coitr^\top_x\< \<M, x\>, *\>$. 
Indeed, the statement ${\tt stream}(M) \bullet {\tt tl}^n[{\tt hd}[K]]$ is reduced to $M$ for any $n$. 
This means that any $n$-th element of ${\tt stream}(M)$ is $M$. 

We note that this stream example can be considered under the call-by-name setting (section~\ref{CbvCbnDCmu}) 
if we restrict coterms in the above example to covalues. 

\begin{prop}
$\Nat$ is dual to $\Stream(\bot)$, that is, 
$(\Nat)^\circ = \Stream(\bot)$, 
$({\tt 0})^\circ = {\tt hd}[(*)^\circ]$, 
and $({\tt succ}\langle M\rangle)^\circ = {\tt tl}[(M)^\circ]$ hold. 
\end{prop}

If $\Stream(\bot)$ is considered under the call-by-name setting 
and $\Nat$ is considered under the call-by-value setting, 
then the duality of the above proposition can be understood as follows. 
The type $\top$ means the singleton set $\{ * \}$.
The type $\bottom$ means the type of 
a program that returns some answer after computation with the input $*$
since $\bottom$ is equivalent to $\neg \top$.
The type $\Nat$ means the infinite disjoint sum $\top + \top + \top + \ldots$.
The type $\Stream(\bottom)$ means the infinite cartesian product
$\bottom \times \bottom \times \bottom \times \ldots$.
Since a term in $\Stream(\bottom)$ is equivalent to a coterm in $\Nat$,
when the term gets some natural number and is computed, it returns
some answer.
When the term gets the natural number ${\tilde n}$,
since ${\tilde n}$ is $*$ in the $n$-th $\top$ in $\top + \top + \ldots$,
the term in the $n$-th $\bottom$ in
$\bottom \times \bottom \times \ldots$ is given the input $*$ and
it is computed to give some answer.

Here we can also consider examples that  include non-deterministic choices. 
Let $M$ and $N$ be terms of same type, $x$ be a fresh variable, $\alpha$ and $\beta$ be fresh covariables. 
We define the non-deterministic choice $\<M\mid N\>$ by 
\[
\<M\mid N\> = ((M\bullet \alpha).\beta\bullet x.(N\bullet \alpha)).\alpha,
\]
where $\alpha$ and $\beta$ are fresh covariables. 
This term has both $(\beta_L)$ and $(\beta_R)$-redexes. 
It is reduced to $M$ if the $(\beta_R)$-redex is chosen, and is reduced to $N$ if the $(\beta_L)$-redex is chosen. 
Thus, $\<M\mid N\>$ can be considered as a non-deterministic choice of either $M$ or $N$. 
This non-deterministic choice $\<M\mid N\>$ is forced to choose $M$ under the call-by-value strategy, 
and is forced to choose $N$ under the call-by-name strategy. 

An example of non-deterministic computation is the list insertion function. 
This function gets a list as its input data, 
and non-deterministically chooses one arbitrary place in the list. 
Then it returns a new list that is obtained by inserting a given element at the place. 

Let $M$ be a term of type $A$, and $K'$ be a coterm of type $\List(A)\land\List(A)$. 
Then we define ${\tt ins}_M[K']$ of type $\List(A)$ by \\[5pt]
\indent
$\begin{array}{l}
{\tt ins}_M[K'] = \Itr^{\List(A)\land\List(A)}_\alpha[[L_1(\alpha),L_2(\alpha)], K'], \\
L_1(\alpha) = x.(\<M::\Nil, \Nil\>\bullet \alpha), \\
L_2(\alpha) = z.(\<\,\<\pi_1(z)::\pi_1\pi_2(z), \pi_1(z)::\pi_2\pi_2(z)\>\mid \<M::\pi_1(z)::\pi_2\pi_2(z), \pi_1(z)::\pi_1\pi_2(z)\>\,\>\bullet \alpha)
\end{array}$\\[5pt]
where 
$x$ occurs in $L_1(\alpha)$ is a fresh variable of type $\top$, and 
$z$ occurs in $L_2(\alpha)$ is a fresh variable of type $A\land (\List(A)\land\List(A))$. 
Then if $Nl$ is a list and $Nl'$ is a list obtained by inserting $M$ in some place of $Nl$, 
then the statement $Nl \bullet {\tt ins}_M[K']$ can be reduced to $\<Nl', Nl\>\bullet K'$. 
We can show this by induction on the length of $Nl$. 
If $Nl$ is $\Nil$, then $Nl'$ is $M::\Nil$. 
The statement $\Nil \bullet {\tt ins}_M[K']$ is reduced to $\<M::\Nil,\Nil\>\bullet K'$. 
If $Nl$ is $N::Nl_0$, then $Nl'$ is either $N::Nl'_0$ or $M::N::Nl_0$, 
where $Nl_0'$ is an inserted list obtained from $Nl_0$. 
The statement $(N::Nl_0) \bullet {\tt ins}_M[K']$ is reduced to 
$(\<N, (Nl_0\bullet {\tt ins}_M[\gamma]).\gamma\>\bullet L_2(\alpha)).\alpha\bullet K'$. 
Then this statement is reduced to 
$(\<N, \<Nl_0, Nl'_0\>\>\bullet L_2(\alpha)).\alpha\bullet K'$ by the induction hypothesis. 
We have $\<\,\<N::Nl'_0, Nl\> \mid \<M::N::Nl_0, Nl\>\,\>\bullet K'$. 
Hence we can obtain $\<M::N::Nl_0, Nl\> \bullet K'$ or $\<M::N::Nl_0, Nl\>\bullet K'$. 

Let $K$ be a coterm of type $\List(A)$. 
Here we define 
\[
{\tt insert}_M[K] = {\tt ins}_M[\Fst[K]]. 
\]
Then the statement $Nl \bullet {\tt insert}_M[K]$ is reduced to $Nl' \bullet K$ 
for any inserted list $Nl'$ obtained from $Nl$. 

\section{The Second-Order Dual Calculus \texorpdfstring{$\DCtwo$}{}}\label{Sect:DCtwo}

We consider the second-order extension $\DCtwo$ of $\Duca$ given by Tzevelekos~\cite{Tzevelekos06}. 
He showed the basic properties of $\DCtwo$, 
such as the substitution lemma and subject reduction. 
Without formal discussion, 
he also mentioned that his translation from $\DC$ into 
the symmetric $\lambda$-calculus can be extended to the second-order case. 
In this section, we give a formal definition of the second-order translation from $\DCtwo$ 
into the second-order symmetric $\lambda$-calculus, and show
the strong normalization of $\DCtwo$ 
by using this translation. 
For this purpose we will use the strong normalization result of 
the second-order symmetric $\lambda$-calculus given in \cite{Par00:01}.

\begin{defi}
An expression is defined to be strongly normalizing if
there does not exist any infinite reduction sequence starting from 
the expression.
\end{defi}

First, we define a second-order extension $\DCtwo$ of $\DC$. 
\begin{defi}[$\DCtwo$]
The types, terms, coterms, and statements of $\DCtwo$ are defined by:
\begin{center}
\begin{tabular}{ll}
Types&
$A \Coloneqq X \mid A\land A \mid A \vee A \mid \neg A \mid \forall X.A \mid \exists X.A$,\\
Terms&
$M \Coloneqq x \mid \langle M, M\rangle \mid \langle M\rangle\Inl \mid \langle M\rangle\Inr 
\mid [K]\Not \mid (S).\alpha \mid \langle M\rangle\A \mid \langle M\rangle\E$,\\
Coterms&
$K \Coloneqq \alpha \mid [K, K] \mid \Fst[K] \mid \Snd[K] 
\mid \Not\langle M\rangle \mid x.(S) \mid \A[K] \mid \E[K]$,\\
Statements&
$S \Coloneqq M\bullet K$.
\end{tabular}
\end{center}
The typing rules and reduction rules (denoted by $\longrightarrow_{\Duca 2}$) of $\DCtwo$ are defined by 
extending the rules of $\Duca$ with the following rules: 
\[
\begin{array}{c@{\qquad}c}
\infer[(\forall R)]{
  \Gamma \Dcvdash \Delta \dcmid \langle M\rangle\A:\forall Z.A
}{
  \Gamma \Dcvdash \Delta \dcmid M:A
}
&
\infer[(\forall L)]{
  \A[K]:\forall X.A \dcmid \Gamma \Dcvdash \Delta
}{
  K:A[B/X] \dcmid \Gamma \Dcvdash \Delta
}\\[\VSKIP]
\infer[(\exists R)]{
  \Gamma \Dcvdash \Delta \dcmid \langle M\rangle\E:\exists X.A
}{
  \Gamma \Dcvdash \Delta \dcmid M:A[B/X]
}
&
\infer[(\exists L)]{
  \E[K]:\exists Z.A \dcmid \Gamma \Dcvdash \Delta
}{
  K:A \dcmid \Gamma \Dcvdash \Delta
}
\end{array}
\]
\[
\begin{array}{ll}
(\beta\forall)
&
\langle M\rangle\A \bullet \A[K] \longrightarrow_{\Duca 2} M\bullet K,\\[\VSKIP]
(\beta\exists)
&
\langle M\rangle\E \bullet \E[K]
  \longrightarrow_{\Duca 2}
  M\bullet K, 
\end{array}
\]
where $Z$ is not free in $\Gamma$ and $\Delta$ in $(\forall R)$ and $(\exists L)$. 
We write $\longrightarrow_{\Duca 2}^+$ to denote the transitive closure 
of $\longrightarrow_{\Duca 2}$. 
\end{defi}
We have
the new constructors $\A$ and $\E$, 
which are trivial witnesses for the quantifiers at the level of expressions,
so that the system has subject reduction.
We choose our $\DC2$ so that it does not contain type information
in expressions, since
our purpose is to show strong normalization of the second-order
dual calculus, and in general 
the strong normalization of the system with type information
is implied by the strong normalization of the system without
type information. 

We sometimes use the symbol $\prove_{\Duca 2}$ instead of the symbol $\prove$ 
that appears in a judgment 
in order to explicitly show it is a judgment of $\DCtwo$. 
That is, 
we write $\Gamma \prove_{\Duca 2} \Delta \dcmid M\colon A$ for 
the judgment $\Gamma \prove \Delta \dcmid M\colon A$. 
Similarly, we write 
$K\colon A \dcmid \Gamma \prove_{\Duca 2} \Delta$ and 
$\Gamma \dcmid S \prove_{\Duca 2} \Delta$. 

We write $\Gamma[B/X]$ for $x_1\colon C_1[B/X], \ldots, x_n\colon C_n[B/X]$ 
if $\Gamma$ is $x_1\colon C_1, \ldots, x_n\colon C_n$. 
We also write $\Delta[B/X]]$ for $\alpha_1\colon D_1[B/X], \ldots, \alpha_m\colon D_m[B/X]$ 
if $\Delta$ is $\alpha_1\colon D_1, \ldots, \alpha_m\colon D_n$. 

\begin{lem}\label{lem:type_substitution_DC2}
The following claims hold. 
\begin{enumerate}[\rm(1)]
\item[\rm(1)]
If $\Gamma \prove_{\Duca 2} \Delta \dcmid M\colon A$, then 
$\Gamma[B/X] \prove_{\Duca 2} \Delta[B/X] \dcmid M\colon A[B/X]$ holds. 
\item[\rm(2)]
If $K\colon A \dcmid \Gamma \prove_{\Duca 2} \Delta$, then 
$K\colon A \dcmid \Gamma[B/X] \prove_{\Duca 2} \Delta[B/X]$ holds. 
\item[\rm(3)]
If $\Gamma \dcmid S \prove_{\Duca 2} \Delta$, then 
$\Gamma[B/X] \dcmid S \prove_{\Duca 2} \Delta[B/X]$ holds. 
\end{enumerate}
\end{lem}
\proof
They are shown simultaneously by induction on expressions. 
\qed

The basic lemmas for $\DC$ and $\DCmu$ are also shown in $\DCtwo$. 
We use Lemma~\ref{lem:type_substitution_DC2} to show weakening lemma. 

\begin{lem}[Weakening lemma]\label{lem:Weakening_DC2}
Let $\Gamma \subseteq \Gamma'$ and $\Delta \subseteq \Delta'$. Then the following hold in $\DCtwo$. 

\begin{enumerate}[\rm(1)]
\item[\rm(1)]
If $\Gamma \prove \Delta \dcmid M\colon A$ is provable, then 
$\Gamma' \prove \Delta' \dcmid M\colon A$ holds. 

\item[\rm(2)]
If $K\colon A \dcmid \Gamma \prove \Delta$ is provable, then 
$K\colon A \dcmid \Gamma' \prove \Delta'$ holds. 

\item[\rm(3)]
If $\Gamma \dcmid S \prove \Delta$ is provable, then 
$\Gamma' \dcmid S \prove \Delta'$ holds. 
\end{enumerate}
\end{lem}
\proof
They are shown simultaneously by induction on $M$, $K$, and $S$. 
We use Lemma~\ref{lem:type_substitution_DC2} when we show the cases of 
$\langle M\rangle\A$ and $\E[K]$. 
We consider these cases. 

The case of $\langle M \rangle\A$. 
Assume $\Gamma \subseteq \Gamma'$, $\Delta \subseteq \Delta'$, and 
$\Gamma \prove \Delta \dcmid \langle M\rangle\A\colon A$ is derivable. 
Since the last rule of the derivation must be $(\forall R)$, we have $A$ is $\forall X.B$ for some $B$, 
the variable $X$ is not free in $\Gamma$ and $\Delta$, and 
$\Gamma \prove \Delta \dcmid M\colon B$ is derivable. 
Then we have $\Gamma \prove \Delta \dcmid M\colon B[Z/X]$ for a fresh type variable $Z$ 
by using Lemma~\ref{lem:type_substitution_DC2}. 
By the induction hypothesis, $\Gamma' \prove \Delta' \dcmid M\colon B[Z/X]$ holds. 
Therefore we obtain $\Gamma' \prove \Delta' \dcmid \langle M\rangle\A\colon \forall Z.(B[Z/X])$ 
by $(\forall R)$ rule, since $Z$ is not free in $\Gamma'$ and $\Delta'$. 

The case of $\E[K]$ is shown similar to the case of $\langle M \rangle\A$. 

The other cases are straightforwardly proved by the induction hypothesis.
\qed

\begin{lem}\label{lem:thinning_DC2}
Let $\Gamma' \subseteq \Gamma$ and $\Delta' \subseteq \Delta$. 
Then the following hold in $\DCtwo$. 
\begin{enumerate}[(\rm(1)]
\item[\rm(1)]
If $FV(M) \subseteq dom(\Gamma')$ and $FCV(M) \subseteq dom(\Delta')$, 
then 
$\Gamma \prove \Delta \dcmid M\colon A$ implies 
$\Gamma' \prove \Delta' \dcmid M\colon A$. 

\item[\rm(2)]
If $FV(K) \subseteq dom(\Gamma')$ and $FCV(K) \subseteq dom(\Delta')$, 
then 
$K\colon A \dcmid \Gamma \prove \Delta$ implies 
$K\colon A \dcmid \Gamma' \prove \Delta'$. 

\item[\rm(3)]
If $FV(S) \subseteq dom(\Gamma')$ and $FCV(S) \subseteq dom(\Delta')$, 
then 
$\Gamma \mid S \prove \Delta$ implies 
$\Gamma' \mid S \prove \Delta'$. 
\end{enumerate}
\end{lem}
\proof
They are shown simultaneously by induction on $M$, $K$, and $S$. 
\qed

\begin{lem}[Substitution lemma]\label{lem:Substitution_DC2}
The following claims hold. 
\begin{enumerate}[\rm(1)]
\item[\rm(1)] 
Suppose $\Gamma \prove_{\Duca 2} \Delta \dcmid N\colon A$ is derivable. 
Then the following hold. 

\begin{enumerate}[\rm(1a)]
\item[\rm(1a)]
If $\Gamma, x\colon A \prove_{\Duca 2} \Delta \dcmid M\colon B$, then 
$\Gamma \prove_{\Duca 2} \Delta \dcmid M[N/x]\colon B$. 

\item[\rm(1b)]
If $K\colon B \dcmid \Gamma, x\colon A \prove_{\Duca 2} \Delta$, then 
$K[N/x] \colon B \dcmid \Gamma \prove_{\Duca 2} \Delta$. 

\item[\rm(1c)]
If $\Gamma, x\colon A \dcmid S \prove_{\Duca 2} \Delta$, then 
$\Gamma \dcmid S[N/x] \prove_{\Duca 2} \Delta$. 
\end{enumerate}

\item[\rm(2)]
Suppose $L\colon A \dcmid \Gamma \prove_{\Duca 2} \Delta$ is derivable. 
Then the following hold. 

\begin{enumerate}[\rm(1a)]
\item[\rm(2a)]
If $\Gamma \prove_{\Duca 2} \Delta, \alpha\colon A \dcmid M\colon B$, then 
$\Gamma \prove_{\Duca 2} \Delta \dcmid M[L/\alpha]\colon B$. 

\item[\rm(2b)]
If $K\colon B \dcmid \Gamma \prove_{\Duca 2} \Delta, \alpha\colon A$, then 
$K[L/\alpha] \colon B \dcmid \Gamma \prove_{\Duca 2} \Delta$. 

\item[\rm(2c)]
If $\Gamma \dcmid S \prove_{\Duca 2} \Delta, \alpha\colon A$, then 
$\Gamma \dcmid S[L/\alpha] \prove_{\Duca 2} \Delta$. 
\end{enumerate}
\end{enumerate}
\end{lem}
\proof
The claims (1a),(1b), and (1c) are shown simultaneously by induction on $M$, $K$, and $S$. 
The claims (2a),(2b), and (2c) are also shown simultaneously by induction on $M$, $K$, and $S$. 
\qed

This system has subject reduction. 

\begin{prop}[Subject reduction of $\DCtwo$]\label{SR_DC2}
The following claims hold. 
\begin{enumerate}[\rm(1)]
\item[(1)] If $\Gamma \prove_{\Duca 2} \Delta \dcmid M\colon A$ and $M \longrightarrow_{\Duca 2} N$, then 
$\Gamma \prove_{\Duca 2} \Delta \dcmid N\colon A$ holds. 

\item[\rm(2)] If $K\colon A \dcmid \Gamma \prove_{\Duca 2} \Delta$ and $K \longrightarrow_{\Duca 2} L$, then 
$L\colon A \dcmid \Gamma \prove_{\Duca 2} \Delta$ holds. 

\item[\rm(3)] If $\Gamma \dcmid S \prove_{\Duca 2} \Delta$ and $S \longrightarrow_{\Duca 2} T$, then 
$\Gamma \dcmid T \prove_{\Duca 2} \Delta$ holds. 
\end{enumerate}
\end{prop}
\proof
They are shown simultaneously by induction on the generation of $\longrightarrow_{\Duca 2}$ 
using Lemma~\ref{lem:type_substitution_DC2}, \ref{lem:Weakening_DC2}, \ref{lem:thinning_DC2}, and \ref{lem:Substitution_DC2}. 
We show the cases of $(\beta\forall)$ and $(\beta\exists)$. 

The case of $(\beta\forall)$. 
Suppose $\Gamma \dcmid \langle M \rangle\A\bullet \A[K] \prove \Delta$ is derivable. 
Then $\Gamma \prove \Delta \dcmid \langle M \rangle\A\colon C$
and 
$\A[K]\colon C \dcmid \Gamma \prove \Delta$
are derivable for some type $C$. 
Since the last rules of these derivation must be $(\forall R)$ and $(\forall L)$, 
we have $C$ is $\forall X.A$ for some $A$, $X$ is not free in both $\Gamma$ and $\Delta$, and
$\Gamma \prove \Delta \dcmid M\colon A$ and 
$K\colon A[B/X] \dcmid \Gamma \prove \Delta$ are derivable for some $B$. 
Then we can obtain 
$\Gamma \prove \Delta \dcmid M\colon A[B/X]$ by Lemma~\ref{lem:type_substitution_DC2}. 
Therefore $\Gamma \dcmid M\bullet K\prove \Delta$ can be derived by $(Cut)$ rule. 

The case of $(\beta\exists)$ is shown similar to the case of $(\beta\forall)$. 

The other cases are straightforwardly proved by the induction hypothesis. 
\qed

\begin{rem}
The trivial witnesses $\A$ and $\E$ are necessary for the subject reduction. 
If we did not have these constructors, the subject reduction would fail.
If we chose the following $(\forall R')$ and $(\forall L')$ 
instead of $(\forall R)$ and $(\forall L)$, 
\[\begin{array}{l}
\infer[(\forall R')]{
  \Gamma \Dcvdash \Delta \dcmid M:\forall Z.A
}{
  \Gamma \Dcvdash \Delta \dcmid M:A
}
\qquad
\infer[(\forall L')]{
  K:\forall X.A \dcmid \Gamma \Dcvdash \Delta
}{
  K:A[B/X] \dcmid \Gamma \Dcvdash \Delta
}
\end{array}\]
then the following would be a counter-example: 
we would have 
$\Gamma \dcmid (x\bullet \Fst[\alpha]).\alpha\bullet \beta \Dcvdash \Delta$
where $\Gamma$ is $x\colon X\land Y$, 
the sequence $\Delta$ is $\beta\colon \forall Z.X$, and $Z \not= X, Y$,
but would not have
$\Gamma \dcmid x\bullet \Fst[\beta]\vdash \Delta$,
though $(x\bullet \Fst[\alpha]).\alpha\bullet \beta$ is reduced to $x\bullet \Fst[\beta]$. 

In $\lambda$-calculus 
the constructor $\A$ is not necessary for subject reduction
while the constructor $\E$ is necessary for it \cite{Tatsuta07}.
In our system,
since $\forall$ and $\exists$ are dual,
the constructor $\A$ is also needed. 
\end{rem}

The duality transformation can be extended from $\DC$ to $\DCtwo$. 
\begin{defi}[Duality Transformation]\label{def:duality_trans_DC2}
The duality transformation 
for types, expressions, and inference rule names of $\Duca 2$ 
is defined by those of $\Duca$ and 
the following equations:
\[\begin{array}{l}
(\forall X.A)^\circ =\exists X.(A)^\circ,
\qquad
(\exists X.A)^\circ=\forall X.(A)^\circ,
\\
(\langle M\rangle\A)^\circ=\E[(M)^\circ], 
\qquad
(\E[K])^\circ = \langle (K)^\circ\rangle\A, 
\\
(\langle M\rangle\E)^\circ=\A[(M)^\circ], 
\qquad
(\A[K])^\circ = \langle (K)^\circ\rangle\E,
\\
(\forall R)^\circ = (\exists L),
\qquad
(\exists L)^\circ = (\forall R),
\qquad
(\forall L)^\circ = (\exists R),
\qquad
(\exists R)^\circ = (\forall L).\vspace{6 pt}
\end{array}\]
\end{defi}

\noindent This duality transformation preserves substitution of types, terms, and coterms. 

\begin{lem}\label{Lemma:Duality_Substitution_DC2}
Let $A$ and $B$ be types, $D$ be an expression, $M$ be a term, and $K$ be a coterm of $\DCtwo$. 
Then the following hold. 
\begin{enumerate}[\rm(1)]
\item[\rm(1)]
$(A[B/X])^\circ = (A)^\circ[(B)^\circ/X]$. 
\item[\rm(2)]
$(D[M/x])^\circ = (D)^\circ[(M)^\circ/x']$. 
\item[\rm(3)]
$(D[K/\alpha])^\circ = (D)^\circ[(K)^\circ/\alpha']$. 
\end{enumerate}
\end{lem}
\proof
The claim (1) is shown by induction on $A$. 
The claims (2) and (3) are shown by induction on $D$. 
\qed

The extended duality transformation preserves typing and reduction. It is an involution in $\DCtwo$. 
\begin{prop}[Duality of \DCtwo]\label{prop:dual_DC2}
The following claims hold. 
\begin{enumerate}[\rm(1)]
\item[\rm(1)]

If $J$ is derived from $J_1,\ldots,J_n$ ($n = 1$ or $2$) by an inference rule $R$, 
then 
$(J)^\circ$ is derived from $(J_n)^\circ,\ldots,(J_1)^\circ$ by the inference rule $(R)^\circ$. 
\item[\rm(2)]
$D \longrightarrow_{\Duca 2} E$ implies  $(D)^\circ \longrightarrow_{\Duca 2} (E)^\circ$. 
\item[\rm(3)]
$((A)^\circ)^\circ = A$, 
\SKIP 
$((D)^\circ)^\circ = D$, 
and 
$((J)^\circ)^\circ = J$ hold. 
\end{enumerate}
\end{prop}
\proof
The claim (1) is proved by case analysis of the inference rules of $\DCtwo$. 
The claim (2) is proved by induction of the generation of $\longrightarrow_{\Duca 2}$ 
using Lemma~\ref{Lemma:Duality_Substitution_DC2}. 
The claim (3) is proved by induction on types and expressions.
\qed

Next we give a definition of the second-order symmetric $\lambda$-calculus $\SL2$.
The symmetric $\lambda$-calculus is introduced by Barbanera and Berardi~\cite{Bar96:01} 
as a classical extension of the $\lambda$-calculus. The strong normalization  of its 
second-order extension $\SL2$ is proved by Parigot~\cite{Par00:01} using the reducibility  method.
The particular system we consider here is an extension of Parigot's system with 
two additional rules ($\eta_r$ and $\eta_l$). 
As discussed in ibid., Parigot's proof works with this variant without problem. 
\begin{defi}[$\SL2$]
We define
the second-order symmetric $\lambda$-calculus $\SL2$.
The types of $\SL2$ are either the special type $\bot$ or m-types (denoted by $\tau, \sigma, \ldots$) given by: 
\[
\tau \Coloneqq X \cmid X^\bot \cmid \tau \times \tau \cmid \tau + \tau
        \cmid \forall X.\tau \cmid \exists X.\tau
\]
where $X, Y, \ldots$ range over type variables.
The types $\forall X.\tau$ and $\exists X.\tau$ bind $X$ in $\tau$. 
The negation $(\tau)^\bot$ of $\tau$ is defined by: 
\[\begin{array}{ll}
(X)^\bot=X^\bot, &
(X^\bot)^\bot=X, 
\\
(\tau \times \sigma)^\bot=(\tau)^\bot + (\sigma)^\bot, &
(\tau + \sigma)^\bot=(\tau)^\bot \times (\sigma)^\bot, 
\\
(\forall X.\tau)^\bot=\exists X.(\tau)^\bot, &
(\exists X.\tau)^\bot=\forall X.(\tau)^\bot.
\end{array}\]
The symbols $x, y, \ldots$, $\alpha,\beta,\ldots$ range over variables. 
The terms of $\SL2$, denoted by $t, u, \ldots$, are defined by 
\[
t \Coloneqq
        x \cmid \Inj_1(t) \cmid \Inj_2(t) \cmid \langle t,t\rangle \cmid t*t \cmid \lambda x.t 
        \cmid \A(t) \cmid \E(t).
\]

The one-step reduction relation $\longrightarrow_{\SL2}$ of $\SL2$ is defined as the compatible closure of the following rules:
\begin{center}
\begin{tabular}{l@{\SKIP}r@{\ $\longrightarrow_{\SL2}$\ }l@{\qquad}l@{\SKIP}r@{\ $\longrightarrow_{\SL2}$\ }l}
$(\beta_r)$&
        $(\lambda x.t)*u$&$t[u/x]$, &
$(\beta_l)$&
        $u*(\lambda x.t)$&$t[u/x]$,\\
$(\beta_{\times+}1)$&
        $\langle t_1,t_2\rangle * \Inj_1(u)$&$t_1*u$,
&
$(\beta_{+\times}1)$&
        $\Inj_1(u) * \langle t_1,t_2\rangle$&$u*t_1$,\\
$(\beta_{\times+}2)$&
        $\langle t_1,t_2\rangle * \Inj_2(u)$&$t_2*u$,
&
$(\beta_{+\times}2)$&
        $\Inj_2(u) * \langle t_1,t_2\rangle$&$u*t_2$,\\
$(\beta_{\forall\exists})$&
        $\A(t) * \E(u)$&$t*u$,
&
$(\beta_{\exists\forall})$&
        $\E(u) * \A(t)$&$u*t$,\\
$(\eta_r)$&
        $\lambda y.(y*t)$&$t$,
&
$(\eta_l)$&
        $\lambda y.(t*y)$&$t$,
\end{tabular}
\end{center}
where $y$ is not free in $t$ in $(\eta_l)$ and $(\eta_r)$. 

A typing context (denoted by $\Gamma$, $\Delta$) is a finite set and 
of the form 
$x_1\colon \tau_1,\ldots,x_n\colon \tau_n$. 
A judgment of $\SL2$ takes either the form $\Gamma \vdash t:\tau$ or $\Gamma \vdash t:\bot$.
The typing rules of $\SL2$ are defined as follows: 
\[\begin{array}{c}
\infer[({\rm Ax})]{
        \Gamma, x:\tau \vdash x:\tau
        }{}
\qquad
\infer[({\rm abs})]{
        \Gamma \vdash \lambda x.t:(\tau)^\bot
        }{
        \Gamma, x:\tau \vdash t:\bot
        }
\qquad
\infer[({\rm app})]{
        \Gamma \vdash t*u:\bot
        }{
        \Gamma \vdash t:(\tau)^\bot
        &
        \Gamma \vdash u:\tau
        }
\\[\VSKIP]
\infer[(+_i)]{
        \Gamma \vdash \Inj_i(t):\tau_1+\tau_2
        }{
        \Gamma \vdash t:\tau_i
        }
\hbox{\ $(i=1,2)$}
\qquad
\infer[(\times)]{
        \Gamma \vdash \langle t,u\rangle:\tau\times \sigma
        }{
        \Gamma \vdash t:\tau
        &
        \Gamma \vdash u:\sigma
        }
\\[\VSKIP]
\infer[(\forall)]{
        \Gamma \vdash \A(t):\forall X.\tau
        }{
        \Gamma \vdash t:\tau
        }
\quad
\hbox{($X$ is not free in $\Gamma$)}
\qquad
\infer[(\exists)]{
        \Gamma \vdash \E(t):\exists X.\tau
        }{
        \Gamma \vdash t:\tau[\sigma/X]
        }
\end{array}\]
\end{defi}

\begin{thm}[Strong normalization of $\SL2$ \cite{Par00:01}]\label{thm:Parigot}
Every typable term is strongly normalizing in $\SL2$.
\end{thm}

We will give a reduction-preserving and type-preserving translation from $\DCtwo$ into $\SL2$.  
Our translation is a second-order extension of 
the translation from $\Duca$ into the symmetric $\lambda$-calculus 
given by Tzevelekos~\cite{Tzevelekos06}.

\begin{defi}
Let $A$ be a type of $\DCtwo$.
The type $(A)^\SDCtoSSL$ of $\SL2$ is defined as follows: 
\[\begin{array}{lll}
(X)^\dag =X, 
&
(A \land B)^\dag=(A)^\dag \times (B)^\dag,
&
(A \vee B)^\dag=(A)^\dag + (B)^\dag,
\\
(\neg A)^\dag=((A)^\dag)^\bot,
&
(\forall X.A)^\dag=\forall X.(A)^\dag,
&
(\exists X.A)^\dag=\exists X.(A)^\dag.
\end{array}\]

\noindent Let $D$ be an expression of $\DCtwo$.
The term $(D)^\dag$ of $\SL2$ is defined by: 
\[\begin{array}{l@{\qquad}l}
(x)^\dag = x,
&
(\alpha)^\dag=\alpha,
\\
((S).\alpha)^\dag = \lambda \alpha.(S)^\dag,
&
(x.(S))^\dag=\lambda x.(S)^\dag,
\\
(\langle M \rangle\A)^\dag=\A((M)^\dag),
&
(\langle M \rangle\E)^\dag=\E((M)^\dag),
\\
(\E[K])^\dag=\A((K)^\dag),
&
(\A[K])^\dag=\E((K)^\dag),
\\
(\langle M\rangle\Inl)^\dag=\Inj_1((M)^\dag),
&
(\Fst[K])^\dag=\Inj_1((K)^\dag),
\\
(\langle M\rangle\Inr)^\dag=\Inj_2((M)^\dag),
&
(\Snd[K])^\dag=\Inj_2((K)^\dag),
\\
(\langle M, N\rangle)^\dag=\langle (M)^\dag, (N)^\dag \rangle,
&
([K, L])^\dag=\langle (K)^\dag, (L)^\dag \rangle,
\\
([K]\Not)^\dag= \lambda x.(x * (K)^\dag),
&
(\Not\langle M\rangle)^\dag=(M)^\dag, 
\\
(M\bullet K)^\dag = (M)^\dag * (K)^\dag.
\span
\end{array}\]
We define the translation of $[K]\Not$ by using 
$\eta$-expansion, 
so that all reductions in $\DCtwo$ are strictly simulated in $\SL2$. 

$(\Gamma)^\dag$
and 
$((\Delta)^\dag)^\bot$ are defined as
$x_1\colon (A_1)^\dag, \ldots, x_n\colon (A_n)^\dag$ 
and 
$\alpha_1\colon ((B_1)^\dag)^\bot, \ldots, \alpha_m\colon ((B_m)^\dag)^\bot$ 
respectively if 
$\Gamma$ is $x_1\colon A_1, \ldots, x_n\colon A_n$, 
and 
$\Delta$ is $\alpha_1\colon B_1, \ldots, \alpha_m\colon B_m$.
For a judgment $J$ of $\DCtwo$,
the judgment $(J)^\dag$ of $\SL2$ is defined as follows:
The judgment $(\Gamma \Dcvdash \Delta \dcmid M\colon A)^\dag$ is
defined as
$(\Gamma)^\dag, ((\Delta)^\dag)^\bot \vdash (M)^\dag\colon (A)^\dag$.
The judgment
$(K\colon A \dcmid \Gamma \Dcvdash \Delta)^\dag$ is defined as
$(\Gamma)^\dag, ((\Delta)^\dag)^\bot \vdash (K)^\dag\colon ((A)^\dag)^\bot$.
The judgment
$(\Gamma \dcmid S \Dcvdash \Delta)^\dag$ is defined as
$(\Gamma)^\dag, ((\Delta)^\dag)^\bot \vdash (S)^\dag\colon \bot$.
\end{defi}

This translation preserves provability and one-step reductions. 
\begin{prop}\label{prop:soundness_dag}
The following claims hold. 
\begin{enumerate}[\rm(1)]
\item[\rm(1)]
If $J$ is provable in $\DCtwo$, then $(J)^\dag$ is provable in $\SL2$.
\item[\rm(2)]
$D \longrightarrow_{\Duca 2} E$ implies $(D)^\dag \longrightarrow_{\SL2} (E)^\dag$. 
\end{enumerate}
\end{prop}
\proof
The claim (1) is shown by induction on the proof of $J$.
The claim (2) is shown by induction on the definition of $\longrightarrow_{\Duca 2}$. 
\qed

We can obtain strong normalization of $\DCtwo$ from the above proposition. 

\begin{thm}[Strong normalization of $\DCtwo$]\label{thm:SN_DCtwo}
Every typable expression is strongly normalizing in $\DCtwo$.
\end{thm}
\proof
Assume there is an infinite reduction sequence 
\[
D = D_0\longrightarrow_{\Duca 2} D_1 \longrightarrow_{\Duca 2} \ldots 
\]
starting from $D$. 
From Proposition~\ref{prop:soundness_dag}, 
the expression $(D)^\dag$ is typable in $\SL2$, and 
\[
(D)^\dag \longrightarrow_{\SL2} (D_1)^\dag \longrightarrow_{\SL2} \ldots 
\]
is an infinite reduction sequence. 
This contradicts Theorem~\ref{thm:Parigot}. 
\qed

\begin{rem}
Tzevelekos~\cite{Tzevelekos06} also gave 
a back translation $(-)^p$ from the symmetric $\lambda$-calculus into $\Duca$. 
As noted in his paper (Section 3, Note 3.5), this translation cannot extend to the second-order case 
since it does not preserve typing judgments for existential quantification. 
This is because the translation $(-)^p$ does not preserve type substitution: $(A[B/X])^p \neq (A)^p[(B)^p/X]$. 
The same argument applies to ours. 
\end{rem}

\section{Strong Normalization \texorpdfstring{of $\DCmu$}{}}

In this section, we prove strong normalization in $\DCmu$. 
We will give a translation from $\DCmu$ into $\DCtwo$ that is 
based on the second-order encoding of inductive and coinductive types. 
Our proof of strong normalization will be done 
by showing the fact that one-step reduction in $\DCmu$ is translated to one or more 
steps reduction in $\DCtwo$. 

We use the following degree of expressions in $\DCmu$ for defining
the second-order coding of inductive and coinductive types.
 
When we try to prove some properties of expressions by induction on
expressions, that induction sometimes does not work,
since the expression contains
$\Map^{X,C}_{A,B,x.M}\{N\}$ that is defined by using induction on $||C||_X$.
In order for solving this,
we will introduce the pair of the summation of $||C||_X$ and the size of an expression
as a measure. 
\begin{defi}
Let $D$ be an expression in $\DCmu$. The number $||D||$ is defined by:
\[\begin{array}{l}
|| x ||=|| \alpha || = 0, 
\\
|| \langle M, N \rangle ||=|| \Coitr^A_x\langle M, N \rangle || = \Max(|| M ||, || N ||), 
\\
|| M\bullet K ||=\Max(|| M ||, || K ||), 
\\
|| [K, L] ||=|| \Itr^A_\alpha[K,L] || = \Max(|| K ||, || L ||), 
\\
|| (S).\alpha ||=|| x.(S) || = || S ||, 
\\
|| \langle M\rangle\Inl ||=|| \langle M\rangle\Inr || = || \Not\langle M\rangle || = || M ||, 
\\
|| \In^{\mu X.A}\langle M\rangle ||=|| M || + || A ||_X + 1, 
\\
|| \Fst[K] ||=|| \Snd[K] || = || [K]\Not || = || K ||, 
\\
|| \Out^{\nu X.A}[K] ||=|| K || + || A ||_X + 1. 
\end{array}\]

The number $|D|$ is defined by: 
\[\begin{array}{l}
| x |=| \alpha | = 0, 
\\
| \langle M, N \rangle |=| \Coitr^A_x\langle M, N \rangle | = | M | + | N | + 1, 
\\
| M\bullet K |=| M | + | K | + 1, 
\\
| [K, L] |=| \Itr^A_\alpha[K,L] | = | K | + | L | + 1, 
\\
| (S).\alpha |=| x.(S) | = | S | + 1, 
\\
| \langle M\rangle\Inl |=| \langle M\rangle\Inr |= | \Not\langle M\rangle | = | \In^{\mu X.A}\langle M\rangle | = | M | + 1, 
\\
| \Fst[K] |=
| \Snd[K] | 
= | [K]\Not | 
= | \Out^{\nu X.A}[K] |
= | K | + 1.
\end{array}\]
The degree $\Deg(D)$ of the expression $D$
is defined as the  pair $(|| D ||, |D|)$.
We also define the order of the degrees by the lexicographic order. 
\end{defi}
The number $|D|$ is the number of constructors in the expression $D$.
The number $||D||$ is the maximum summation of $(||A||_X+1)$ for 
$\In^{\mu X.A}\langle M\rangle$ and
$\Out^{\nu X.A}[K]$ in paths in $D$.
 For example,
$\Deg(\In^{\mu X.\neg X \lor X}\langle 
( \langle [x.(\langle x\rangle\Inr\bullet \gamma)]\Not\rangle\Inl\bullet \gamma ).\gamma
\rangle) = (4,7)$. 
We have $||E|| \le ||D||$ and $|E| < |D|$ when
the expression $E$ is a proper subexpression of $D$.
The degree satisfies the following properties. 

\begin{lem}\label{lem:evaluation_of_map}
The following claims hold.

\begin{enumerate}[\rm(1)]
\item[\rm(1)]
 $|| D || = || (D)^\circ ||$ and $| D | = | (D)^\circ | $ hold.
\item[\rm(2)]
$||\,\Map^{X.A}_{B,C,x}\{M, N\}\,|| \le || M || + || N || + || A ||_X$ holds.
\item[\rm(3)]
$\Deg(\In^{\mu X.A}\langle M\rangle) > 
\Deg(\, \Map^{X.A}_{\mu X.A,Y,\alpha}\{\, x.(y\bullet (x@\alpha)), \beta \,\})$ holds.
\end{enumerate}
\end{lem}
\proof
The claims of (1) are shown by induction on $D$. 
The claim (2) is shown by induction on $|| A ||_X$.  
The claim (3) is proved by using (2). 
\qed

We present the second-order encoding for $\DCmu$.
We will write $\lambda(x,\alpha).S$ for $\lambda x.((S).\alpha)$. 
Then $(\lambda(x,\alpha).S)\bullet (N@K)$ is reduced to $S[N/x][K/\alpha]$. 

\begin{defi}[Translation $\overline{(-)}$ from $\DCmu$ into $\DCtwo$]
Let $A$ be a type of $\DCmu$. 
The type $\overline{A}$ of $\DCtwo$ is defined as follows: 
\[
\overline{X} =X, 
\qquad
\overline{A \land B} =\overline{A} \land \overline{B}, 
\qquad
\overline{\neg A} =\neg\overline{A}, 
\qquad
\overline{A \vee B} =\overline{A} \vee \overline{B}, 
\]
\[
\overline{\mu X.A} =\forall X.((\overline{A}\supset X)\supset X), 
\qquad
\overline{\nu X.A} =\exists X.( \neg( \neg\overline{A} \land X) \land X), 
\]
where $\supset$ is defined in Definition \ref{def:imp}.
For an expression $D$ of $\DCmu$,
the expression $\overline{D}$ of $\DCtwo$ is defined by induction on $\Deg(D)$ as follows. 
For the expressions $D$ of the same degree, we first define
$\overline D$ for $D$ such that $D$ is not of the form
$\overline{\Out^{\nu X.A}[K]}$ or
$\overline{\Coitr^A_x\langle M,N\rangle}$,
and we next define 
$\overline D$ for $D$ such that $D$ is of the form
$\overline{\Out^{\nu X.A}[K]}$ or
$\overline{\Coitr^A_x\langle M,N\rangle}$.
\[
\begin{array}{c@{\qquad\qquad}c}
\overline{x}=x, 
&
\overline{\alpha}=\alpha,
\\
\overline{(S).\alpha}=(\overline{S}).\alpha,
&
\overline{x.(S)}=x.(\overline{S}),
\\
\overline{\langle M,N\rangle}=\langle \overline{M}, \overline{N}\rangle,
&
\overline{[K, L]}=[\overline{K}, \overline{L}],
\\
\overline{\langle M\rangle\Inl}=\langle\overline{M}\rangle\Inl,
&
\overline{\Fst[K]}=\Fst[\overline{K}],
\\
\overline{\langle M\rangle\Inr}=\langle\overline{M}\rangle\Inr,
&
\overline{\Snd[K]}=\Snd[\overline{K}],
\\
\overline{[K]\Not}=[\overline{K}]\Not,
&
\overline{\Not\langle M\rangle}=\Not\langle\overline{M}\rangle,
\end{array}
\]
\begin{align*}
\overline{M\bullet K} &= \overline{M}\bullet \overline{K},
\\
\overline{\Itr^A_\alpha[K,L]}
&=
\A[\,(\,\lambda (x,\alpha).(x\bullet \overline{K})\,)@\overline{L}\,],
\\
\overline{\Out^{\nu X.A}[K]}
&=
  (
         \overline{
	            \In^{\mu X.(A)^\circ}\langle (K)^\circ \rangle
	            }
 )^\circ,
\\
\overline{\In^{\mu X.A}\langle M\rangle}
&=
\langle\,
\lambda (y,\beta).
                (
		y\bullet 
		(\,
		({\mathcal Q}_Y[X.A]\bullet{\mathcal R}_M\{y, \gamma\}).\gamma@\beta
		\,)
                )
		\,\rangle\A, 
\\
\overline{\Coitr^A_x\langle M,N\rangle}
&=
 (
         \overline{
	            \Itr^{(A)^\circ}_{x'}[(M)^\circ, (N)^\circ]
	            }
 )^\circ, 
\end{align*}
where  
${\mathcal Q}_Y[X.A]$ is defined as
$\lambda y.\lambda(z,\beta).
        (\,
	z\bullet \overline{
	  \Map^{X.A}_{\mu X.A,Y,\alpha}\{\,
	  x.(y\bullet (x@\alpha)), \beta
	  \,\}
	  }
	\,)$,
and 
${\mathcal R}_M\{N, K\}$ is defined as
$(\,
\lambda(x,\alpha).(x\bullet \A[N@\alpha])
\,)@(\overline{M}@K)$.

We also define the translation of judgments.
The context $\overline{\Gamma}$ is defined as
$x_1\colon\overline{A_1},\ldots, x_n\colon\overline{A_n}$ if 
$\Gamma$ is $x_1\colon A_1,\ldots, x_n\colon A_n$.
The cocontext
$\overline{\Delta}$ is defined as
$\alpha_1\colon\overline{B_1},\ldots, \alpha_m\colon\overline{B_m}$ if 
$\Delta$ is $\alpha_1\colon B_1,\ldots, \alpha_m\colon B_m$. 
The judgment
$\overline{\Gamma \vdash \Delta \dcmid M\colon A}$ is defined as 
$\overline{\Gamma} \vdash \overline{\Delta} \dcmid \overline{M}\colon \overline{A}$.
The judgment 
$\overline{K\colon A \dcmid \Gamma \vdash \Delta}$ is defined as
$\overline{K}\colon \overline{A} \dcmid \overline{\Gamma} \vdash \overline{\Delta}$.
The judgment
$\overline{\Gamma \dcmid S \vdash \Delta}$ is defined as
$\overline{\Gamma} \dcmid \overline{S} \vdash \overline{\Delta}$.
\end{defi}

The next lemma shows that
this translation commutes with $(-)^\circ$. 
\begin{lem}\label{lemma:commute}
$(\,\overline{A}\,)^\circ = \overline{(A)^\circ}$, 
$(\,\overline{D}\,)^\circ = \overline{(D)^\circ}$, and 
$(\,\overline{J}\,)^\circ = \overline{(J)^\circ}$ hold. 
\end{lem}
\proof
The claim for $A$ is proved by induction on $A$. 

The claim for $D$ is proved by induction on $D$. 
The cases of 
$\In^{\mu X.A}\langle M\rangle$, $\Out^{\nu X.A}[K]$, 
$\Itr^A_\alpha[K, L]$, and $\Coitr^A_x\langle M, N \rangle$ are shown by 
the definition of the translation and the dualities of $\DCmu$ and $\DCtwo$. 
The case of $\In^{\mu X.A}\langle M\rangle$ is shown as follows: 
$(\overline{\In^{\mu X.A}\langle M\rangle})^\circ = 
(\overline{\In^{\mu X.((A)^\circ)^\circ}\langle ((M)^\circ)^\circ\rangle})^\circ = 
\overline{\Out^{\nu X.(A)^\circ}[(M)^\circ]} = 
\overline{\,(\In^{\mu X.A}\langle M\rangle)^\circ\,}$. 
We can show the cases $\Out^{\nu X.A}[K]$, 
$\Itr^A_\alpha[K, L]$, and $\Coitr^A_x\langle M, N \rangle$ similarly. 
The other cases are straightforwardly proved by the induction hypothesis.

The claim for $J$ is proved by using the claims for $A$ and $D$. 
\qed

The translation $\overline{(-)}$  preserves substitution. 
\begin{lem}\label{lemma:subst_term}
$\overline{A[B/X]} = \overline{A}[\overline{B}/X]$, 
$\overline{D[N/x]} = \overline{D}[\overline{N}/x]$, and 
$\overline{D[L/\alpha]} = \overline{D}[\overline{L}/\alpha]$ hold. 
\end{lem}
\proof
The first claim
is shown by induction on $A$. 
The second and the third claims are shown simultaneously by induction on $\Deg(D)$. 
For the expressions $D$ of the same degree, we first show the claims
for $D$ such that $D$ is not of the form
$\overline{\Out^{\nu X.A}[K]}$ or
$\overline{\Coitr^A_x\langle M,N\rangle}$,
and we next show the claims
for $D$ such that $D$ is of the form
$\overline{\Out^{\nu X.A}[K]}$ or
$\overline{\Coitr^A_x\langle M,N\rangle}$.

We consider the cases of $\In^{\mu X.A}\<M\>$, $\Itr^A_\alpha[K,L]$,
$\Out^{\nu X.A}[K]$ and $\Coitr^A_x\langle M, N \rangle$. 
The other cases are straightforwardly proved by the induction hypothesis.

The second claim of the case $\In^{\mu X.A}\<M\>$ is shown in the following way. 
By the induction hypothesis, we have 
$\overline{\Map^{X.A}_{\mu X.A,Y,\alpha}\{\, x.(y\bullet (x@\alpha)), \beta \,\}}[\overline{N}/x] = 
\overline{\Map^{X.A}_{\mu X.A,Y,\alpha}\{\, x.(y\bullet (x@\alpha)), \beta \,\}[N/x]}$ 
since $\Deg(\, \Map^{X.A}_{\mu X.A,Y,\alpha}\{\, x.(y\bullet (x@\alpha)), \beta \,\}) 
< \Deg(\In^{\mu X.A}\<M\>) $ by Lemma~\ref{lem:evaluation_of_map}~(3). 
By Lemma~\ref{lem:freevar_of_map}~(2a), 
$\overline{\Map^{X.A}_{\mu X.A,Y,\alpha}\{\, x.(y\bullet (x@\alpha)), \beta \,\}[N/x]} = 
\overline{\Map^{X.A}_{\mu X.A,Y,\alpha}\{\, x.(y\bullet (x@\alpha)), \beta \,\}}$. 
Hence we have $({\mathcal Q}_Y[X.A])[\overline{N}/x] = {\mathcal Q}_Y[X.A]$. 
By the induction hypothesis, we have 
${\mathcal R}_M\{y, \gamma\}[\overline{N}/x] = {\mathcal R}_{M[N/x]}\{y, \gamma\}$ 
since $\Deg(M) < \Deg(\In^{\mu X.A}\<M\>)$. 
Therefore, $\overline{\In^{\mu X.A}\<M\>}[\overline{N}/x]$ is equal to 
$\langle\,
\lambda (y,\beta).
                (
		y\bullet 
		(\,
		(({\mathcal Q}_Y[X.A])[\overline{N}/x]\bullet({\mathcal R}_M\{y, \gamma\})[\overline{N}/x]).\gamma@\beta
		\,)
                )
		\,\rangle\A$. 
Then it is equal to 
$\langle\,
\lambda (y,\beta).
                (
		y\bullet 
		(\,
		({\mathcal Q}_Y[X.A]\bullet{\mathcal R}_{M[N/x]}\{y, \gamma\}.\gamma@\beta
		\,)
                )
		\,\rangle\A$. 
The last term is equal to $\overline{(\In^{\mu X.A}\<M\>)[N/x]}$ by the definition of $\overline{(-)}$. 

The second claim of the case $\Itr^A_\alpha[K,L]$ is shown in the following way. 
The coterm $\overline{\Itr^A_\alpha[K,L]}[\overline{N}/x]$ is equal to 
$\A[\,(\,\lambda (y,\alpha).(y\bullet \overline{K}[\overline{N}/x])\,)@\overline{L}[\overline{N}/x]\,]$. 
By the induction hypothesis, it is equal to 
$\A[\,(\,\lambda (y,\alpha).(y\bullet \overline{K[N/x]})\,)@\overline{L[N/x]}\,]$. 
Hence it is equal to $\overline{(\Itr^A_\alpha[K,L])[N/x]}$ by the definition of $\overline{(-)}$. 

The second claim of the case $\Out^{\nu X.A}[K]$ is shown in the following way. 
Since $||K|| = ||(K)^\circ||$ and $|K| = |(K)^\circ|$ by Lemma~\ref{lem:evaluation_of_map} (1), 
we have $\Deg(\Out^{\nu X.A}[K]) = \Deg(\In^{\mu X.(A)^\circ}\langle (K)^\circ \rangle)$. 
Hence 
$\overline{\In^{\mu X.(A)^\circ}\langle (K[N/x])^\circ \rangle} 
= 
\overline{\In^{\mu X.(A)^\circ}\<(K)^\circ\>}[\overline{(N)^\circ}/x']$  holds
by  Lemma~\ref{lem:Duality_Substitution_DCmu},
since the third claim for 
$\In^{\mu X.(A)^\circ}\langle (K)^\circ \rangle$ is already shown
before this case.
Then we can obtain the claim of this case as follows: 

$\overline{(\Out^{\nu X.A}[K])[N/x]} 
= 
(\,\overline{\In^{\mu X.(A)^\circ}\<(K[N/x])^\circ \>}\,)^\circ
= 
(\,\overline{\In^{\mu X.(A)^\circ}\<(K)^\circ\>}[\overline{(N)^\circ}/x']\,)^\circ
=
(\,\overline{\In^{\mu X.(A)^\circ}\<(K)^\circ\>}\,)^\circ[(\overline{(N)^\circ})^\circ/x]
=
\overline{\Out^{\nu X.A}[K]}[((\overline{N})^\circ)^\circ/x]
= 
\overline{\Out^{\nu X.A}[K]}[\overline{N}/x]
$. 

The third claim of this case is shown similarly. 

The second and third claims of the case $\Coitr^A_x\langle M, N\rangle$ 
is shown in the similar way to the case of $\Out^{\nu X.A}[K]$. 
\qed

Note that the second and third claims of the above lemma cannot
be proved straightforwardly by induction on $D$. For example,
for proving the case of $\In^{\mu X.A}\<M\>$ in the second claim,
we need induction hypothesis for
$\Map^{X.A}_{\mu X.A,Y,\alpha}\{\,
	  x.(y\bullet (x@\alpha)), \beta
	  \,\}
$
but it is not a subterm of $\In^{\mu X.A}\<M\>$.

The next proposition says
the translation $\overline{(-)}$ preserves provability. 
\begin{prop}\label{prop:DCMUtoSDC_provability}
If $J$ is provable in $\DCmu$, then $\overline{J}$ is provable in $\DCtwo$. 
\end{prop}
\proof
This is shown by induction on the degree of the principal expression in $J$. 
We show the cases of $\In^{\mu X.A}\langle M \rangle$, $\Out^{\nu X.A}[K]$, $\Itr^A_\alpha[K,L]$, 
and $\Coitr^A_x\langle M, N \rangle$. 

The case of $\Itr^A_\alpha[K,L]$ is shown by the induction hypothesis and Lemma~\ref{lemma:subst_term}. 
The cases of $\Coitr^A_x\langle M, N \rangle$ and $\Out^{\nu X.A}[K]$ are 
shown by the induction hypothesis and the dualities of $\DCmu$ and $\DCtwo$. 

We prove the case of $\In^{\mu}\langle M \rangle$. 
We write $\mu$, $A[B]$, and $\overline{A}[C]$ 
as abbreviations of $\mu X.A$, $A[B/X]$, and $\overline{A}[C/X]$ respectively. 
This case is shown by the following three steps: 
(a)  we show $\mathcal{R}_M\{y,\gamma\}\colon (\overline{\mu}\supset Y)\supset \overline{A}[\overline{\mu}] \supset \overline{A}[Y] \dcmid \overline{\Gamma}, y\colon \overline{A}[Y]\supset Y \prove \overline{\Delta}, \gamma\colon \overline{A}[Y]$ is derivable, where $\mathcal{R}_M\{y,\gamma\}$ is 
$(\lambda(x,\alpha).(x\bullet \A[y@\alpha])\,)@(\overline{M}@\gamma)$. 
Next, (b) we show $\quad \prove \quad \dcmid \mathcal{Q}_Y[X.A]\colon (\overline{\mu}\supset Y) \supset \overline{A}[\overline{\mu}] \supset \overline{A}[Y]$ is derivable, where 
${\mathcal Q}_Y[X.A]$ is 
$\lambda y.\lambda(z,\beta).
        (\,
	z\bullet \overline{
	  \Map^{X.A}_{\mu X.A,Y,\alpha}\{\,
	  x.(y\bullet (x@\alpha)), \beta
	  \,\}
	  }
	\,)$. 
Finally, (c) we can easily show $\overline{\Gamma} \prove \overline{\Delta} \dcmid \In^{\mu}\langle M\rangle\colon \mu$ from (a) and (b). 

The claim (a) is shown in the following way. 
Suppose $\Gamma \prove \Delta \dcmid \In^{\mu}\langle M \rangle\colon \mu$ is derivable. 
Then we have the derivation of $\Gamma \prove \Delta \dcmid M\colon A[\mu]$. 
By the induction hypothesis and Lemma~\ref{lemma:subst_term}, we obtain 
$\overline{\Gamma} \prove \overline{\Delta} \dcmid \overline{M}\colon \overline{A}[\overline{\mu}]$. 
Then we have a derivation of 
$\overline{M}@\gamma\colon \overline{A}[\overline{\mu}] \supset \overline{A}[Y] \dcmid \overline{\Gamma} \prove \overline{\Delta}, \gamma\colon \overline{A}[Y]$ by $(\supset L)$ rule. 
On the other hand, we can show 
$y\colon \overline{A}[Y]\supset Y  \prove \quad \dcmid \lambda(x,\alpha).(x\bullet \A[y@\alpha])\colon \overline{\mu} \supset Y$. 
Then we have 
$\mathcal{R}_M\{y,\gamma\}\colon (\overline{\mu}\supset Y)\supset \overline{A}[\overline{\mu}] \supset \overline{A}[Y] \dcmid \overline{\Gamma}, y\colon \overline{A}[Y]\supset Y \prove \overline{\Delta}, \gamma\colon \overline{A}[Y]$. 

The claim (b) is shown as follows. 
We can show 
$\Map^{X.A}_{\mu,Y,\alpha}\{\,x.(y\bullet (x@\alpha)), \beta\,\}\colon A[\mu] \dcmid y\colon \mu\supset Y \prove \beta\colon A[Y]$ in $\DCmu$ by using Lemma~\ref{lem:Map},  the judgment
$x.(y\bullet (x@\alpha))\colon \mu \dcmid y\colon \mu\supset Y \prove \alpha\colon Y, \beta\colon A[Y]$,  and 
$\beta \colon A[\mu] \dcmid y\colon \mu\supset Y \prove \beta\colon A[Y]$. 
By Lemma~\ref{lem:evaluation_of_map} (3), 
we have $\Deg(\In^{\mu X.A}\langle M \rangle) > \Deg(\Map^{X.A}_{\mu,Y,\alpha}\{\,x.(y\bullet (x@\alpha)), \beta\,\})$. 
Hence $\overline{\Map^{X.A}_{\mu,Y,\alpha}\{\,x.(y\bullet (x@\alpha)), \beta\,\}}\colon \overline{A}[\overline{\mu}] \dcmid y\colon \overline{\mu}\supset Y \prove \beta\colon \overline{A}[Y]$ is derivable by induction hypothesis and Lemma~\ref{lemma:subst_term}. 
Therefore we obtain 
$\  \prove \quad \dcmid \mathcal{Q}_Y[X.A]\colon (\overline{\mu}\supset Y) \supset \overline{A}[\overline{\mu}] \supset \overline{A}[Y]$. 

The other cases are straightforwardly proved by the induction hypothesis.
\qed

The translation $\overline{(-)}$ maps one-step reduction to one or more steps 
of reduction. 

\begin{prop}\label{prop:DCMUtoSDC_reduction}
For expressions $D$ and $E$ of $\DCmu$,
the relation 
$D \longrightarrow_{\Duca\mu\nu} E$ implies $\overline{D} \longrightarrow^+_{\Duca 2} \overline{E}$. 
\end{prop}
\proof
First we show the claim without $(\beta \mu)$ nor $(\beta\nu)$ 
by induction on $\longrightarrow_{\Duca\mu\nu}$ with Lemma~\ref{lemma:subst_term}. 
Next, by using this and Lemma~\ref{lemma:subst_term},
we show the claim of this proposition
by induction on $\longrightarrow_{\Duca\mu\nu}$.
We consider cases according to the reduction rule.

The case of $(\beta\mu)$ is shown as follows: 
Suppose we have $\overline{\In^{\mu X.A}\langle M\rangle\bullet \Itr^B_\alpha[K,L]}$. 
This is equal to 
$\langle 
\lambda (y,\beta).
                (
		y\bullet 
		( 
		(\mathcal{Q}_Y[X.A]\bullet\mathcal{R}_M\{y, \gamma\}).\gamma@\beta
		 )
                )
		 \rangle\A
\bullet 
\A[ ( \lambda (x.\alpha).(x\bullet \overline{K}) )@\overline{L} ]$. 
It is reduced to 
$( 
\lambda (y,\beta).
                (
		y\bullet 
		( 
		(\mathcal{Q}_Y[X.A]\bullet\mathcal{R}_M\{y, \gamma\}).\gamma@\beta
		 )
                )
		 )
\bullet 
( ( \lambda (x.\alpha).(x\bullet \overline{K}) )@\overline{L} )$, 
and then we have 
$(
	\lambda (x.\alpha).(x\bullet \overline{K}) 
)
        \bullet 
		( 
                	( 
                                \mathcal{Q}_Y[X.A]\bullet\mathcal{R}_M\{
                                                                \lambda (x.\alpha).(x\bullet \overline{K}) , \gamma
                                                                \}
                        ).\gamma@\overline{L}
		 )$ 
by more than one step reduction. 
Since 
$\lambda (x.\alpha).(x\bullet \overline{K})$
equals 
$\lambda x.((x\bullet \overline{K}).\alpha)$, 
we have 
$
(( 
        \mathcal{Q}_Y[X.A]\bullet\mathcal{R}_M\{
                                         \lambda (x.\alpha).(x\bullet \overline{K}) , \gamma
                                          \}
).\gamma
\bullet 
\overline{K}).\alpha \bullet \overline{L}$ 
by $(\beta\supset)$. 
This is reduced to 
$(( 
        \mathcal{Q}_Y[X.A]\bullet\mathcal{R}_M\{
                                         \lambda (x.\alpha).(x\bullet \overline{K}) , \overline{K}
                                          \}).\alpha \bullet \overline{L}$ 
by $(\beta R)$. 
Here 
$\mathcal{R}_M\{
                                         \lambda (x.\alpha).(x\bullet \overline{K}) , \overline{K}
                                          \}$ 
is equal to 
$(
\overline{\lambda(y,\beta).(y\bullet \Itr^B_\alpha[K, \beta])}@(\overline{M}@\overline{K})
)$. 
Hence we can reduce 
$\mathcal{Q}_Y[X.A]
\bullet
\mathcal{R}_M\{ 
             \lambda (x.\alpha).(x\bullet \overline{K}) , \overline{K}
              \}$ 
to 
$\overline{M} \bullet \overline{\Map^{X.A}_{\mu, Y, \alpha}\{\Itr^B_\alpha[K,\beta], K\}}$ 
by using Lemma~\ref{lemma:subst_term} and the first claim.
Therefore the previously obtained expression 
$(( 
        \mathcal{Q}_Y[X.A]\bullet\mathcal{R}_M\{
                                         \lambda (x.\alpha).(x\bullet \overline{K}) , \overline{K}
                                          \}).\alpha \bullet \overline{L}$ 
is reduced to 
$( 
\overline{M} \bullet \overline{\Map^{X.A}_{\mu X.A, Y, \alpha}\{\Itr^B_\alpha[K,\beta], K\}}
        ).\alpha \bullet \overline{L}$. 
This is equal to 
$\overline{
	  (M
          \bullet 
	  \Map^{X.A}_{\mu X.A,Y,\beta}\{ 
                \Itr^B_\alpha[K,\beta], K
	   \}).\alpha \bullet L
        }$. 

The case of $(\beta\nu)$ is shown by using the duality of $(\beta\nu)$ and $(\beta\mu)$, 
the duality of $\DCtwo$, and Lemma~\ref{lemma:commute}. 

Other cases are shown straightforwardly.
\qed

Finally, we complete a proof of strong normalization of $\DCmu$.
\begin{thm}[Strong normalization of $\DCmu$]\label{thm:SN_iDuca}
Every typable expression of $\DCmu$ is strongly normalizing.
\end{thm}

\proof
Assume that 
$D$ is typable in $\DCmu$ and there is an infinite reduction sequence 
\[
D \longrightarrow_{\Duca\mu\nu} D_1 \longrightarrow_{\Duca\mu\nu} \ldots
\]
starting from $D$.
Then 
$\overline{D}$ is typable in $\DCtwo$
by Proposition \ref{prop:DCMUtoSDC_provability} and 
\[
\overline{D} \longrightarrow^+_{\Duca 2} \overline{D_1} \longrightarrow^+_{\Duca 2} \ldots 
\]
is an infinite reduction sequence starting from $\overline D$
by Proposition \ref{prop:DCMUtoSDC_reduction}. 
This contradicts Theorem~\ref{thm:SN_DCtwo}. 
\qed

\section{
The call-by-value and call-by-name 
\texorpdfstring{$\DCmu$}{Dual Calculus with inductive and coinductive types}
} \label{CbvCbnDCmu}

The motivation for introducing the dual calculus in \cite{Wad03:01} 
was to show the duality between call-by-value and call-by-name. 
In this section, we follow this motivation. 
That is, we will extend the duality to inductive and coinductive types by 
introducing the call-by-value and call-by-name variants of $\DCmu$. 
These variants also satisfy the important properties such as strong normalization and the Church-Rosser property. 

We recall the definition of the call-by-value and call-by-name $\DC$. 
The call-by-value and call-by-name dual calculus use the notion of values and covalues. 
They are defined as follows. 

\begin{defi}[Values and covalues of $\DC$ \cite{Wad03:01}]
The values (denoted by $V, W,\ldots$) and covalues (denoted by $P, Q, \ldots$) of $\DC$ 
are defined by the following grammar: 
\[\begin{array}{l}
V ::= x \mid \langle V,V\rangle \mid \langle V\rangle\Inl \mid \langle V\rangle\Inr \mid [K]\Not, 
\\
P ::= \alpha \mid [P,P] \mid \Fst[P] \mid \Snd[P] \mid \Not\<M\>, 
\end{array}\]
where $M$ is a term and $K$ is a coterm of $\DC$. 
\end{defi}

The types, expressions, and typing rules of the call-by-value and call-by-name $\DC$ are the same as them of $\DC$. 
The call-by-value reduction relation of $\DC$ is defined as follows. 

\begin{defi}[Call-by-value reduction rules of $\DC$]
The call-by-value reduction relation $\longrightarrow^v_{\tt DC}$ of $\DC$ 
is defined from the following rules. 
\[
\begin{array}{ll}
(\beta\land_1) _v&
\langle V,W\rangle\bullet \Fst[K]  \longrightarrow^v_{\tt DC} V\bullet K, \\
(\beta\land_2)_ v&
\langle V,W\rangle\bullet \Snd[K] \longrightarrow^v_{\tt DC} W\bullet K, \\
(\beta\vee_1)_v&
\langle V\rangle\Inl \bullet [K,L] \longrightarrow^v_{\tt DC} V\bullet K, \\
(\beta\vee_2)_v&
\langle W\rangle\Inr \bullet [K,L] \longrightarrow^v_{\tt DC} W\bullet L, \\
(\beta\neg)_v&
[K]\Not \bullet \Not\langle M\rangle \longrightarrow^v_{\tt DC} M\bullet K,\\
(\beta R)_v&
(S).\alpha \bullet K \longrightarrow^v_{\tt DC} S[K/\alpha],\\
(\beta L)_v&
 V \bullet x.(S) \longrightarrow^v_{\tt DC} S[V/x],\\
(\varsigma \land_1)_v&
\<\nvalue,N\> \longrightarrow^v_{\tt DC} (\nvalue\bullet x.(\<x,N\>\bullet \alpha)).\alpha, \\
(\varsigma \land_2)_v&
\<V, \nvalue\> \longrightarrow^v_{\tt DC} (\nvalue\bullet x.(\<V,x\>\bullet \alpha)).\alpha, \\
(\varsigma \vee_1)_v&
\<\nvalue\>\Inl \longrightarrow^v_{\tt DC} (\nvalue\bullet x.(\<x\>\Inl\bullet \alpha)).\alpha, \\ 
(\varsigma \vee_2)_v&
\<\nvalue\>\Inr \longrightarrow^v_{\tt DC} (\nvalue\bullet x.(\<x\>\Inr\bullet \alpha)).\alpha, \\
(\eta R)^+_v&
M \longrightarrow^v_{\tt DC} (M\bullet \alpha).\alpha,\quad \mbox{and}\\
(\eta L)^+_v&
K \longrightarrow^v_{\tt DC} x.(x\bullet K), 
\end{array}
\]
where $\nvalue$ is not a value, 
and $x$ and $\alpha$ in 
$(\varsigma\land_1)_v$, $(\varsigma\land_2)_v$, 
$(\varsigma\vee_1)_v$, $(\varsigma\vee_2)_v$, $(\eta L)^+_v$ and $(\eta R)^+_v$ 
are fresh. 
\end{defi}

An example of use of $\varsigma$-rules is
\[
\<(S).\alpha\>\Inl \red^v_\Duca ((S).\alpha \bullet x.(\<x\>\Inl \bullet \beta)).\beta \red^v_\Duca (S[x.(\<x\>\Inl \bullet \beta)/\alpha]).\beta.
\]
This system is obtained from
the call-by-value dual calculus given in \cite{Wad03:01}
by removing the implication.

We note that the original system in \cite{Wad03:01} includes implication types, 
values for implication, and a call-by-value $\beta$-rule for implication. 
However, as mentioned in \cite{Wad03:01}, 
an implication $A \supset B$ can be defined as $\neg(A\land \neg B)$ under call-by-value. 
Hence each value for implication can be replaced a value in terms of other connectives, 
and the reduction rule for implication can be simulated by the other $\beta$-rules. 

The rules $(\varsigma\land_1)_v$, $(\varsigma\land_2)_v$, 
$(\varsigma\vee_1)_v$, and $(\varsigma\vee_2)_v$ are 
the separated forms 
of the rule $(\varsigma)$ given in \cite{Wad03:01},
 and our rules  are equivalent to his rule. 
However, we prefer this separated form since this form is easy to 
add $\varsigma$-rules for inductive and coinductive types later. 

The symbol $+$ used in $(\eta L)^+_v$ and $(\eta R)^+_v$ means $\eta$-expansion rules. 
 When we extend call-by-value and call-by-name calculi with inductive and
coinductive types later in this section, 
we will use the reduction $(\eta R)$ and $(\eta L)$ instead of
the above expansion $(\eta R)_v^+$ and $(\eta L)_v^+$ for the following 
reasons.
In \cite{Wad03:01}, $\eta$-rules requires side conditions to avoid infinite reduction sequence: 
``expansions $(\eta L)$ and $(\eta R)$ should be applied only to a term $M$ or coterm $K$ that is not the
immediate subject of a cut''. 
However, two problems still remain about $\eta$-expansion rules. 
One problem is that a value becomes non-value by the $\eta$-expansion: 
For example, a value $x$ is expanded to a non-value $(x\bullet \alpha).\alpha$ by $(\eta R)^+_v$-rule. 
The second problem is that infinite reduction sequences occur with $\varsigma$-rule: 
For example, 
$\<x\>\Inl \bullet \beta$ is reduced to 
$\<(x\bullet \alpha).\alpha\>\Inl\bullet \beta$ by $(\eta R)^+_v$. 
Since $(x\bullet \alpha).\alpha$ is not a value, it can be reduced to 
$((x\bullet \alpha).\alpha \bullet y.(\<y\>\Inl\bullet \gamma)).\gamma \bullet \beta$ by $(\varsigma\vee_1)_v$. 
Then, we have $\<x\>\Inl \bullet \beta$ again by $(\beta L)_v$ and $(\beta R)_v$-rules. 
Tzevelekos~\cite{Tzevelekos06} assumed additional conditions on $\eta$-expansion rules, 
and showed strong normalization and the Church-Rosser properties 
of the call-by-value and call-by-name $\DC$ under his conditions. 
However, his approach does not solve the first problem. 
One simple solution for the both problems is to replace $\eta$-expansion by $\eta$-reduction. 
For this reason, we will adopt $\eta$-reduction in our call-by-value and call-by-name systems. 

The dual calculus considered in \cite{Wad05:01} has $\eta$-rules for conjunction, disjunction, and negation. 
These rules could be defined naturally because the system 
in \cite{Wad05:01} was based on equations. 
However, we cannot define these $\eta$-rules naively in the call-by-value and call-by-name reduction systems of $\DC$ 
since these rules break the Church-Rosser property: 
The call-by-value $(\eta\vee)$-rule defined in \cite{Wad05:01} is 
$[x.(\<x\>\Inl\bullet K), y.(\<y\>\Inr\bullet K)] = K$, where $K$ has type $A\vee B$. 
Suppose that we add $(\eta\vee)$-reduction rule 
$[x_1.(\<x_1\>\Inl\bullet K), x_2.(\<x_2\>\Inr\bullet K)] \to^v_{\Duca} K$  to the call-by-value $\DC$. 
Then the statement $[x_1.(\<x_1\>\Inl\bullet y.(z\bullet \alpha)), x_2.(\<x_2\>\Inr\bullet y.(z\bullet \alpha))]$ 
has two normal forms $[x_1.(z\bullet \alpha), x_2.(z\bullet \alpha)]$ and $y.(z\bullet \alpha)$. 
Suppose that we add $(\eta\vee)$-expansion rule 
$K \to^v_{\Duca} [x_1.(\<x_1\>\Inl\bullet K), x_2.(\<x_2\>\Inr\bullet K)]$  to the call-by-value $\DC$. 
The statement $x\bullet y.(z\bullet \alpha)$
\quad (the variable $z$ and the covariable $\alpha$ have type $X$, and the variables $x$ and $y$ have type $A\vee B$) 
is reduced to $z\bullet \alpha$ by $(\beta_L)$-rule. 
The statement $x\bullet y.(z\bullet \alpha)$ is also expanded to 
$x\bullet [x_1.(\<x_1\>\Inl\bullet y.(z\bullet \alpha)), x_2.(\<x_2\>\Inr\bullet y.(z\bullet \alpha))]$ 
by $(\eta\vee)$-rule, 
and then it is reduced to $x\bullet [x_1.(z\bullet \alpha), x_2.(z\bullet \alpha)]$ by $(\beta_L)_v$-rule. 
These two results are never confluent since the first one $z\bullet \alpha$ cannot produce 
a coterm of the form $[K,L]$, and the bracket $[..]$ in 
the second one $x\bullet [x_1.(z\bullet \alpha), x_2.(z\bullet \alpha)]$ cannot be eliminated. 

The call-by-name reduction relation of $\DC$ is defined as follows. 

\begin{defi}[Call-by-name reduction rules of $\DC$]
The call-by-name reduction relation $\longrightarrow^n_{\tt DC}$ of $\DC$ 
is defined from the following rules. 
\[
\begin{array}{ll}
(\beta\land_1) _n&
\langle M,N\rangle\bullet \Fst[P]  \longrightarrow^n_{\tt DC} M\bullet P, \\
(\beta\land_2)_ n&
\langle M,N\rangle\bullet \Snd[P] \longrightarrow^n_{\tt DC} N\bullet P, \\
(\beta\vee_1)_n&
\langle M\rangle\Inl \bullet [P,Q] \longrightarrow^n_{\tt DC} M\bullet P, \\
(\beta\vee_2)_n&
\langle M\rangle\Inr \bullet [P,Q] \longrightarrow^n_{\tt DC} M\bullet Q, \\
(\beta\neg)_n&
[K]\Not \bullet \Not\langle M\rangle \longrightarrow^n_{\tt DC} M\bullet K,\\
(\beta R)_n&
(S).\alpha \bullet P \longrightarrow^n_{\tt DC} S[P/\alpha],\\
(\beta L)_n& 
 M \bullet x.(S) \longrightarrow^n_{\tt DC} S[M/x],\\
(\varsigma \land_1)_n&
\Fst[\ncovalue] \longrightarrow^n_{\tt DC} x.((x\bullet \Fst[\alpha]).\alpha \bullet \ncovalue), \\
(\varsigma \land_2)_n&
\Snd[\ncovalue] \longrightarrow^n_{\tt DC} x.((x\bullet \Snd[\alpha]).\alpha \bullet \ncovalue), \\
(\varsigma \vee_1)_n&
[\ncovalue,L] \longrightarrow^n_{\tt DC} x.((x\bullet [\alpha,L]).\alpha \bullet \ncovalue), \\
(\varsigma \vee_2)_n&
[P,\ncovalue] \longrightarrow^n_{\tt DC} x.((x\bullet [P,\alpha]).\alpha \bullet \ncovalue), \\
(\eta R)^+_n&
M \longrightarrow^n_{\tt DC} (M\bullet \alpha).\alpha,\quad \mbox{and}\\
(\eta L)^+_n&
K \longrightarrow^n_{\tt DC} x.(x\bullet K), 
\end{array}
\]
where $\ncovalue$ is not a covalue, 
and $x$ and $\alpha$ in 
$(\varsigma\land_1)_n$, $(\varsigma\land_2)_n$, 
$(\varsigma\vee_1)_n$, $(\varsigma\vee_2)_n$, $(\eta L)^+_n$ and $(\eta R)^+_n$ 
are fresh. 
\end{defi}

This system is obtained from
the call-by-name dual calculus given in \cite{Wad03:01}
by removing the implication.

As mentioned in \cite{Wad03:01}, 
an implication $A \supset B$ can be defined as $\neg A\vee B$ under call-by-name. 
Hence, covalues for implication, and a call-by-name reduction rules for implication 
given in the original system can be replaced in terms of other connectives. 

The call-by-value reduction and 
the call-by-name reduction are dual strategies in $\DC$. 

\begin{prop}[Duality between call-by-value and call-by-name in $\DC$ \cite{Wad03:01}]
Let $D$ and $E$ be expressions of $\DC$. Then, \ 
$D \longrightarrow^v_{\tt DC} E$ iff $(D)^\circ \longrightarrow^n_{\tt DC} (E)^\circ$, 
where $(-)^\circ$ is the duality transformation defined in the section~2. 
\end{prop}

Now we will introduce the call-by-value and call-by-name variants of $\DCmu$. 
We first consider a call-by-value restriction of $\DCmu$ (called {\it weak call-by-value $\DCmu$}) which is given by 
simply restricting the reduction rules of $\DCmu$. 
This restricted system satisfies both strong normalization and the Church-Rosser properties. 
However, this system is rather weak since it lacks the $\varsigma$-rules. 
The call-by-value $\DCmu$ (denoted by ${\tt CBV}$ $\Duca\mu\nu$) is obtained 
by adding the $\varsigma$-rules to the weak call-by-value $\DCmu$. 
The weak call-by-name $\DCmu$ and the call-by-name $\DCmu$ (denoted by ${\tt CBN}$ $\Duca\mu\nu$) 
are also considered. The call-by-name $\DCmu$ is the dual system of the call-by-value $\DCmu$. 

We first define the notion of values and covalues in $\DCmu$. 

\begin{defi}[Values and covalues of  $\DCmu$]
The values (denoted by $V, W, \ldots$) and 
the covalues (denoted by $P, Q, \ldots$) of $\DCmu$ are defined by the following grammar: 
\[\begin{array}{l}
V ::= x \mid \langle V,V\rangle \mid \langle V\rangle\Inl \mid \langle V\rangle\Inr \mid [K]\Not 
\mid \In^{\mu X.A}\langle V\rangle \mid \Coitr^A_x\langle M,V\rangle, 
\\
P ::= \alpha \mid [P,P] \mid \Fst[P] \mid \Snd[P] \mid \Not\<M\>
\mid \Out^{\nu X.A}[P] \mid \Itr^A_\alpha[K,P],
\end{array}\]
where $M$ is a term and $K$ is a coterm of $\DCmu$. 
\end{defi}

The set of values of $\DCmu$ is a subset  of terms of $\DCmu$. 
The set of covalues of $\DCmu$ is a subset  of coterms of $\DCmu$. 
Note that the above definition is a straightforward extension of the definition of values and covalues in $\DC$. 

The set of values and covalues are closed under substitution of values and covalues, respectively. 

\begin{lem}\label{lem:subst_value}
Let $V$ and $W$ be values, and $P$ and $Q$ be covalues of $\DCmu$. 
The following claims hold. 
\begin{enumerate}[\rm(1)]
\item[\rm(1)]
$V[W/x]$ is a value of $\DCmu$. 
\item[\rm(2)]
$P[Q/\alpha]$ is a covalue of $\DCmu$. 
\end{enumerate}
\end{lem}
\proof
They are straightforwardly proved by induction on $V$ and $P$.
\qed

The types, expressions, and typing rules of 
the weak call-by-value and the weak call-by-name $\DCmu$ are the same as them of $\DCmu$. 
The reduction relation of the weak call-by-value $\DCmu$ is given as follows. 

\begin{defi}[Reduction rules of the weak call-by-value $\DCmu$]\rm
The reduction relation $\wCBVred$
of the weak call-by-value $\DCmu$ is defined as the compatible closure of the reduction rules 
$(\beta\land_1)_v$, $(\beta\land_2)_v$, $(\beta\vee_1)_v$, $(\beta\vee_2)_v$, $(\beta\neg)_v$, 
$(\beta R)_v$, $(\beta L)_v$, 
and the following reduction rules: \\

$\begin{array}{ll}
(\beta\mu)_v&
\In^{\mu X.C}\langle V\rangle \bullet \Itr^A_\alpha[K,L]
\wCBVred
(V\bullet \Map^{X.C}_{\mu X.C, A,\beta}\{\, \Itr^A_\alpha[K,\beta],\, K\,\}).\alpha \bullet L,
\\
(\beta\nu)_v&
\Coitr^A_x\langle M,V\rangle \bullet \Out^{\nu X.C}[K]
\wCBVred
V \bullet x.(\Map^{X.C}_{A,\nu X.C,z}\{\,\Coitr^A_x\langle M, z\rangle,\, M\,\}\bullet K), 
\\
(\eta R)_v&
(M\bullet \alpha).\alpha \wCBVred M, 
\\
(\eta L)_v&
x.(x\bullet K) \wCBVred K, 
\end{array}$\\

\noindent
where $x$ and $\alpha$ are fresh in $(\eta L)_v$ and $(\eta R)_v$, respectively. 
\end{defi}

The reduction relation of the weak call-by-name $\DCmu$ is given as follows. 

\begin{defi}[Reduction rules of the weak call-by-name $\DCmu$]\rm
The reduction relation $\wCBNred$
of the weak call-by-name $\DCmu$ is defined as the compatible closure of the reduction rules 
$(\beta\land_1)_n$, $(\beta\land_2)_n$, $(\beta\vee_1)_n$, $(\beta\vee_2)_n$, $(\beta\neg)_n$, 
$(\beta R)_n$, $(\beta L)_n$, 
and the following reduction rules: \\

$\begin{array}{ll} 
(\beta\mu)_n&
\In^{\mu X.C}\langle M\rangle \bullet \Itr^A_\alpha[K,P]
\wCBNred
(M\bullet 
\Map^{X.C}_{\mu X.C, A,\beta}\{\, \Itr^A_\alpha[K,\beta],\, K\,\}).\alpha \bullet P,\\
(\beta\nu)_n&
\Coitr^A_x\langle M,N\rangle \bullet \Out^{\nu X.C}[P]
\wCBNred
M \bullet 
x.(\Map^{X.C}_{A,\nu X.C,z}\{\,\Coitr^A_x\langle M, z\rangle,\, M\,\} \bullet P), \\
(\eta R)_n&
(M\bullet \alpha).\alpha \wCBNred M, \\
(\eta L)_n&
x.(x\bullet K) \wCBNred K, 
\end{array}$\\

\noindent
where $x$ and $\alpha$ are fresh in $(\eta L)_n$ and $(\eta R)_n$, respectively. 
\end{defi}

The weak call-by-value reduction and the weak call-by-name reduction are dual strategies. 

\begin{prop}[Duality between weak call-by-value and weak call-by-name in $\DCmu$]\label{prop:dual_wCBVwCBN}
Let $D$ and $E$ be expressions of $\DCmu$. Then, \ 
$D \wCBVred E$ iff $(D)^\circ \wCBNred (E)^\circ$, 
where $(-)^\circ$ is the duality transformation of $\DCmu$ defined in the section~3. 
\end{prop}

The rules $(\beta\neg)_v$, $(\beta R)_v$, $(\eta L)_v$, and $(\eta R)_v$ are the same as 
$(\beta\neg)$, $(\beta R)$, $(\eta L)$, and $(\eta R)$-rules of $\DCmu$, respectively. 
The rules 
$(\beta\land_1)_v$, 
$(\beta\land_2)_v$, 
$(\beta\vee_1)_v$, 
$(\beta\vee_2)_v$, 
$(\beta L)_v$, 
$(\beta\mu)_v$, and $(\beta\nu)_v$ 
are just restrictions of the rules 
$(\beta\land_1)$, 
$(\beta\land_2)$, 
$(\beta\vee_1)$, 
$(\beta\vee_2)$, 
$(\beta L)$, 
$(\beta\mu)$, and $(\beta\nu)$, respectively. 
The situation of the call-by-name case is similar to the call-by-value case. 
Hence, we can easily obtain the following proposition. 

\begin{prop}
Let $D$ and $E$ be expressions in $\DCmu$. Then the following claims hold. 
\begin{enumerate}[\rm(1)]
\item[\rm(1)]
If $D \wCBVred E$, then $D \longrightarrow_{\Duca\mu\nu} E$. 
\item[\rm(2)]
If $D \wCBNred E$, then $D \longrightarrow_{\Duca\mu\nu} E$. 
\end{enumerate}
\end{prop}

From the above proposition and the strong normalization result of $\DCmu$ (Theorem~\ref{thm:SN_iDuca}), 
we have the strong normalization of the weak call-by-value and the weak call-by-name reduction relations. 

\begin{prop}[Strong normalization of the weak CBV and CBN $\DCmu$]\label{prop:SN_wCBVwCBN}
We have the following. 
\begin{enumerate}[\rm(1)]
\item[\rm(1)]
Every typable expression is strongly normalizing in the weak call-by-value $\DCmu$. 
\item[\rm(2)]
Every typable expression is strongly normalizing in the weak call-by-name $\DCmu$. 
\end{enumerate}
\end{prop}

The reduction relations $\wCBVred$ and $\wCBNred$ of $\DCmu$ satisfy the Church-Rosser property. 
We first recall the definition of the Church-Rosser property. 

\begin{defi}[Church-Rosser property]\rm
Let $A$ be a set and $\rightarrow$ be a reduction relation on $A$. 
We write $b \leftarrow a \rightarrow c$ if both $a \rightarrow b$ and $a \rightarrow c$ hold. 
We also write $b \rightarrow a \leftarrow c$ if both $b \rightarrow a$ and $c \rightarrow a$ hold. 
\begin{enumerate}[(1)]
\item[(1)]
The reduction relation $\to$ satisfies the diamond property if, for all $a, b, c \in A$,  the relation
$b \leftarrow a \rightarrow c$ implies that there exists $d \in A$ such that $b \rightarrow d \leftarrow c$. 
\item[(2)]
The reduction relation $\to$ satisfies the Church-Rosser property if $\to^*$ satisfies the diamond property, 
where $\to^*$ is the reflexive transitive closure of $\to$. 
\end{enumerate}
\end{defi}

From now on, we concentrate to show the Church-Rosser property of $\wCBVred$. 
The Church-Rosser property of $\wCBNred$ can be obtained from 
the result of $\wCBVred$ and the duality (Proposition~\ref{prop:dual_wCBVwCBN}). 
In order to show the Church-Rosser property of $\wCBVred$, 
we will use the parallel reduction technique. 
The definition of the parallel reduction relation is given as follows. 

\begin{defi}[Parallel reduction of the weak call-by-value $\DCmu$]\rm
The parallel reduction relation (denoted by $\pared$) of the weak call-by-value $\DCmu$ 
is defined inductively from the following rules. 

$x \pared x$ and $\alpha \pared \alpha$ for any variable $x$ and covariable $\alpha$. 

$\< M,N\> \pared \<M',N'\>$ \quad
if $M \pared M'$ and $N \pared N'$. 

$[K,L] \pared [K',L']$ \quad
if $K \pared K'$ and $L \pared L'$. 

$\< M\>\Inl \pared \<M'\>\Inl$, \ $\< M\>\Inr \pared \<M'\>\Inr$, 
 and \ $\Not\<M\> \pared \Not\<M'\>$ \quad
if $M \pared M'$. 

$\Fst[K] \pared \Fst[K']$, \ $\Snd[K] \pared \Snd[K']$, 
and \ $[K]\Not \pared [K']\Not$ \quad
if $K \pared K'$. 

$\In^{\mu X.A}\< M\> \pared \In^{\mu X.A}\<M'\>$\quad
if $M \pared M'$. 

$\Out^{\nu X.A}[K] \pared \Out^{\nu X.A}[K']$\quad
if $K \pared K'$. 

$\Coitr^A_x\< M,N\> \pared \Coitr^A_x\<M',N'\>$\quad
if $M \pared M'$ and $N \pared N'$. 

$\Itr^A_\alpha[K,L] \pared \Itr^A_\alpha[K',L']$\quad
if $K \pared K'$ and $L \pared L'$. 

$M\bullet K \pared M'\bullet K'$\quad
if $M \pared M'$ and $K \pared K'$. 

$(S).\alpha \pared (S').\alpha$\ and\ $x.(S) \pared x.(S')$\quad
if $S \pared S'$. 

$M\bullet x.(S) \pared S'[V/x]$\quad
if $M \pared V$ and $S \pared S'$. 

$(S).\alpha\bullet K \pared S'[K'/\alpha]$\quad
if $K \pared K'$ and $S \pared S'$. 

$\<M,N\>\bullet \Fst[K] \pared V\bullet K'$\quad
if $M \pared V$, $N \pared W$, and $K \pared K'$.  

$\<M,N\>\bullet \Snd[K] \pared W\bullet K'$\quad
if $M \pared V$, $N \pared W$, and $K \pared K'$.  

$\<M\>\Inl\bullet [K,L] \pared V\bullet K'$\quad
if $M \pared V$, $K \pared K'$. 

$\<M\>\Inl\bullet [K,L] \pared V\bullet L'$\quad
if $M \pared V$, $L \pared L'$. 

$[K]\Not\bullet \Not\<M\> \pared M'\bullet K'$\quad
if $M \pared M'$ and $K \pared K'$. 

$\In^{\mu X.C}\langle M\rangle \bullet \Itr^A_\alpha[K,L]
\pared
(V\bullet \Map^{X.C}_{\mu X.C, A,\beta}\{\, \Itr^A_\alpha[K',\beta],\, K'\,\}).\alpha \bullet L'$\quad
if $M \pared V$, $K \pared K'$, and $L \pared L'$. 

$\Coitr^A_x\langle M,N\rangle \bullet \Out^{\nu X.C}[K]
\pared
V \bullet x.(\Map^{X.C}_{A,\nu X.C,z}\{\,\Coitr^A_x\langle M', z\rangle,\, M'\,\}\bullet K')$\quad
if $M \pared M'$, $N \pared V$, and $K \pared K'$. 

$(M\bullet \alpha).\alpha \pared M'$\quad
if $M \pared M'$ and $\alpha$ is not free in $M$. 

$x.(x\bullet K) \pared K'$\quad
if $K \pared K'$ and $x$ is not free in $K$. 

\end{defi}

The parallel reduction of the weak call-by-value $\DCmu$ satisfies the following basic properties. 

\begin{lem}\label{lem:pared_basic}
Let $M$ be a term, $V$ and $V'$ be values, $K$ and $K'$ be coterms, 
and $D$ and $D'$ be expressions of $\DCmu$. 
Then the following hold. 
\begin{enumerate}[\rm(1)]
\item[\rm(1)]
Suppose $D \pared E$. If $D$ is a term, then $E$ is also a term. 
If $D$ is a coterm, then $E$ is also a coterm. 
If $D$ is a statement, then $E$ is also a statement. 
If $D$ is a value, then $E$ is also a value. 
\item[\rm(2)]
$D \pared D$. 
\item[\rm(3)]
If $M \pared V$ and $D \pared D'$, then $D[M/x] \pared D'[V/x]$. 
\item[\rm(4)]
If $K \pared K'$ and $D \pared D'$, then $D[K/\alpha] \pared D'[K'/\alpha]$. 
\end{enumerate}
\end{lem}
\proof
The claim (1) is shown by induction on the definition of $\pared$. 
The claim (2) is shown by induction on $D$. 

The claim (3) is shown by induction on $D \pared D'$ with Lemma~\ref{lem:subst_value}. 
We show the case that $N_0\bullet y.(T_0) \pared T_1[W/y]$ is derived from 
$N_0 \pared W$ and $T_0\pared T_1$. 
By the induction hypothesis, we have $N_0[M/x] \pared W[V/x]$ and $T_0[M/x] \pared T_1[V/x]$. 
By Lemma~\ref{lem:subst_value}, $W[V/x]$ is a value. 
Hence we have 
$(N_0\bullet y.(T_0))[M/x] = (N_0[M/x])\bullet y.(T_0[M/x]) 
\pared T_1[V/x][W[V/x]/y] = T_1[W/y][V/x]$. 
The other cases are straightforwardly proved by the induction hypothesis.

The claim (4) is shown by induction on $D \pared D'$. 
\qed

\begin{lem}\label{lem:pared_CR}
Let $D$ and $D'$ be expressions of $\DCmu$. Then the following claims hold. 
\begin{enumerate}[\rm(1)]
\item[\rm(1)]
If $D \wCBVred E$, then $D \pared E$. 
\item[\rm(2)]
If $D \pared E$, then $D \wCBVred^* E$. 
\item[\rm(3)]
The parallel reduction relation $\pared$ satisfies the diamond property, that is, 
if the relation $D_1 \Leftarrow D \pared D_2$ holds, 
then there exists $E$ such that $D_1 \pared E \Leftarrow D_2$. 
\end{enumerate}
\end{lem}
\proof
The claim (1) is shown by induction on the definition of $\wCBVred$. 
The claim (2) is shown by induction on the definition of $\pared$. 

The claim (3) is shown by induction on $D$. 
We show the case that $D$ is the shape of $(S).\alpha\bullet x.(T)$, $D_1$ is $S'[L/\alpha]$, and $D_2$ is $T'[V/x]$ 
with the conditions $S \pared S'$, $T \pared T'$, $x.T \pared L$, and $(S).\alpha \pared V$. 
Recall that a critical pair in $\DCmu$ occurs in this shape. 
This case is most important to see that this critical pair is avoided in the weak call-by-value $\DCmu$ .

From the definition of the parallel reduction and $(S).\alpha \pared V$, 
we have $S = M\bullet \alpha$, $M \pared V$, and $\alpha$ is not free in $M$. 
Then, from $M\bullet \alpha = S \pared S'$, we have the following two cases: \ 
(i)\ $S' = M'\bullet \alpha$ and $M \pared M'$ for some $M'$, or \ 
(ii)\ $M = (S_0).\beta$, $S' = {S_0}'[\alpha/\beta]$, and $S_0 \pared {S_0}'$ for some $S_0$ and ${S_0}'$. 
From the condition $x.(T) \pared L$, we also have the following two cases: \ 
(a)\ $T = x\bullet K$, $x$ is not free in $K$, and $K \pared L$ for some $K$ and $L$, or\ 
(b)\ $L = x.(T'')$ and $T \pared T''$ for some $T''$. 

The case of (i). \ 
We have $D_1 = (S')[L/\alpha] = (M'\bullet \alpha)[L/\alpha] = M'\bullet L$. 
By the induction hypothesis and $V \Leftarrow M \Rightarrow M'$, 
there exists a term $W$ such that $V \Rightarrow W \Leftarrow M'$. 
From Lemma~\ref{lem:pared_basic}~(1), $W$ is a value. 
We then consider the subcases (a) and (b). 

The subcase of (a). \ 
From the condition $K \Rightarrow L$ and $T = x\bullet K$, we have 
$(x\bullet L) \Leftarrow T \Rightarrow T'$. 
By the induction hypothesis, 
there exists a statement $\tilde{T}$ such that $(x\bullet L) \Rightarrow \tilde{T} \Leftarrow T'$. 
Hence we have $D_1 = (x\bullet L)[M'/x] \Rightarrow \tilde{T}[W/x] \Leftarrow T'[V/x] = D_2$ 
from $M' \Rightarrow W \Leftarrow V$ and Lemma~\ref{lem:pared_basic}~(3). 

The subcase of (b). \ 
By the induction hypothesis and $T' \Leftarrow T \Rightarrow T''$, 
there exists $\tilde{T}$ such that $T' \Rightarrow \tilde{T} \Leftarrow T''$. 
Hence we have $D_2 = T'[V/x] \Rightarrow \tilde{T}[W/x]$ 
by Lemma~\ref{lem:pared_basic}~(3) and $V \Rightarrow W$. 
We also have $D_1 = M'\bullet L = M'\bullet x.(T'') \Rightarrow \tilde{T}[W/x]$ 
from $M'\Rightarrow W$ and $T''\Rightarrow \tilde{T}$. 
Therefore $D_1 \Rightarrow \tilde{T}[W/x] \Leftarrow D_2$ holds. 

The case of (ii). \ 
We first claim that, for any $S$ and $V$, if $(S).\alpha \pared V$, then 
there is some $M$ such that 
$S = M \bullet \alpha$, $M\pared V$, and $\alpha$ is not free in $M$. 
This claim is easily obtained from the definition of the parallel reduction. 
In this case, we have $V \Leftarrow M = (S_0).\beta \Rightarrow ({S_0}').\beta$. 
By the induction hypothesis and Lemma~\ref{lem:pared_basic}~(1), 
there is a value $W$ such that $V \Rightarrow W \Leftarrow ({S_0}').\beta$. 
Then, there exists a $N$ such that ${S_0}' = N\bullet \beta$, $N \Rightarrow W$, and $\beta$ is not free in $N$ 
from the above claim. 
Hence we have $D_1 = S'[L/\alpha] = {S_0}'[\alpha/\beta][L/\alpha] = {S_0}'[L/\beta] = (N\bullet \beta)[L/\beta] 
= N\bullet L$. We then consider the subcases (a) and (b). 

The subcase of (a). \ 
By the induction hypothesis, 
there exists a statement $\tilde{T}$ such that $(x\bullet L) \Rightarrow \tilde{T} \Leftarrow T'$. 
Hence we have $D_1 = (x\bullet L)[N/x] \Rightarrow \tilde{T}[W/x] \Leftarrow T'[V/x] = D_2$ 
from $N \Rightarrow W \Leftarrow V$ and Lemma~\ref{lem:pared_basic}~(3). 

The subcase of (b). \ 
By the induction hypothesis and $T' \Leftarrow T \Rightarrow T''$, 
there exists $\tilde{T}$ such that $T' \Rightarrow \tilde{T} \Leftarrow T''$. 
Hence we have $D_2 = T'[V/x] \Rightarrow \tilde{T}[W/x]$ 
by Lemma~\ref{lem:pared_basic}~(3) and $V \Rightarrow W$. 
We also have $D_1 = N\bullet L = N\bullet x.(T'') \Rightarrow \tilde{T}[W/x]$ 
from $N\Rightarrow W$ and $T''\Rightarrow \tilde{T}$. 
Therefore $D_1 \Rightarrow \tilde{T}[W/x] \Leftarrow D_2$ holds. 

The other cases are also proved by the induction hypothesis.\smallskip
\qed

\noindent From Lemma~\ref{lem:pared_CR}, we can obtain the Church-Rosser property of the weak call-by-value $\DCmu$. 

\begin{prop}
The reduction relations $\wCBVred$ and $\wCBNred$ of $\DCmuV0$ satisfy the Church-Rosser property. 
\end{prop}
\proof
We first show the Church-Rosser property of $\wCBVred$. 
Suppose that $D \wCBVred^* D'$ and 
$D \wCBVred^* D''$ hold. 
We will show that 
there exists some $E$ such that $D' \wCBVred^* E$ 
and $D'' \wCBVred^* E$. 
 We have 
$D = D_{00} \wCBVred D_{01} 
\wCBVred \ldots \wCBVred D_{0n} = D'$ 
and 
$D \wCBVred D_{10} \wCBVred \ldots \wCBVred D_{1m} = D''$ 
for some $n, m\ge 0$. 
By Lemma~\ref{lem:pared_CR}~(1), 
$D \pared D_{01} \pared \ldots \pared D_{0n} = D'$ and 
$D \pared D''_{10} \pared \ldots \pared D''_{1m} = D''$ hold. 
By the diamond property of $\pared$, 
there exists $D_{(i+1)(j+1)}$ such that $D_{i(j+1)} \Rightarrow D_{(i+1)(j+1)} \Leftarrow D_{(i+1)j}$ 
for each $0 \le i \le n-1$ and $0 \le j \le m-1$. 
Hence we have 
$D' = D_{0n} \pared D_{1n} \pared \ldots\pared D_{mn}$ and $D'' = D_{m0} \pared D_{m1} \pared \ldots \pared D_{mn}$. 
By Lemma~\ref{lem:pared_CR}~(2), we can replace $\pared$ by $\wCBVred^*$. 
Therefore, we have $D' \wCBVred^* D_{mn}$ and $D'' \wCBVred^* D_{mn}$. 

The Church-Rosser property of $\wCBNred$ is shown 
by the former result and the duality between $\wCBVred$ and $\wCBNred$ 
(Proposition~\ref{prop:dual_wCBVwCBN}). 
\qed

We will next define the call-by-value and the call-by-name $\DCmu$, 
which we call $\CBVDCmu$ and $\CBNDCmu$. 
The types, expressions, and typing rules of $\CBVDCmu$ and $\CBNDCmu$ 
are the same as them of $\DCmu$. 

The reduction relation of $\CBVDCmu$ is obtained by adding $\varsigma$-rules to the weak call-by-value $\DCmu$. 

\begin{defi}[Reduction relation of $\CBVDCmu$]\rm
The reduction relation $\CBVred$  of $\CBVDCmu$ is defined by 
the compatible closure of the reduction rules of the weak call-by-value $\DCmu$ 
and $(\varsigma \land_1)_v$, $(\varsigma \land_2)_v$, $(\varsigma \vee_1)_v$, $(\varsigma \vee_2)_v$, 
and the following reduction rules: 
\[
\begin{array}{ll}
(\varsigma\mu)_v&
\In^{\mu X.C}\<\nvalue\>
\CBVred
(\nvalue \bullet x.(\In^{\mu X.C}\<x\>\bullet \alpha)).\alpha,\\
(\varsigma\nu)_v&
\Coitr^A_y\langle M,\nvalue\rangle
\CBVred
(\nvalue \bullet x.(\Coitr^A_y\langle M,x\rangle \bullet \alpha)).\alpha,
\end{array}
\]
where $\nvalue$ is not a value of $\DCmu$, 
and the variable $x$ and the covariable $\alpha$ in 
$(\varsigma\mu)_v$, $(\varsigma\nu)_v$ are fresh. 
\end{defi}

We sometimes write $(\beta)_v$ to mean 
$(\beta\land_1)_v$, $(\beta\land_2)_v$, $(\beta\vee_1)_v$, $(\beta\vee_2)_v$, 
$(\beta\neg)_v$, $(\beta\mu)_v$, $(\beta\nu)_v$, $(\beta L)_v$, or $(\beta R)_v$-rule. 
We write $(\eta)_v$ to mean $(\eta L)_v$ or $(\eta L)_v$-rule. 
We also write $(\varsigma)_v$ to mean $(\varsigma\land_1)_v$, $(\varsigma\land_2)_v$, 
$(\varsigma\vee_1)_v$, $(\varsigma\vee_2)_v$, $(\varsigma\mu)_v$, or $(\varsigma\nu)_v$-rule. 

The reduction relation of $\CBNDCmu$ is obtained by adding $\varsigma$-rules to the weak call-by-name $\DCmu$. 

\begin{defi}[Reduction relation of $\CBNDCmu$]\rm
The reduction relation $\CBNred$  of $\CBNDCmu$ is defined by 
the compatible closure of the reduction rules of the weak call-by-name $\DCmu$ 
and $(\varsigma \land_1)_n$, $(\varsigma \land_2)_n$, $(\varsigma \vee_1)_n$, $(\varsigma \vee_2)_n$, 
and the following reduction rules: 
\[
\begin{array}{ll}
(\varsigma\mu)_n&
\Itr^A_\beta[K,\ncovalue]
\CBNred
x.((x \bullet \Itr^A_\beta[K,\alpha]).\alpha \bullet \ncovalue),\\
(\varsigma\nu)_n&
\Out^{\nu X.C}[\ncovalue]
\CBNred
x.((x \bullet \Out^{\nu X.C}[\alpha]).\alpha\bullet \ncovalue),
\end{array}
\]
where $\ncovalue$ is not a covalue of $\DCmu$, 
and the variable $x$ and the covariable $\alpha$ in 
$(\varsigma\mu)_n$, $(\varsigma\nu)_n$ are fresh. 
\end{defi}

From the above definitions, $\CBVDCmu$ includes the weak call-by-value $\DCmu$, 
and $\CBNDCmu$ includes  the weak call-by-name $\DCmu$. 
That is, the following lemma holds. 

\begin{lem}\label{lem:wCBVinCBV}
Let $D$ and $E$ be expressions of $\DCmu$. Then, the following claims hold. 
\begin{enumerate}[\rm(1)]
\item[\rm(1)]
If $D \wCBVred E$, then $D \CBVred E$. 
\item[\rm(2)]
If $D \wCBNred E$, then $D \CBNred E$. 
\end{enumerate}
\end{lem}

The call-by-value $\DCmu$ is dual to the call-by-name $\DCmu$. 

\begin{prop}[Duality between call-by-value and call-by-name in $\DCmu$]\label{prop:dual_CBVCBN}
Let $D$ and $E$ be expressions of $\DCmu$. Then, \ 
$D \CBVred E$ iff $(D)^\circ \CBNred (E)^\circ$, 
where $(-)^\circ$ is the duality transformation of $\DCmu$ defined in the section~3. 
\end{prop}

The call-by-value and call-by-name $\DCmu$ satisfy both the Church-Rosser and strong normalization properties. 
We will concentrate to show these properties of $\CBVDCmu$. 
The proof will be performed by giving a transformation from $\CBVDCmu$ into the weak call-by-value $\DCmu$. 
The transformation $(-)^\snow$ given as follows. 

\begin{defi}\rm
Let $D$ be a expression of $\DCmu$. 
The expression $(D)^\snow$ of $\DCmu$ is defined inductively as follows. 
\[
\begin{array}{l}
(x)^\snow = x, \\
(\<V,W\>)^\snow = \<(V)^\snow, (W)^\snow\>, \\
(\<V,\nvalueb\>)^\snow = ((\nvalueb)^\snow \bullet y.(\<(V)^\snow, y\>\bullet \alpha)).\alpha, \\
(\<\nvalue,W\>)^\snow = ((\nvalue)^\snow \bullet x.(\<x,(W)^\snow\>\bullet \alpha)).\alpha, \\
(\<\nvalue,\nvalueb\>)^\snow 
= 
\Bigl((\nvalue)^\snow\bullet x.\bigl(((\nvalueb)^\snow\bullet y.(\<x,y\>\bullet \beta)).\beta\bullet \alpha\bigr)\Bigr).\alpha, \\
(\<V\>\Inl)^\snow = \<(V)^\snow\>\Inl, \\
(\<\nvalue\>\Inl)^\snow = ((\nvalue)^\snow \bullet x.(\<x\>\Inl\bullet \alpha)).\alpha, \\
(\<V\>\Inr)^\snow = \<(V)^\snow\>\Inr, \\
(\<\nvalue\>\Inr)^\snow = ((\nvalue)^\snow \bullet x.(\<x\>\Inr\bullet \alpha)).\alpha, \\
([K]\Not)^\snow = [(K)^\snow]\Not, \\
(\In^{\mu X.A}\<\nvalue\>)^\snow = ((\nvalue)^\snow \bullet x.(\In^{\mu X.A}\<x\>\bullet \alpha)).\alpha, \\
(\In^{\mu X.A}\<V\>)^\snow = \In^{\mu X.A}\<(V)^\snow\>, \\
(\Coitr^A_z\<M,V\>)^\snow = \Coitr^A_z\<(M)^\snow, (V)^\snow\>, \\
(\Coitr^A_z\<M,\nvalueb\>)^\snow 
= ((\nvalueb)^\snow \bullet y.(\Coitr^A_z\<(M)^\snow, y\>\bullet \alpha)).\alpha, \\
((S).\alpha)^\snow = ((S)^\snow).\alpha, \\
(\alpha)^\snow = \alpha, \\
([K,L])^\snow = [(K)^\snow, (L)^\snow], \\
(\Fst[K])^\snow = \Fst[(K)^\snow], \\
(\Snd[K])^\snow = \Snd[(K)^\snow], \\
(\Not\<M\>)^\snow = \Not\<(M)^\snow\>, \\
(\Out^{\nu X.A}[K])^\snow = \Out^{\nu X.A}[(K)^\snow], \\
(\Itr^A_\gamma[K,L])^\snow = \Itr^A_\gamma[(K)^\snow, (L)^\snow], \\
(x.(S))^\snow = x.((S)^\snow),  \quad\mbox{and}\\
(M\bullet K)^\snow = (M)^\snow\bullet (K)^\snow, 
\end{array}
\]
 where $V$ and $W$ are values, 
$\nvalue$ and $\nvalueb$ are not values, and $x$,  $y$, $\alpha$, $\beta$ are fresh. 
\end{defi}

We need the redundant definition of $(\<\nvalue,\nvalueb\>)^\snow$ 
for a technical reason, and it 
is necessary in order to show Proposition~\ref{prop:transCBV}. 

The transformation $(-)^\snow$ preserves typing. 
\begin{prop}\label{prop:transCBV_typing}
Let $M$ be a term, $K$ be a coterm, and $S$ be a statement of $\DCmu$. 
The following claims hold. 
\begin{enumerate}[\rm(1)]
\item[\rm(1)]
If $\Gamma \vdash_{\Duca\mu\nu} \Delta \mid M:A$ is provable, 
then $\Gamma \vdash_{\Duca\mu\nu} \Delta \mid (M)^\snow:A$ holds. 
\item[\rm(2)]
If $K:A \mid \Gamma \vdash_{\Duca\mu\nu} \Delta$ is provable, 
then $(K)^\snow:A \mid \Gamma \vdash_{\Duca\mu\nu} \Delta$ holds. 
\item[\rm(3)]
If $\Gamma \mid S \vdash_{\Duca\mu\nu} \Delta$ is provable, 
then $\Gamma \mid (S)^\snow \vdash_{\Duca\mu\nu} \Delta$ holds. 
\end{enumerate}
\end{prop}
\proof
They are shown simultaneously by induction on $M$, $K$, and $S$. 
\qed

The transformation $(-)^\snow$ satisfies the following basic properties. 

\begin{lem}\label{lem:transCBV_basic}
Let $V$ be a value,  $M$ and $N$ be terms, $D$ be an expression of $\DCmu$. 
Then the following claims hold. 
\begin{enumerate}[\rm(1a)]
\item[\rm(1)]
$M$ is a value  iff $(M)^\snow$ is a value. 
\item[\rm(2a)]
$\<(M)^\snow, (N)^\snow\> \wCBVred^* (\<M,N\>)^\snow$. 
\item[\rm(2b)]
$\<(M)^\snow\>\Inl \wCBVred^* (\<M\>\Inl)^\snow$, 
and $\<(M)^\snow\>\Inr \wCBVred^* (\<M\>\Inr)^\snow$. 
\item[\rm(2c)]
$\In^{\mu X.A}\<(M)^\snow\> \wCBVred^* (\In^{\mu X.A}\<M\>)^\snow$. 
\item[\rm(2d)]
$\Coitr^A_z\<(M)^\snow,(N)^\snow\> \wCBVred (\Coitr^A_z\<M,N\>)^\snow$. 
\item[\rm(3a)]
$\Map^{X.A}_{B,C,x}\{(M)^\snow, (N)^\snow\} \wCBVred^* (\Map^{X.A}_{B,C,x}\{M, N\})^\snow$. 
\item[\rm(3b)]
$\Map^{X.A}_{B,C,\alpha}\{(K)^\snow, (L)^\snow\} \wCBVred^* (\Map^{X.A}_{B,C,\alpha}\{K, L\})^\snow$. 
\item[\rm(4)]
$D \CBVred^* (D)^\snow$. 
\end{enumerate}
\end{lem}
\proof
The claim (1) is shown by the definition of $(-)^\snow$. 
The claims (2a), (2b), (2c), and (2d) are shown by (1) and $\varsigma$-rules. 
The claims (3a) and (3b) are shown by induction on $||C||_X$ using (2a), (2b), (2c), and (2d). 
The claim (4) is shown by induction on $D$. 
\qed

The transformation $(-)^\snow$ preserves substitution of a value for a variable, and of a coterm for a covariable. 

\begin{lem}\label{lem:transCBV_subst}
$(D[V/x])^\snow = (D)^\snow[(V)^\snow/x]$ 
and 
$(D[K/\alpha])^\snow = (D)^\snow[(K)^\snow/\alpha]$. 
\end{lem}
\proof
The former claim is shown by induction on $D$ using  Lemma \ref{lem:transCBV_basic} (1). 
The latter one is shown by induction on $D$. 
\qed

The transformation $(-)^\snow$ translates one step reduction of $\CBVred$ into 
zero or more steps reduction of $\wCBVred$. 

\begin{prop}\label{prop:transCBV}
$D \CBVred E$ implies $(D)^\snow \wCBVred^* (E)^\snow$. 
In particular, if $D \CBVred E$ by $(\beta)_v$ or $(\eta)_v$, 
then $(D)^\snow \wCBVred^+ (E)^\snow$ holds. 
\end{prop}
\proof
The claim is shown by induction on the definition of $\CBVred$. 
We show the cases of $(\beta L)_v$, $(\beta\mu)_v$, and $(\varsigma\land_1)_v$. 

The case of $(\beta L)_v$ is proved by Lemma~\ref{lem:transCBV_subst}.
We have $(V\bullet x.(S))^\snow = (V)^\snow \bullet x.((S)^\snow) \wCBVred (S)^\snow[(V)^\snow/x]$. 
By Lemma~\ref{lem:transCBV_subst}, the last statement is $(S[V/x])^\snow$. 

The case of $(\beta\mu)_v$ is proved by Lemma~\ref{lem:transCBV_basic}~(3b).
We have $(\In^{\mu X.C}\langle V\rangle \bullet \Itr^A_\alpha[K,L])^\snow
= 
\In^{\mu X.C}\langle (V)^\snow \rangle \bullet \Itr^A_\alpha[(K)^\snow,(L)^\snow]
\wCBVred
((V)^\snow \bullet \Map^{X.C}_{\mu X.C, A,\beta}\{\, \Itr^A_\alpha[(K)^\snow,\beta],\, (K)^\snow\,\}).\alpha \bullet (L)^\snow
= 
((V)^\snow \bullet \Map^{X.C}_{\mu X.C, A,\beta}\{\, (\Itr^A_\alpha[K,\beta])^\snow,\, (K)^\snow\,\}).\alpha \bullet (L)^\snow$. 
By Lemma~\ref{lem:transCBV_basic}~(3b), the last statement is reduced to 
$((V)^\snow \bullet (\Map^{X.C}_{\mu X.C, A,\beta}\{\, \Itr^A_\alpha[K,\beta],\, K\,\})^\snow).\alpha \bullet (L)^\snow$ by $\wCBVred^*$. 
Therefore this statement is 
$((V \bullet \Map^{X.C}_{\mu X.C, A,\beta}\{\, \Itr^A_\alpha[K,\beta],\, K\,\}).\alpha \bullet L)^\snow$. 

The case of $(\varsigma\land_1)_v$ is proved by the definition of $(-)^\snow$. 
We consider the subcase of 
$\<\nvalue,\nvalueb\> \CBVred (\nvalue\bullet x.(\<x,\nvalueb\>\bullet \alpha)).\alpha$, 
where $\nvalue$ and $\nvalueb$ are not values. 
Hence we have $(\<\nvalue,\nvalueb\>)^\snow 
= 
\Bigl((\nvalue)^\snow\bullet x.\bigl(((\nvalueb)^\snow\bullet y.(\<x,y\>\bullet \beta)).\beta\bullet \alpha\bigr)\Bigr).\alpha 
=
\bigl((\nvalue)^\snow\bullet x.\bigl((\<x,\nvalueb\>)^\snow\bullet \alpha\bigr)\bigr).\alpha 
= 
\bigl(\nvalue\bullet x.\bigl(\<x,\nvalueb\>\bullet \alpha\bigr)\bigr).\alpha\bigr)^\snow $. 
The other subcase of $(\varsigma\land_1)_v$ 
 for $\<{\cal N},V\>$ with a non-value $\cal N$
is shown in the similar way. 

The other cases are also shown by the induction hypothesis.
\qed

Then we can show the Church-Rosser property of $\CBVred$ and $\CBNred$. 

\begin{thm}
The reduction relations $\CBVred$ of $\CBVDCmu$ 
and $\CBNred$ of $\CBNDCmu$ satisfy the Church-Rosser property. 
\end{thm}
\proof
We first show the Church-Rosser property of $\CBVred$.

Assume that $D \CBVred^* D'$ and $D \CBVred^* D''$ hold. 
By Proposition~\ref{prop:transCBV}, we have
$(D)^\snow \wCBVred^* (D')^\snow$ and $(D)^\snow \wCBVred^* (D'')^\snow$. 
By the Church-Rosser property of $\wCBVred$, 
there exists $E$ such that $(D')^\snow \wCBVred^* E$ and $(D'')^\snow \wCBVred^* E$. 
Therefore, by Lemma~\ref{lem:wCBVinCBV}~(1) and Lemma~\ref{lem:transCBV_basic}~(4), 
we have $D' \CBVred^* (D')^\snow \CBVred^* E$ and $D'' \CBVred^* (D'')^\snow \CBVred^* E$. 

The Church-Rosser property of $\CBNred$ is shown 
by the former result and the duality between $\CBVred$ and $\CBNred$ 
(Prop~\ref{prop:dual_CBVCBN}). 
\qed

We will prove strong normalization of $\CBVDCmu$ and $\CBNDCmu$. 
This property is shown by using the strong normalization result of 
the weak call-by-value and the weak call-by-name $\DCmu$ (Proposition~\ref{prop:SN_wCBVwCBN}). 

We define the following rank of expressions in $\DCmu$. 
This rank is used to show that there is no infinite sequence of $\varsigma$-rules. 

\begin{defi}\rm
Let $D$ be an expression in $\DCmu$. The rank $r(D)$ of $D$ is defined by: 
\[
\begin{array}{l}
r(x) = r(\alpha) = 0,  \\
r([K]\Not) = r(\Fst[K]) = r(\Snd[K]) = r(\Out^{\nu X.A}[K]) = r(K), \\
r([K,L]) = r(\Itr^A_\alpha[K,L]) = r(K)+r(L), \\
r(\Not\<M\>) = r(M), \\
r(\<\nvalue,\nvalueb\>) = r(\nvalue)+r(\nvalueb)+2, \\
r(\<\nvalue,V\>) = r(\<V,\nvalue\>) = r(\nvalue)+r(V) + 1, \\
r(\<V,W\>) = r(V) + r(W), \\
r(\<\nvalue\>\Inl) = r(\<\nvalue\>\Inr) = r(\In^{\mu X.A}\<\nvalue\>) = r(\nvalue) + 1, \\
r(\<V\>\Inl) = r(\<V\>\Inr) = r(\In^{\mu X.A}\<V\>) = r(V), \\
r(\Coitr^A_x\<M,\nvalueb\>) = r(M) + r(\nvalueb) + 1, \\
r(\Coitr^A_x\<M,V\>) = r(M) + r(V), \\
r(x.(S)) = r((S).\alpha) = r(S), \quad \mbox{and}\\
r(M\bullet K) = r(M) + r(K), 
\end{array}
\]
where $V$ and $W$ are values, and $\nvalue$ and $\nvalueb$ are not values. 
\end{defi}

The rank $r(D)$ counts the number of redexes of $(\varsigma\land_1)_v$, $(\varsigma\land_2)_v$, 
$(\varsigma\vee_1)_v$, $(\varsigma\vee_2)_v$, $(\varsigma\mu)_v$, and $(\varsigma\nu)_v$-rules. 
We write $D \longrightarrow_{\varsigma_v} E$ when $D$ is reduced to $E$ by one step $(\varsigma)_v$-reduction. 

\begin{lem}\label{lem:varsigma}
Let $D$ and $E$ be expressions of $\DCmu$. Then, the following claims hold. 
\begin{enumerate}[\rm(1)]
\item[\rm(1)]
If $D \longrightarrow_{\varsigma_v} E$, then $r(D) > r(E)$. 
\item[\rm(2)]
There is no infinite sequence of $(\varsigma)_v$-reduction.
\end{enumerate}
\end{lem}
\proof
The claim (1) is shown by induction on $D$. The claim (2) is shown by (1). 
\qed

We then show strong normalization of $\CBVDCmu$ and $\CBNDCmu$. 

\begin{thm}[Strong normalization of $\CBVDCmu$ and $\CBNDCmu$]
The following claims hold. 
\begin{enumerate}[\rm(1)]
\item[\rm(1)]
Every typable expression is strongly normalizing in $\CBVDCmu$. 
\item[\rm(2)]
Every typable expression is strongly normalizing in $\CBNDCmu$. 
\end{enumerate}
\end{thm}
\proof
We first show the call-by-value case. 
Assume that $D$ is typable in $\DCmu$ and there is an infinite reduction sequence 
\[
D \CBVred D_1 \CBVred \ldots
\]
starting from $D$.
Then $(D)^\snow$ is typable by Proposition~\ref{prop:transCBV_typing}, 
and 
we have 
\[
(D)^\snow \wCBVred^* (D_1)^\snow \wCBVred^* \ldots
\]
by Proposition~\ref{prop:transCBV}. 
From the strong normalization result of the weak call-by-value $\DCmu$ (Proposition~\ref{prop:SN_wCBVwCBN}), 
there is some $D_k$ such that 
\[
(D_k)^\snow = (D_{k+1})^\snow = \ldots. 
\]
By the latter part of Proposition~\ref{prop:transCBV}, 
we have the following infinite sequence of $(\varsigma)_v$-reduction:
\[
D_k \longrightarrow_{\varsigma_v} D_{k+1}\longrightarrow_{\varsigma_v} \ldots.
\]
This contradicts Lemma~\ref{lem:varsigma}~(2). 

The call-by-name case is proved by strong normalization of $\CBVDCmu$ and 
the duality between $\CBVDCmu$ and $\CBNDCmu$ 
(Proposition~\ref{prop:dual_CBVCBN}). 
\qed

\section{Conclusion}

We have introduced the non-deterministic system $\DCmu$ by extending
the dual calculus given in \cite{Wad03:01} with inductive types and
coinductive types.  Besides the same duality of the original dual
calculus, we have shown the duality of inductive and coinductive
types, by giving the involution that maps terms and coterms for
inductive types to coterms and terms of coinductive types respectively
and vice versa, and maps their reduction rules to each other.  We have
proved its strong normalization by translating it into the
second-order dual calculus $\DC2$.

The second-order dual calculus $\DC2$ also have been introduced. 
Its strong normalization have been shown 
by translating it into the second-order symmetric lambda calculus.

We have finally introduced the call-by-value system $\CBVDCmu$ 
and the call-by-name system $\CBNDCmu$
of the dual calculus with inductive and coinductive types.
We have
shown the duality of call-by-value and call-by-name with inductive
and coinductive types,
their Church-Rosser property, and their strong normalization. 
Their strong normalization have been shown by translating them into $\DCmu$.

The first author introduced the call-by-value and call-by-name dual calculi 
with recursive types~\cite[section 4.2]{Kim07:02}. 
In these systems, a recursive type ${\tt rec}\ X.A$ can be defined for any type $A$. 
If we assume that ${\tt rec}\ X.A$ can be defined only if every $X$ positively occurs in $A$, 
then we can define two provability-preserving transformations 
from the dual calculi with recursive types into $\DCmu$. 
The one translates a recursive type to an inductive type, 
and the other translates a recursive type to a coinductive type. 
We could not straightforwardly show that these transformations 
preserve reductions (or equations)  since some additional rules 
such as $\eta$-rules for connectives seem to be required.  
This problem would be future work.

The duality of call-by-value and call-by-name in $\lambda\mu$-calculus
is shown by using the dual calculi in \cite{Wad05:01}.
Since our systems $\CBVDCmu$ and $\CBNDCmu$ are extensions of his dual
calculi, we could show the duality of call-by-value and call-by-name
in $\lambda\mu$-calculus with inductive and coinductive definitions, by using
our systems $\CBVDCmu$ and $\CBNDCmu$. It would be future work.

A reduction-based duality between call-by-value and call-by-name
in the $\lambda\mu$-calculi was presented in \cite{Kim07:01},
by refining Wadler's result~\cite{Wad05:01}. 
Extending the result given in \cite{Kim07:01} with
inductive and coinductive types would be future work.

Our systems use the iteration for inductive types.
An extension of the iteration to primitive recursion would be
future work.

A CPS translation from the dual calculus to $\lambda$-calculus was
given in \cite{Wad03:01}.
Extending this CPS translation to the systems with inductive and
coinductive types would be future work.

\section*{Acknowledgment}

We would like to thank Professor Philip Wadler
for discussions and suggestions.
We would also like to thank Dr. Alwen Tiu,
and Professor Dieter Spreen for discussions.
We would also like to thank anonymous referees for valuable comments.


\end{document}